%% file: main.tex
\newif\ifFullVersion
\newif\ifShortVersion
\newif\ifAllProofs
\newif\ifIACRVersion

\FullVersiontrue

\ifIACRVersion
\documentclass[envcountsect, envcountsame,runningheads]{llncs}
\else
\documentclass{llncs}
\fi

\usepackage[T1]{fontenc}
\usepackage[english]{babel}
\usepackage[utf8x]{inputenc}

\usepackage[section]{placeins}
\usepackage{epsfig,epstopdf,url,array}
\usepackage{mathtools}
\usepackage{color}
\usepackage{url}
\usepackage{hyperref}
\usepackage{tabularx,booktabs,multirow}
\usepackage{pgf}
\usepackage{tikz}
\usetikzlibrary{arrows,automata,cd, graphs, positioning, fit}
\usepackage{enumitem}
\usepackage{epigraph}
\usepackage{bm} %
\usepackage{bbm} %
\usepackage{float}
\usepackage{xspace}
\usepackage{xcolor,colortbl}
\usepackage{stmaryrd}
\usepackage{comment}
\usepackage{subcaption}
\usepackage[ruled]{algorithm2e}
\usepackage{algpseudocode}
\usepackage{orcidlink}
\usepackage{pdflscape} %
\usepackage{array}     %
\usepackage{adjustbox}

\usepackage[colorinlistoftodos]{todonotes}
\definecolor{darkpink}{HTML}{AA336A}

\input{symbols}

\usepackage[innerleftmargin=5pt,innerrightmargin=5pt]{mdframed} %

\makeatletter
\renewcommand*{\l@title}[2]{}
\renewcommand*{\l@author}[2]{}
\makeatother
\begin{document}

\title{Batch, Pack, and Prove: More Efficient Verifiable Computation for CKKS}

\author{
	Ignacio Cascudo \inst{1} \orcidlink{0000-0001-5520-5386}
	\and
	Anamaria Costache \inst{2} \orcidlink{0000-0001-8793-6116}
	\and
	Daniele Cozzo \inst{1} \orcidlink{0000-0001-5289-3769}
	\and
	Dario Fiore \inst{1} \orcidlink{0000-0001-7274-6600}
	\and
	Antonio Guimar\~aes \inst{1} \orcidlink{0000-0001-5110-6639}
	\and
	Eduardo Soria-Vazquez \inst{3} \orcidlink{0000-0002-4882-0230}
}
\ifIACRVersion
\authorrunning{Cascudo, Costache, Cozzo, Fiore, Guimar\~aes, and Soria-Vazquez}
\fi

\institute{
	IMDEA Software Institute, Madrid, Spain. \\
	\email{firstname.lastname@imdea.org}
	\and
	École Polytechnique, Paris, France.\\
	\email{lastname@lix.polytechnique.fr}
	\and
	Technology Innovation Institute, Abu Dhabi, UAE.\\
	\email{firstname.lastname@tii.ae}
}

\maketitle

\begin{abstract}
An important challenge in homomorphic encryption (HE) is ensuring the integrity of the
encrypted computations. Although verifiable computation can address this in
theory, making it practical is far from trivial. [Cascudo et al.~CRYPTO 2025] recently proposed a
solution to this problem for the CKKS scheme. In particular, they propose new succinct arguments
that efficiently prove arithmetic circuits and range checks over the ring $\Rq =
\ZZ_q[Y]/(Y^N+1)$, and combine them to prove computations on CKKS ciphertexts. Despite
the important progress, their work faces scalability challenges: the size of the polynomial
commitment proofs, a large $\bigO{N}$ overhead in range proofs, and the lack of an explicit method
for efficient---sublinear in circuit size---verification.

In this paper we present new techniques that remove all these bottlenecks. First, we provide a more
efficient arithmetization of CKKS computations. Second, we introduce BatchFold, a new polynomial
commitment for $\Rq$ that supports batched evaluations and achieves shorter proofs. Third, we
design packing, a novel algebraic technique that eliminates the need to commit to massive polynomials
and removes the $\bigO{N}$ overhead in range proofs. Finally, we introduce SparsePack, an efficient
protocol for
proving the evaluations of sparse, circuit-dependent polynomials, leveraging both packing and
batched evaluations. We evaluate the efficiency of our solution on a generic neural network
architecture, achieving proofs several orders of magnitude smaller and faster than prior approaches.
\end{abstract}

\ifFullVersion
\setcounter{tocdepth}{2}
\tableofcontents
\newpage
\fi

\input{introduction}

\input{preliminaries}

\input{packing}

\input{new-range-proof-rq}

\input{ac_plusplus}

\input{sparse-poly}

\input{basefold.tex}

\input{experiments.tex}

\subsubsection*{Acknowledgements}
We would like to thank the anonymous reviewers of Asiacrypt 2026 for the insightful comments and discussions that allowed us to improve the presentation in Section \ref{sec:proof-friendly-CKKS} and our results in Section \ref{sec:PIOP_AC}.
This work is partially funded by the grants PID2022-142290OB-I00, JDC2022-049711-I, JDC2024-055789-I, and CEX2024-001471-M, funded by MCIN/AEI/10.13039/501100011033/ FEDER, UE and ESF+, and it is part of the PICOCRYPT project that has received funding from the European Research Council (ERC) under the European Union's Horizon 2020 research and innovation programme (Grant agreement No. 101001283), and of the VERIFHE project funded by the European Union (GA: 101287502). Views and opinions expressed are however those of the author(s) only and do not necessarily reflect those of the European Union or the European Research Council. Neither the European Union nor the granting authority can be held responsible for them.

\ifIACRVersion
	\bibliographystyle{splncs04}
\else
\bibliographystyle{alpha}
\fi
\bibliography{main}

\appendix

\section{Additional Preliminaries}\label{apx:preliminaries}

\headingb{PIOP Efficiency.}
We consider the following efficiency metrics:
prover's (resp. verifier's) running time $\mathsf{T}_{\iopP}$ (resp. $\mathsf{T}_{\iopV}$);
number of rounds $\mathsf{r}$;
number of oracles sent by the prover $\mathsf{o}$;
size $\mathsf{s}$ of the proof oracles (i.e., the sum of length of all the polynomials sent by the prover);
number of verifier's queries $\mathsf{q}$ to the oracles.

\input{polycom_def}
\input{rbr-soundness}
\input{sumcheck}
\input{pfckks}

\input{apx-packing}

 \input{apx-range-proof}

\input{new-PIOP-range-proof}

\input{app_GKR}

\input{apx-sparsepack}

\input{appendix-basefold}

\input{extra-results}

\end{document}

%% file: symbols.tex
\usepackage{amsmath}
\usepackage{amsfonts,amssymb}
\usepackage{mathtools, nccmath}
\usepackage[advantage,
adversary,
asymptotics,
complexity,
events,
keys,
lambda,
landau,
mm,
operators,
primitives,
probability,
notions,
sets]{cryptocode}

\usepackage{wasysym}

\createprocedureblock{procb}{center ,boxed}{}{}{}

\newcommand{\headingb}[1]{\medskip\noindent{\bf #1}}

\newcommand{\pres}{\ensuremath{\mathsf{p}}} %

\newcommand{\Tvar}{\mathtt{T}}
\newcommand{\Xvar}{\mathtt{X}}
\newcommand{\Yvar}{\mathtt{Y}}
\newcommand{\Zvar}{\mathtt{Z}}
\newcommand{\vTvar}{\vec{\Tvar}}
\newcommand{\vXvar}{\vec{\Xvar}}
\newcommand{\vYvar}{\vec{\Yvar}}
\newcommand{\vZvar}{\vec{\Zvar}}

\newcommand{\pXmult}[1][]{\mathtt{Xmult}_{#1}}
\newcommand{\prescon}[1][]{\mathtt{rescon}_{#1}}
\newcommand{\pbdcon}[1][]{\mathtt{bdcon}_{#1}}
\newcommand{\pcopycon}[1][]{\mathtt{copycon}_{#1}}
\newcommand{\pevk}[1][]{\mathtt{evk}_{#1}}
\newcommand{\pcomb}[1][]{\mathtt{C}_{#1}}
\newcommand{\peq}{\mathtt{eq}}

\newcommand{\tgate}[1][]{\widetilde{\mathtt{gate}}_{#1}}
\newcommand{\Vi}[1][i]{\ensuremath{\tilde{V}_{#1}}}

\newcommand{\tXmult}[1][]{\widetilde{\mathtt{Xmult}}_{#1}}
\newcommand{\trescon}[1][]{\widetilde{\mathtt{rescon}}_{#1}}
\newcommand{\tbdcon}[1][]{\widetilde{\mathtt{bdcon}}_{#1}}
\newcommand{\tcopycon}[1][]{\widetilde{\mathtt{copycon}}_{#1}}
\newcommand{\tevk}[1][]{\widetilde{\mathtt{evk}}_{#1}}
\newcommand{\tcomb}[1][]{\widetilde{\mathtt{C}}_{#1}}
\newcommand{\teq}{\widetilde{\mathtt{eq}}}

\newcommand{\booktable}[2][]{\ensuremath{\mathbf{T}^{#1}_{#2}}}
\newcommand{\FuncEval}{\texttt{FunctionEvaluations}}
\newcommand{\SparseFuncEval}{\texttt{SparseFuncEval}}
\newcommand{\Precomp}{\texttt{Precompute}}
\newcommand{\Sumcheck}{\texttt{Sumcheck}}
\newcommand{\SumcheckX}{\texttt{Sumcheck\_X}}
\newcommand{\SumcheckProduct}{\texttt{SumcheckProduct}}
\newcommand{\SparseSumcheckProduct}{\texttt{SparseSumcheckProduct}}
\newcommand{\InitCopyY}{\texttt{Init\_copy\_Y}}
\newcommand{\InitCopyX}{\texttt{Init\_copy\_X}}

\newcommand{\InitX}{\texttt{Init\_X}}

\newcommand{\chall}[1]{\vec{r_{#1}}}

\def\polRing{\hat\Ring}
\def\modpol{I}
\def\pack{\ensuremath{\mathsf{pack}}}
\def\unpack{\ensuremath{\mathsf{unpack}}}
\def\ueval{\ensuremath{\mathsf{uEval}}}
\def\relusum{\rel_{usum}}
\def\relueval{\rel_{ueval}}
\def\relspeval{\rel_{speval}}
\newcommand{\piopusum}{\iopPi_{usum}}
\newcommand{\piopsumzq}{\iopPi^{0}_{sum}}
\newcommand{\piopueval}{\iopPi_{ueval}}

\def\reliptree{\rel_{iptree}}
\def\ExceptionalSet{\hat{S}}
\def\Esetz{\ExceptionalSet_0}
\def\Eset{\ExceptionalSet_N}
\def\EsetN{\ExceptionalSet_N}
\def\Esetd{\ExceptionalSet_{d}}

\def\eq{\tilde{\mathsf{eq}}}
\def\tr{{\mathsf{tr}}}
\def\np{a}
\def\val{\mathsf{val}}
\def\spidx{\mathsf{ind}}
\newcommand\mat[1]{\mathbf{#1}}
\def\pol{\mathsf{pol}}
\def\packpol{\overline{\mathsf{pol}}}
\def\lnp{\alpha}
\newcommand\spack{\textsf{SparsePack}\xspace}

\def\E{{\cal E}}

\def\O{{\cal O}}

\def\Ring{{R}}

\def\Rq{\Ring_q}
\def\Rqml{\Rq^{(\leq 1)}}
\def\Zqml{\ZZ_{q}^{(\leq 1)}}

\def\Gq{H_q}
\def\Hq{H_q}

\def\Gqml{\Gq^{(\leq 1)}}
\def\Hqml{\Hq^{(\leq 1)}}
\def\Aq{\Esetd}
\def\relgrange{\hat\rel_{range}}
\newcommand{\piopGqrange}{\hat\iopPi_{range}}

\def\ZZ{\mathbb{Z}}
\def\NN{\mathbb{N}}

\def\Zq{\ZZ_q}

\def\rel{\mathcal{R}}
\def\lang{\mathcal{L}}

\def\r1cswitgen{\ensuremath{\mathsf{WitGen}}}

\def\ind{\mathbbm{i}}
\def\stm{\mathbbm{x}}
\def\wit{\mathbbm{w}}

\newcommand{\ora}[1]{\llbracket{#1}\rrbracket}

\def\relac{\rel_{AC}}

\def\relsum{\rel_{sum}}
\def\relrange{\rel_{range}}

\newcommand{\defeq}{\coloneqq}

\newcommand{\normty}[1]{\norm{#1}_\infty}

\newcommand{\bit}{\{0,1\}}

\newcommand{\BigO}{\mathcal{O}}

\def\secpar{\lambda}

\def\cm{\ensuremath{\mathsf{cm}}}
\def\opn{\ensuremath{\mathsf{opn}}}
\def\ck{\ensuremath{\mathsf{ck}}}

\newcommand{\iopP}{\mathcal{P}}
\newcommand{\iopV}{\mathcal{V}}
\newcommand{\iopPi}{\Pi}
\newcommand{\piopckks}{\iopPi_{CKKS}}
\newcommand{\piopsum}{\iopPi_{sum}}
\newcommand{\piopac}{\iopPi_{AC}}
\newcommand{\pioprange}{\iopPi_{range}}

\newcommand{\piopiptree}{\iopPi_{iptree}}

\newcommand{\PC}{\ensuremath{\mathsf{PC}}}
\newcommand{\Com}{\ensuremath{\mathsf{Com}}}
\newcommand{\VerCom}{\ensuremath{\mathsf{VerCom}}}

\newcommand{\ProveEval}{\ensuremath{\mathsf{ProveEval}}}
\newcommand{\VerEval}{\ensuremath{\mathsf{VerEval}}}

\def\Setup{{\sf Setup}}

\def\Enc{{\sf Enc}}
\def\Dec{{\sf Dec}}

\def\Eval{{\sf Eval}}

\def\evk{{\sf evk}}

\def\sk{{\sf sk}}
\def\ct{{\sf ct}}

\newcommand{\eval}{\ensuremath{\mathsf{Eval}}}

\renewcommand{\adv}{\ensuremath{\mathsf{Adv}}}

\newcommand{\keyswitch}{\ensuremath{\mathsf{KeySwitch}}}

\newcommand{\premult}{\ensuremath{\mathsf{PreMult}}}

\newcommand{\hescale}{\ensuremath{\mathsf{Rescale}}}

\newcommand{\crt}{\ensuremath{\mathsf{CRT}}}
\newcommand{\pw}{\ensuremath{\mathsf{PW}}}

\renewcommand{\vec}[1]{\boldsymbol{#1}}

\newcommand{\vC}{\mathbf{C}}

\newcommand{\va}{\mathbf{a}}
\newcommand{\vb}{\mathbf{b}}
\newcommand{\vc}{\mathbf{c}}

\newcommand{\vd}{\mathbf{d}}

\newcommand{\vm}{\mathbf{m}}

\newcommand{\vr}{\mathbf{r}}
\newcommand{\vs}{\mathbf{s}}
\newcommand{\vt}{\mathbf{t}}

\newcommand{\vvv}{\mathbf{v}}
\newcommand{\vw}{\mathbf{w}}
\newcommand{\vX}{\mathbf{X}}
\newcommand{\vx}{\mathbf{x}}
\newcommand{\vy}{\mathbf{y}}
\newcommand{\vz}{\mathbf{z}}
\newcommand{\0}{\mathbf{0}} %

\newcommand{\Input}{\ensuremath{\mathsf{Input}}}

\mathchardef\mhyphen="2D

\newcommand{\Output}{{\textbf{Output}}}

\newcommand{\BI}{\begin{itemize}[topsep=\globaltopsep]}
	\newcommand{\EI}{\end{itemize}}

\newcommand{\BE}{\begin{enumerate}[topsep=\globaltopsep]}
	\newcommand{\EE}{\end{enumerate}}

\newcommand{\figlab}[1]{\label{fig:#1}}
\newcommand{\figref}[1]{Figure~\ref{fig:#1}}

\usepackage{epsfig,pgfplots}
\pgfplotsset{}

	\newenvironment{Boxfig}[4][]{
		\begin{figure}[tbp!]
			
			\newcommand{\FigCaption}{#2}
			\newcommand{\FigLabel}{#3}
			\vspace{-0.25cm}
			\begin{center}
				\begin{small}
					\begin{tabular}{@{}|@{~~}l@{~~}|@{}}
						\hline
						\rule[-1.5ex]{0pt}{1ex}\begin{minipage}[b]{.96\linewidth}
							\vspace{1ex}
							\smallskip
							\begin{center}
								\textbf{#4}
								\vspace{-0.1cm}
							\end{center}
						}
						{%
						\end{minipage}\\
						\hline
					\end{tabular}
				\end{small}
				\vspace{-0.25cm}
				\caption{\FigCaption}
				\label{fig:\FigLabel}
			\end{center}
			\vspace{-.4cm}
		\end{figure}
	}

	\renewcommand{\paragraph}[1]{\medskip\noindent {\em #1}}
	\newcommand{\ignore}[1]{}

	\newcommand{\verylongrightarrow}[1]     %
	{\setlength{\unitlength}{.01in}     %
		\begin{picture}(#1,1) \put(0,0){\thicklines\vector(1,0){#1}} \end{picture}}

	\newcommand{\lncs}[1]{#1}
	\newcommand{\full}[1]{}

	\newcommand{\pprotocol}[5]{
		{\begin{figure}[#4]
				\lncs{
				}
				\begin{center}
					\fbox{
						\small
						\hbox{\quad
							\full{\begin{minipage}{6.5in}}
								\lncs{\begin{minipage}{6.3in}}
									#5
									\lncs{
										\vspace{-.45cm}
									}
									\caption{\label{#3} #2}
								\end{minipage}
								\quad
							}
						}
					\end{center}
					\lncs{
					}
				\end{figure}
		} }

\newcommand{\dig}[2]{\ensuremath{\mathsf{dig}^{(#1)}_{#2}}}
\newcommand{\toint}{\ensuremath{\mathsf{to\mbox{-}int}}}
\newcommand{\tobin}{\ensuremath{\mathsf{bin}}}
\newcommand{\readts}{\ensuremath{\mathsf{read\_ts}}}
\newcommand{\finalts}{\ensuremath{\mathsf{final\_cts}}}
\newcommand{\idx}{\ensuremath{\mathsf{idy}}}

\def\WS{\ensuremath{\mathsf{WS}}}
\def\RS{\ensuremath{\mathsf{RS}}}
\def\sfS{\ensuremath{\mathsf{S}}}

\def\reps{\ensuremath{\mathsf{rp}}}
\def\Evalcommon{\ensuremath{\Eval^{\mathsf{com}}}}
\def\wHam{\ensuremath{w_{\mathsf{Ham}}}}
\def\dmin{\ensuremath{d_{\mathsf{min}}}}

%% file: introduction.tex
\section{Introduction}\label{sec:introduction}
Homomorphic encryption (HE) \cite{STOC:Gentry09} allows to compute directly over
encrypted data, a crucial feature for data privacy in numerous applications.
A notable use case is secure outsourcing of computation, allowing clients to delegate
computations to untrusted servers, without revealing their inputs. For instance, in the context
of Machine Learning as a Service (MLaaS), users can send encrypted queries to a
server holding a
model and receive encrypted results (e.g., predictions or classifications) in return.

However, {\em HE alone does not guarantee the integrity of the computation}. While data
remains private, clients have no assurance that the encrypted results returned by the server are
correct. This creates a dangerous gap in HE-based solutions that leaves users vulnerable to
maliciously manipulated outputs. In fact, in certain deployment scenarios, it has been shown
that malicious use of the malleability feature of HE can lead to key-recovery
attacks~\cite{EC:LiMic21,USENIX:GNSJ24,CCS:CCPSS24}. Tools to enforce the integrity of
the computation can be crucial in providing solutions that counter such
attacks~\cite{EPRINT:AAMP24}.

Consequently, a line of research (see below for details on related works)
 addresses this problem by adding integrity properties to HE using another fundamental
primitive: succinct non-interactive arguments of knowledge (SNARKs).
SNARKs allow to prove the correctness of a computation in a way that can be efficiently
checked. Thus one can use a SNARK to prove a homomorphic computation over HE
ciphertexts, yielding what is called \emph{verifiable HE} (vHE).

Despite the tremendous progress in SNARKs research that brought them to a practical level
that currently enables numerous applications, their use for vHE is not efficient in
practice. This is because HE and SNARKs typically work over algebraic structures that are
incompatible with each other. SNARKs work over large finite fields and are very efficient in
proving arithmetic statements in this context, whereas HE works over polynomial rings with
composite moduli, e.g., $\Rq=\ZZ_q[Y]/(Y^{N}+1)$, where $q$ is a product of primes,
and critically uses highly non-arithmetic operations for noise management.
On top of that, due to the large size of HE ciphertexts (a single $\Rq$ element can be
100-600KB), the actual homomorphic computation to be proven is much larger than its description over plaintexts.
This situation makes general-purpose SNARKs non-ideal candidates for vHE due to overheads
originating from emulation of ring arithmetic, non-algebraic noise-management operations, and
the size of ciphertexts.

A recent work by Cascudo et al.~\cite{C:CCCFGS25} took an important step toward reconciling
the aforementioned arithmetic incompatibilities by introducing a new framework for building
SNARKs for verifiable HE, focusing on the CKKS scheme \cite{AC:CKKS17}. Their approach
begins with a proof-friendly version of CKKS whose performance is only slightly degraded
compared to the original scheme, and whose homomorphic computations can be
described with just two relations over $\Rq$: arithmetic circuit satisfiability and range
checks. They then design a SNARK
for these relations, by combining polynomial interactive oracle proofs (PIOPs) for them and a suitable polynomial commitment (PC). A key feature of their framework is that
their PIOPs work over the ring $\Rq$, not the usual prime fields, and natively supports the
HE
arithmetic without emulation overheads.
This approach shows promise, and \cite{C:CCCFGS25} provides initial evidence of efficiency
gains over general-purpose SNARKs. Their evaluation fully confirms the practicality of the
proof-friendly CKKS HE, but targets only two of the proving building blocks (sumcheck over
$\Rq$ and a variant of the Brakedown~\cite{C:GLSTW23} polynomial commitment) and
indicates application-level efficiency estimates only for a small example consisting of {\em one}
CKKS multiplication followed by noise flooding.

\medskip
In this work, we continue this line of research on building verifiable computation (VC) for CKKS with
the aim of achieving solutions that can, for the first time, scale to realistic HE applications.
Our starting point is the recent work of Cascudo et al.~\cite{C:CCCFGS25}. Despite the fact that
their work represents significant progress over the state-of-the-art, several limitations remain
that prevent their approach from scaling in practice. Most of them stem from the fact
that~\cite{C:CCCFGS25} designed their framework to be efficient with respect to circuit size, but
did not fully address the fact that, by themselves, HE parameters can be very large and
affect concrete performance, even if we view them as constants (for a fixed security level).
In more detail:

\begin{itemize}[leftmargin=*]

\item For a circuit with $\bigO{n}$ HE operations on an RLWE scheme over the ring
$\Rq=\ZZ_q[Y]/(Y^{N}+1)$, their polynomial commitment has $\bigO{\sqrt{n}N\log(q)}$-sized proofs
and verification
time. Although asymptotically efficient (sublinear in $n$), the $N\log(q)$ term already
is in the order of hundreds of kilobytes -- even for very small CKKS parameters.
As the circuit grows, this cost quickly becomes impractical, effectively
limiting~\cite{C:CCCFGS25} to tiny HE circuits.

   \item In dealing with noise maintenance operations, their protocol for range checks yields an
   $\bigO{N}$ overhead to be paid by the
   polynomial commitment. Crucially, this overhead comes {\em on top} of the cost for
   committing to %
   vectors of ciphertexts that inherently have their own
   $\bigO{N}$ size.
   Considering that $N$ is typically around $2^{13}$ or larger, this overhead is
   significant. For
   instance, to prove that a vector of, say, $n=2^{10}$ elements satisfies the range check, they
   need to additionally commit to at least four vectors of length $2^{23}$.

   \item Their protocol for arithmetic circuits is based on a custom version of GKR
   \cite{STOC:GolKalRot08}, and the main contribution of~\cite{C:CCCFGS25} is to efficiently
   integrate the noise-management operations within the protocol.
      As typical in GKR, verifying for a circuit $C$ either requires the verifier to do
      $\bigO{|C|}$ work to evaluate some polynomials that describe the circuit, or
      store commitments to these polynomials, and delegate their evaluation to the
      prover.
      \cite{C:CCCFGS25} does not address the concrete efficiency of this component.
      While this could be solved in theory with standard techniques, these -- when adapted for a
      large ring like $\Rq$ -- would be prohibitively expensive. For instance, an approach similar
      to Spartan~\cite{C:Setty20} would imply proofs in the order of GBs (see
      Section~\ref{sec:implementation}).

\end{itemize}

\headingb{Our contributions.}
To achieve practically efficient verifiable HE for non-trivial circuits, we extend the framework of Cascudo et al.~\cite{C:CCCFGS25} with new techniques that overcome its main limitations. Our approach yields substantial efficiency improvements on generic neural networks, demonstrating the feasibility of verifiable HE for MLaaS applications.

\medskip
\noindent
{\em BatchFold.}
To reduce the $\bigO{\sqrt{n}}$ proof size of Brakedown commitments used in
\cite{C:CCCFGS25}, we build on BaseFold~\cite{C:ZeiCheFis24}, which achieves $\bigO{\log^2n}$
proofs and $\bigO{n \log n}$ proving time.
Our main contribution (see Section \ref{sec:basefold}) is a novel {\em batching technique}.
For $M$ independently committed polynomials $f_i$ with $n$ coefficients over a finite field,
we can prove multiple evaluations ($y_i = f_i(x_i)$),
reducing proof size and proving time from $O(M\log^2 n)$, using naive repetition, to
$O(M\log n+\log^2 n)$. Our batching
technique has a twofold impact. The first obvious one is that we can batch
evaluation proofs when compiling the PIOP into a SNARK.
The second advantage is improving the complexity of our $\Rq$ version of BaseFold {\em even for evaluating a single polynomial}.
In a nutshell, a PC for $\Rq$ is built by exploiting the CRT splitting of $R_q$ into a product of
fields. Crucially, we use the fact that, for each prime factor $p$ of $q$, there are $N/4$ isomorphic fields of characteristic $p$ in the splitting, and thus we can apply batching over the $N/4$ polynomials in the CRT decomposition of the input polynomial. For a $q$, product of $L+1$ different primes, we have a proof size of $O(NL\log n+L\log^2 n)$ field elements (from fields of similar sizes) instead of $O(NL\log^2 n)$.

\medskip
\noindent
{\em Packing and Unpacking in PIOPs.}
The PIOP for range checks of \cite{C:CCCFGS25} proves that each evaluation $\tilde v(\vec b)$ in the boolean hypercube of a multilinear polynomial $\tilde v \in \Rq[\vec X]$ is an element of $\Rq=\ZZ_q[Y]/(Y^N + 1)$ with coefficients $<B$. To do this, \cite{C:CCCFGS25} introduces a decomposition technique that,  $\forall \vec b \in \bin^{\ell}$,
embeds the $N$ coefficients of $\tilde v(\vec b)$ into the evaluations $\tilde h(\vec b, \vec k)$ of a freshly generated $\tilde h\in R_q[\vec X, \vec Z]$ of $\ell + \log N$ variables.
This approach incurs an $\bigO{N}$ overhead, especially for committing to $\tilde h$, an
object $N$ times bigger than $\tilde v$.\footnote{This decomposition and its overhead is
reminiscent of the binary decomposition that creates additional (witness) bits and embeds
them in a finite field.}

In our paper, we introduce \emph{packing},  a novel algebraic technique that allows  PIOPs to prove statements about ``unpacked'' versions of multilinear polynomials.
Analogously to the above decomposition, the unpacking of a polynomial $\tilde p \in \Rq[\vec X]$ is a polynomial $\tilde u \in \Rq[\vec X, \vec Z]$ with $\log N$ more variables and whose evaluations are all in $\ZZ_q$. The crucial difference though is that:
(i) {\em we do not commit to $\tilde u$}, we only store and commit to $\tilde p$ in its dense (aka packed) form,
(ii) the PIOP can work with an efficient sparse representation of $\tilde u$ that has no
$\bigO{N}$ artificial overhead, and
(iii) queries to $\tilde u$ can be efficiently computed from queries to $\tilde p$.
For this reason, we call $\tilde u$ a virtual unpacked oracle.
We refer to Section \ref{sec:packing} for details on the technique.
We discuss here two main applications.

The first one is a new PIOP for range checks that avoids the $\bigO{N}$ overhead
of~\cite{C:CCCFGS25}, achieving not only an asymptotic improvement but also concrete orders of magnitude faster computing time. Specifically for the largest circuit of our experimental benchmark, we estimate \cite{C:CCCFGS25}'s range checks to take 1h20m in 48 threads, while ours takes less than 6 minutes single-threaded.

Beyond range checks, we use packing to compress sparse polynomials over $\Rq$, such as the multilinear extensions of $\ZZ_q$-vectors in GKR.
In particular, we use packing to construct \spack, a new protocol for the evaluation of sparse
polynomials, i.e., having sublinear nonzero evaluations in the hypercube. When applied to our
case (MLEs of wiring predicates), it outperforms state-of-the-art solutions~\cite{C:Setty20} featuring proofs that, in one of our benchmarks, are $41\times$ shorter and $10\times$ faster to prove (see Section \ref{sec:implementation}).

\paragraph{Better arithmetization and preprocessing.}
One bottleneck of \cite{C:CCCFGS25} is handling the polynomials that describe the GKR circuit.
These are as large as the circuit and hence cannot be computed by the verifier.
Delegating this work to the prover is costly since, for a circuit of size
$2^{\ell}$, these polynomials are of size $2^{3 \ell}$, despite having  only
$\bigO{2^{\ell}}$ nonzero evaluations.
To address this, we introduce a new arithmetization for CKKS circuits that improves upon \cite{C:CCCFGS25} in several ways.
We design closed-form polynomials for the wiring predicates related
to CKKS homomorphic multiplication which the verifier can compute efficiently.
We also substantially reduce the number and size of the polynomials that encode the structure of the target FHE circuit (which is arbitrary).
Finally, we use \spack to prove the evaluations of these sparse polynomial (see above).

\paragraph{Experiments.}
The framework of~\cite{C:CCCFGS25} enabled, for the first time and with minimal emulation overhead, verifiability for
complete HE multiplications with non-linear noise management operations. %
However, the authors only presented estimates for a single ciphertext multiplication.
Our improvements can scale beyond that: we provide experiments for small Neural Networks of multiplicative depth 2,
containing between 48 and 192 total ciphertext multiplications each.
We measure performance for
our new packed range proofs and batched polynomial commitment scheme, and estimate it for our sparse commitments, showing that these NNs can be proven in the order of minutes single-threaded.
This represents a major improvement over~\cite{C:CCCFGS25}, which estimated 25 seconds for a single multiplication in 48 threads.
Additionally, we present execution time and proof size for commitments over
$\Rq$ using Batchfold, showing a factor of 2.8 (resp. 9.5) improvement in execution time (resp. proof size) compared to Basefold.

\headingb{Related work.}\label{sec:related-work}
Literature on verifiable computation for FHE can be divided in two main approaches.
The first one, formalized in~\cite{CCS:FioGenPas14}, consists in designing VC schemes for proving computations on FHE ciphertexts. With the exception of
recent work of Cascudo et al.~\cite{C:CCCFGS25}, prior works in this line
suffer from either supporting only variants of HE schemes with
limited applicability~\cite{CCS:FioGenPas14,PKC:FioNitPoi20,PKC:BCFK21}, or otherwise incurring large
overheads for emulating ring or noise-maintenance
operations~\cite{JC:GanNitSor23,vFHE-RWC,vFHE-arxiv}.
The second approach computes proofs homomorphically over the plaintext space \cite{C:GarGoeWan24,HELIOPOLIS,EPRINT:GBKSV24}.
While this approach can be faster than proving ciphertext computations, it has
inherent limitations, notably:
it cannot achieve public verifiability, since proof verification requires decryption\footnote{Public verifiability, by definition, cannot rely on secret material. Nevertheless, any designated verifier scheme can be turned into a “semi-publicly verifiable” one by making the verifier provide a publicly verifiable ZK proof of their execution of the verification algorithm. This idea is instantiated with proofs of decryption in \cite{EPRINT:GBKSV24}.};
it cannot support approximate schemes such as CKKS (due to their lack of a ring structure on the
plaintext space) and it is vulnerable to verification oracle attacks.

%% file: preliminaries.tex
\newcommand{\detks}[2]{\vw^{(#1)}_{\text{ks}, #2}}
\newcommand{\detrmd}[1]{\vw^{(#1)}_{\text{rmd}}}
\newcommand{\detquot}[1]{\vw^{(#1)}_{\text{quot}}}
\newcommand{\layerinput}[2]{\vc^{(#1)}_{#2}}
\newcommand{\layeroutput}[2]{\vc''^{(#1)}_{#2}}
\newcommand{\layerpremult}[2]{\vd^{(#1)}_{#2}}
\newcommand{\layerinterm}[2]{\vc'^{(#1)}_{#2}}
\newcommand{\layerinputall}{\vx}
\newcommand{\detksall}{\vw_{\text{ks}}}
\newcommand{\detrmdall}{\vw_{\text{rmd}}}
\newcommand{\detquotall}{\vw_{\text{quot}}}
\newcommand{\layeroutputall}{\vy}

\section{Preliminaries}\label{sec:preliminaries}

For a positive integer $n$, $[n]$ denotes the set $\{0, \ldots, n-1\}$.
We denote vectors $\vvv$ in bold, start indexing at $0$ and denote its $i$-th coefficient by $\vvv[i]$ or $v_i$.
We denote by $\vvv_{[B]}$ the truncation of $\vvv$ to its first $B$ terms, $\vvv_{[A, B)}$ for its terms in the range $[A,B)$ (i.e. $\vvv_{[A,B)} = (v_A, \ldots, v_{B-1})$) and $\vvv_{[A, B]}$ for its terms in the range $[A,B]$.
We define $|\vvv|$ the length of $\vvv$ and $\#(\vvv) = \ceil{\log_2(|\vvv|)}$.
Let $\toint: \bin^{*} \to \ZZ$ be the function that, on input a binary vector $\vec j \in\bin^{n}$, for any positive integer $n$, returns the integer $\sum_{k \in[n]} j_k \cdot 2^k$. Let $\tobin: \ZZ \to \bin^{*}$ be its inverse.
Given a finite set $S$, we write $a \sample S$ to mean that $a$ is sampled uniformly at
random from $S$
and $e \leftarrow \chi$ to mean that $e$ is sampled according to some distribution $\chi$.

We often invoke the infinity norm $\norm{\cdot}_\infty$.
Given a vector $\vvv$ of elements in $R_q := \ZZ_q[Y]/(Y^N+1)$, we apply the norm
coefficient-wise: the expression $\norm{\vvv}_\infty < B$ means that all the monomial coefficients of
$\vvv$ are within the range $[0, B-1]$.

An {\em indexed relation} $\rel$ is a set of tuples $(\ind,\stm;\wit)$ where $\ind$ is the index, $\stm$ the statement and $\wit$ the witness.
The language $\lang(\rel)$ associated to a relation $\rel$ is the set of pairs $(\ind,\stm)$ for which there exists a witness $\wit$ such that $(\ind,\stm;\wit) \in \rel$.

\subsection{Background on Rings}
We restrict ourselves to finite commutative rings with unity, which we often denote by $\Ring$.
Given $a \in \Ring$, $\left[ a \right]_{b}$ denotes the reduction of $a$ modulo $b$.
If $a \in \Ring(Y)$ is an element of a polynomial ring, $a[j]$ (resp. $a[\vec j]$) denotes its $j$-th (resp. $\toint(\vec j)$-th) coefficient (in such representation) and
$\left[ a \right]_{b}$ denotes the coefficient-wise modular reduction.
For a vector $\vvv \in \Ring^{n}$, $\left[ \vvv \right]_p$ denotes modular reduction of each component of $\vvv$.

\begin{definition}%
	An \emph{exceptional set} is a subset $\ExceptionalSet \subset \Ring$ with the property that $a-b$ is an
	invertible element of $\Ring$ for each pair of distinct elements $a, b \in \ExceptionalSet$.
\end{definition}

We denote by $\Ring[X_1, \ldots, X_\ell]$ the polynomials in $\ell$ variables and coefficients in $\Ring$
and by $\Ring^{(\leq d)}[X_1, \ldots, X_\ell]$ the subset of polynomials of individual degree $\leq d$.

\begin{lemma}[Generalized Schwartz-Zippel]\label{lemma:generalizedSZ}
	Let $f \in \Ring[X_1, \dots, X_\ell]$ be a non-zero polynomial of total degree $d$.
	Let $\ExceptionalSet \subset \Ring$ be an exceptional set. Then,
	$
	\Pr_{(r_1, \dots, r_\ell) \gets \ExceptionalSet  ^\ell} \left[ f(r_1, \dots, r_\ell ) = 0 \right] \leq \frac{d}{\vert \ExceptionalSet
	\vert }
	$.
\end{lemma}

The \emph{multilinear extension} (MLE) of a function $f: \bin^{\ell} \to \Ring$ is the (unique) multilinear polynomial $\tilde f:\Ring^{\ell} \to \Ring$ such that
$\tilde f(\vec b) = f(\vec b)$ for all $\vec b \in \bin^{\ell}$.
For any $\ell \geq 1$, let $eq:\bin^{\ell} \times \bin^{\ell} \to \bin$ be the equality predicate (i.e., $eq(\vec a, \vec b)=1$ iff $\vec a =  \vec b$) and let $\eq$ be its MLE,
which has closed-form representation $\eq(\vec X, \vec Y) = \prod_{j=1}^{\ell} (X_j Y_j + (1 - X_j) (1 -
Y_j))$ computable in time $\bigO{\ell}$~\cite{C:Thaler13}.
The MLE of any function $f : \bin^{\ell} \to \Ring$ is defined by the polynomial
$\tilde f(X_1, \ldots, X_\ell) =  \sum_{\vec b \in \bin^{\ell}} f(\vec b) \cdot \eq(\vec b, X_1, \ldots, X_\ell).$ %
Analogously, the MLE of a vector $\vec v \in \Ring^{2^{\ell}}$ (which for convenience we index in binary) is the polynomial
$\tilde v(X_1, \ldots, X_\ell) =  \sum_{\vec b \in \bin^{\ell}} \vec v[\vec b] \cdot \eq(\vec b, X_1, \ldots, X_{\ell}).$

\input{piop}

\subsection{Proof-friendly CKKS and its arithmetization}\label{sec:proof-friendly-CKKS}
In this work we use the proof-friendly version of the CKKS homomorphic encryption scheme~\cite{AC:CKKS17} introduced in~\cite{C:CCCFGS25}.
For a full description of the scheme, see Appendix~\ref{sec:pfckks} or~\cite{C:CCCFGS25}.
Here we only discuss the main technical points that are useful to build our proof system.

\medskip
\noindent{\em Choice of the ring and exceptional sets.}
Fast CKKS implementations use the ciphertext ring $R_q = \mathbb{Z}_q[Y]/(Y^N + 1)$, where $N$ is a power of two and $q = \prod_{i=0}^L p_i$ for primes $p_i$. In proof-friendly CKKS, $p_i=2aN/d+1$ for an odd integer $a$ and a power of two $d$. Thus, $Y^N+1$ factors over $\mathbb{Z}_{p_i}$ into $k = N/d$ irreducible polynomials $Y^d - \zeta_i^{(2j+1)}$, where $\zeta_i$ is a primitive $2N/d$-th root of unity. By the CRT, $R_q \cong \prod_{i=0}^L \prod_{j=0}^{k-1} R_{i,j}$, with each $R_{i,j} = \mathbb{Z}_{p_i}[Y]/(Y^d - \zeta_i^{(2j+1)}) \cong \mathbb{F}_{p_i^d}$.

This structure yields \textit{exceptional sets} large enough for soundness, bounded by the smallest field component. We use two sets, $\Esetz \subset \mathbb{Z}_q$ and $\EsetN \subset R_q$, of sizes $p_{min}$ and $p_{min}^d$ respectively (where $p_{min}$ is the smallest $p_i$). For example, setting $d=4$ (resp.\ $d=2$) for 32-bit (resp.\ 64-bit) primes makes $p_{min}^{-d} \approx 2^{-128}$, i.e., negligible for security parameter $\secpar=128$; using $\Esetz$ may instead require repetitions to amplify soundness. Concretely, we take $\EsetN$ diagonal: each of its elements has the same exceptional element of $\FF_{p_i^d}$ in all the $k$ components $R_{i,j}$ of each $p_i$ (Appendix~\ref{app:sparse_perf} discusses it further).

Since $R_q$ decomposes into copies of $\mathbb{F}_{p_i^d}$, we cannot use full NTTs to accelerate HE operations and instead use \textit{incomplete NTTs}.
We note that efficient implementations of incomplete NTTs have been widely explored in the literature, and can even lead to performance improvements in some
cases~\cite{EUROSP:HYSPOM25,fast_incomplete_ntt,TCHES:LyuSei19}. In~\cite{C:CCCFGS25}, the
authors incur a slowdown of just 20\%  compared to ``regular CKKS'' due to the use of incomplete NTTs.
Since we make use of their scheme unmodified, our construction also inherits their slowdown. We
note this modification in the ciphertext ring has no bearing upon the underlying plaintext ring or
slots. We adopt the standard plaintext encoding $\CC^{N/2} \rightarrow \CC[X]/(X^N +
1)$~\cite{AC:CKKS17}, but one could also use others~\cite{CCS:JPKKKC24,TCHES:Kim25}.

\medskip
\noindent{\em Remaining in the same ring $\Rq$.}
Standard levelled CKKS uses a moduli chain $q := q_L > \dots > q_0$ (where $q_i =
q_{i+1}/p_{i+1}$). In contrast, proof-friendly CKKS operates entirely over a single ring $R_{q_L} =
R_q$.
Operations normally altering the ring (e.g., modulus switching) have their results re-embedded back into
$R_{q_L}$ via the map $\iota_l : R_{q_l} \hookrightarrow R_{q_L}, \quad x \mapsto \tilde{x} \cdot z_l$, where $\tilde{x} \in R_{q_L}$ is a lift of $x$ and $z_l
\defeq \sum_{i=0}^{l} Q_i\hat{Q}_i = \crt(\underbrace{1, \dots, 1}_{l+1~\text{times}}, 0, \dots, 0) \in
R_{q_L}$ with $Q_{i} = q_L/p_i$ and $\hat{Q}_{i} = \left[(q_L/p_i)^{-1}\right]_{p_i}$ defining the CRT map.
As discussed
in~\cite{JC:GanNitSor23,C:CCCFGS25}, this abstraction effectively emulates modular reductions and quotient ring arithmetic without
impacting performance.

\medskip
\noindent{\em Arithmetization of proof-friendly CKKS.}
At the core of \cite{C:CCCFGS25} is a procedure that translates a computation $\vC'(\vx) = \vy$, where $\vC'$ is a circuit of (proof-friendly) CKKS additions and multiplications (see sec.\ref{sec:pfckks}), into a relation over $R_q$.
Ciphertext additions are trivially described by additions over $R_q$.
Other operations such as key switching and rounding (required for ciphertext multiplication) cannot be expressed by arithmetic operations over $R_q$ without significant overhead.
\noindent
For $\keyswitch$, the prover inputs $w_{\text{ks}, 0}, \dots, w_{\text{ks}, l} \in R_q$, satisfying
$\norm{w_{\text{ks}, i}}_\infty < p_i$ and such that they recompose to the original value in $R_q$
via the CRT map (which is an arithmetic operation over $R_q$).

Regarding $\hescale$, suppose we have an element $x \in R_{q_l}$, in its RNS representation
$x = ([x]_{p_0}, [x]_{p_1}, \ldots, [x]_{p_l}, [0]_{p_l+1}, \ldots, [0]_{p_L})$. The division and rounding
of $x$ by the current level $l$ can be proved via Euclidean division by $p_l$.
The prover's inputs are $w_{\text{rmd}} = ([w_{\text{rmd}}]_{p_0}, \ldots, [w_{\text{rmd}}]_{p_L})$
and $w_{\text{quot}} = ([w_{\text{quot}}]_{p_0}, \ldots, [w_{\text{quot}}]_{p_L})$. The range
proof shows that $\norm{w_{\text{rmd}}}_\infty < p_l$ and that $ \norm{w_{\text{quot}}}_\infty < q_l/p_l$.
Finally, recalling that $z_l = \crt(\underbrace{1, \dots, 1}_{l+1~\text{times}}, 0, \dots, 0) \in
R_{q_L}$, we check that $x\cdot z_l = (w_{\text{quot}}\cdot p_l + w_{\text{rmd}})\cdot z_l$.
This shows that proof-friendly CKKS multiplications can be expressed as a sequence of arithmetic
operations over $R_q$ and a finite number of range checks relations for $R_q$ elements.

\noindent
One can apply this argument to a generic circuit $\vC'$ organized into $D$ multiplicative layers of width $W$.
\footnote{Layers are numbered from $0$ to $D-1$, with inputs at level $0$ and outputs at level $D-1$. Without loss of generality, one can assume that all layers have the same width.}
Denoting by $\detks{j}{0}, \dots, \detks{j}{j}, \detrmd{j}, \detquot{j}$ the vectors of non-deterministic inputs for level $j$, the whole computation can be described in terms of two oracle relations:
\begin{align*}
	& \relrange = \left\{ (\ind, \stm; \wit) := (B, \ora{\tilde{\vw}}; \tilde\vw) ~:~ \normty{\tilde\vw} < B \right\} \\
	& \rel_{AC} = \left\{ (\ind, \stm; \wit) := (\vC, (\ora{\tilde{\vx}}, \ora{\tilde{\vy}}, \ora{\tilde{\vw}}); (\tilde\vx, \tilde\vy, \tilde\vw)) ~:~ \vC(\vx, \vy, \vw) = 0 \right\},
\end{align*}
where $\vC$ is an arithmetic circuit over $R_q$ of constant-depth, obtained by flattening the CKKS
circuit $\vC'$. For the details, see \cite{C:CCCFGS25}. Importantly, since all of the RNS primes have
the same bit length, one can batch all the range proofs for $\{ \detks{i}{j}, \detrmd{i}\}_{i,j}$ together
by approximating the bounds to $2^u$, their nearest power of two. Let $\ora{\tilde{\layerinputall}},
\ora{\tilde{\layeroutputall}}, \ora{\tilde{\vw}_{2^u}}, \ora{\tilde{\vw}_l}$ be the oracles of the MLEs of
the inputs $\layerinputall$, outputs $\layeroutputall := \detquot{D-1}$, prover's values for RNS decompositions and remainders $\vw_{2^u} := \detks{0}{0}
\Vert \dots \Vert \detks{0}{D-1}\Vert \dots \Vert \detks{D-1}{0} \Vert \detrmd{0} \Vert \dots \Vert
\detrmd{D-1}$ and prover's values for quotients $\vw_l := \detquot{l}$, for $l=0, \dots, D-2$.
The relation that describes the computation $\vC'$ on CKKS ciphertexts can be compactly written as:
{\small
	\begin{align*}
		\rel_{CKKS}  =  \left\{ (\vC, (\ora{\tilde{\layerinputall}}, \ora{\tilde{\vw}_{2^u}}, \{ \ora{\tilde{\vw}_l} \}_{l=0}^{D-2}, \ora{\tilde{\layeroutputall}});
		( \layerinputall, \vw_{2^u}, \{ \vw_l \}_{l=0}^{D-2}, \layeroutputall ) ) ~:~ \right.\\
		\left. \underbrace{\vC(\layerinputall, \layeroutputall, \vw_{2^u}, \{ \vw_l \}_{l=0}^{D-2}  ) = \0}_{\rel_{AC}} \wedge  \underbrace{\normty{\vw_{2^u}} < 2^u
	\wedge \bigwedge_{l=0}^{D-2} \normty{\vw_l} < \frac{q_{L-l}}{p_{L-l}} \wedge \normty{\layeroutputall} < \frac{q_{L-D+1}}{p_{L-D+1}}}_{\relrange} \right\}
	\end{align*}
}%

A modular PIOP $\piopckks$ for $\rel_{CKKS}$ can be built from PIOPs $\pioprange$ for $\relrange$ and $\piopac$ for $\rel_{AC}$ using a simple `and' composition (see Appendix \ref{sec:modular-piopckks}).
Also, notice that we include the circuit $\vC$ as part of the index $\ind$; this allows the verifier to preprocess $\vC$ once and later verify proofs in time $\mathsf{poly}(\log|\vC|)$.

%% file: piop.tex
\subsection{Polynomial Interactive Oracle Proofs for Rings}\label{sec:piopdef}
We recall polynomial interactive oracle proofs (PIOP) for rings~\cite{C:CCCFGS25} using the notion of
(indexed) oracle relations of~\cite{EC:CBBZ23}.

\begin{definition}[Oracle Relations]
An indexed oracle relation $\rel$ is an indexed relation in which the index $\ind$ and the statement $\stm$ include pointers to oracle polynomials with coefficients in a ring $\Ring$, and the polynomials are part of the witness $\wit$.
A pointer to a polynomial $p$ is denoted by $\ora p$.
\end{definition}

\begin{definition}[Polynomial IOPs]
Let $\Ring$ be a ring and $\rel$ be an indexed oracle relation for (multivariate) polynomials with coefficients in $\Ring$.
A polynomial IOP for $\rel$ is an interactive protocol between a prover $\iopP$ and a verifier
$\iopV$, compactly denoted by $\langle \iopP(\ind,\stm,\wit), \iopV(\ind,\stm) \rangle$. In each round
the prover $\iopP(\ind,\stm,\wit)$ sends oracle polynomials and/ or ring elements, and the verifier
$\iopV(\ind,\stm)$ sends random challenges.
Each oracle $\ora p$ specifies the number of variables $\ell$ and the degree in each variable. The verifier can query all the oracles, the ones sent by the prover and the ones included in $(\ind,\stm)$, at arbitrary points $\vec r \in \Ring^{\ell}$. At the end of the execution, $\iopV$ accepts or rejects. See Appendix \ref{apx:preliminaries} for the efficiency metrics.

\underline{Correctness:} For any honest execution of $\langle \iopP(\ind,\stm,\wit), \iopV(\ind,\stm) \rangle$ on $(\ind,\stm;\wit) \in \rel$, the verifier accepts.

\underline{$\delta$-Soundness:} A PIOP is $\delta$-sound if for any unbounded adversary $\adv$ and any $(\ind,\stm) \notin \lang(\rel)$,
\ifFullVersion
\[
	\Pr \left[ \langle \adv(\ind,\stm), \iopV(\ind,\stm) \rangle = 1\right] \leq \delta
\]
\fi

\underline{$\delta$-Knowledge Soundness:} A PIOP is $\delta$-knowledge-sound if there exists a PPT oracle machine $\ext$ (which can query all the oracle polynomials at arbitrary points) such that for any adversary $\adv$ and any $(\ind,\stm)$
\[
	\Pr \left[ b = 1 ~\wedge~ (\ind,\stm;\wit) \notin\rel :
                         \langle \adv(\ind,\stm), \iopV(\ind,\stm) \rangle = b,~ \wit \gets \ext^{\adv}(\ind,\stm)
                \right] \leq \delta.
\]
\end{definition}
PIOPs can also satisfy the notion of \emph{round-by-round (RBR) soundness}
\cite{STOC:CCHLRR19}, which allows a tighter quantification of the loss incurred when applying the Fiat-Shamir
transformation, and a better analysis of protocols with parallel repetition (see Appendix \ref{sec:rbr-soundness} for more details).

\begin{lemma}[{\cite[Lemma 2.3]{EC:CBBZ23}}]\label{lemma:sound2ksound}
Consider a $\delta$-sound PIOP for oracle relations $\rel$ such that for all $(\ind, \stm, \wit) \in \rel$,
$\wit$ consists only of polynomials such that the statement contains oracles to these polynomials.
 The PIOP has $\delta$-knowledge-soundness, and the extractor runs in time $\bigO{|\wit|}$.
\end{lemma}

\noindent
We generalize the notion of {\em virtual oracles} from \cite{EC:CBBZ23}.
\begin{definition}[Virtual oracles]\label{def:virtual-oracles}
Given a function $g$ and oracles $\{\ora{p_1}, \ldots, \ora{p_k}\}$, a virtual oracle $\ora{p^*}  = \ora{g(p_1, \ldots, p_k)}$ consists of the list of oracles and functions $ev_{g}, h_1, \ldots, h_k$ such that, for any $\vec x$,
$ev_{g}(p_1(h_1(\vec x)), \ldots, p_k(h_k(\vec x)), \vec x) = p^{*}(\vec x)$.
\end{definition}
In other words, the functions associated to the virtual oracle allow one to compute a query to the virtual oracle $\ora{p^{*}}$ at $\vec x$ by querying each oracle $\ora{p_j}$ on some $\vec x_j = h_j(\vec x)$ and then computing $p^{*}(\vec x)$ from their results.

\headingb{Sumcheck protocol.}
Let $\Ring$ be a ring, $p : \Ring^{\ell} \to \Ring$ be an $\ell$-variate polynomial of degree at most $d$ in each variable.
The sumcheck (oracle) relation $\relsum$ is the set of tuples $(y; p)$ (resp. $((y, \ora p); p)$) such that $y = \sum_{\vec b \in\bin^{\ell}} p(\vec b)$.
The sumcheck protocol~\cite{FOCS:LFKN90} is an interactive proof (resp. a PIOP) for $\relsum$
where the verifier needs to only compute (resp. query) $p$ at a point $(r_1, \ldots, r_{\ell})$. While
originally defined for finite fields, it also generalizes to rings $\Ring$ if the random challenges are
sampled from an exceptional set $\ExceptionalSet \subset \Ring$.
We denote the sumcheck PIOP by $\piopsum$; for completeness we present it in Appendix \ref{sec:piop}.
Also, we denote by $\piopsumzq$ a variant of the PIOP $\piopsum$ changed so that the verifier samples random challenges in an exceptional set $\Esetz \subset \ZZ_q$.

\headingb{Polynomial commitments and compilation to SNARKs.}
Prior work shows how to compile PIOPs into SNARKs via polynomial commitments (PC),
a cryptographic primitive that allows committing to a polynomial $p$ and later proving
that $y=p(a)$ with a succinct proof.
In our work, we use an extension of this paradigm that relies on PCs with {\em batched evaluation proofs}.
In a nutshell, this is a generalization of PCs that allow generating a {\em single} proof about the evaluations of multiple, say $M$, polynomials at multiple points,
i.e., $\forall i \in [M] : y_{i} = p_i(a_i)$, and has the advantage that the resulting proof size is smaller
than concatenating $M$ proofs.
For space reasons, see Appendix~\ref{sec:pcdef} for more formal details.

%% file: packing.tex
\section{PIOPs for Unpacked Polynomials}\label{sec:packing}
In this section we give a set of useful results that allow us to construct PIOPs that can prove
statements about ``unpacked'' versions of multilinear polynomials in $\Rq[\vec X]$, which can be seen
as a way to define statements about their $\ZZ_q$-coefficients.
Our techniques generalize to any ring $\Ring$ and polynomial ring $\polRing := \Ring[Y]/\modpol(Y)$,
where $\modpol(Y) \in \Ring[Y]$ of degree $N$.
For ease of exposition, we limit the presentation to our rings of interest $\Ring = \ZZ_q$ and $\polRing = \Rq = \ZZ_q[Y]/(Y^{N} +1)$.

\subsection{Packing and Unpacking polynomials in $\Rq$}
Let $\Rq = \ZZ_q[Y]/(Y^{N} +1)$ for $N=2^{\nu}$.
We consider multivariate polynomial rings $\Rq[X_1, \ldots, X_{\ell}]$ and $\Rq[X_1, \ldots, X_{\ell}, Z_1, \ldots, Z_\nu]$.
For brevity, we denote $\vec X = (X_1, \ldots, X_\ell)$ and $\vec Z = (Z_1, \ldots, Z_\nu)$.
We define two useful functions, $\pack$ and $\unpack$, that allow switching between two different
representations of a multilinear polynomial in $\Rq[\vec X]$.
The function $\unpack$ takes a multilinear polynomial $\tilde p \in R_q[\vec X]$ and returns a multilinear polynomial $\tilde u \in R_q[\vec X, \vec Z]$, of degree 0 in $Y$, which can be naturally identified with $\ZZ_q[\vec X, \vec Z]$, and is defined as follows:
\begin{eqnarray}
\unpack: & \Rqml[\vec X] &\to \Rq^{(\leq 1)}[\vec X, \vec Z] \label{eq:unpack} \\
& \tilde p(\vec X) &\mapsto \tilde u(\vec X, \vec Z)  =      \sum_{\vec b \in \bin^{\ell}} \sum_{\vec k \in
\bin^{\nu}} \tilde p(\vec b)[\vec k] \cdot \eq((\vec b, \vec k), (\vec X, \vec Z)) \nonumber. %
\end{eqnarray}
$\unpack$ essentially extracts all the coefficients in the variable $Y$ of the evaluations of $\tilde p$ in the hypercube, and encodes these coefficients under the basis
$\{\eq(\vec k, \vec Z)\}_{\vec k \in \bin^{\nu}}$
instead of $\{Y^{k}\}_{k \in [N]}$.
While $\unpack$ intuitively goes from (a subset of) $\ZZ_q[\vec X, Y]$ to $\ZZ_q[\vec X, \vec Z]$, we stress that its
outputs are embedded in $\Rq[\vec X, \vec Z]$.
We informally call polynomials in the image of $\unpack$ {\em unpacked polynomials}.

The function $\pack: R_q[\vec X, \vec Z] \to R_q[\vec X]$ instead is defined as follows:
\begin{equation}\label{eq:pack}
u(\vec X, \vec Z) \mapsto p(\vec X)  =   \sum_{\vec k \in \bin^{\nu}} u(\vec X, \vec k) \cdot Y^{\toint(\vec
k)}.
\end{equation}
Notice that in general $\pack$ is not the inverse of $\unpack$, yet for any multilinear $\tilde p$ one can see that $\tilde p  = \pack(\unpack(\tilde p))$.
In particular, given a multilinear polynomial $\tilde u(\vec X, \vec Z)$ that has degree $0$ in $Y$,
$\pack$ ``packs'' $\tilde u$ into a dense polynomial with fewer variables, hence the name.

\headingb{Evaluating unpacked polynomials.}
A key property of unpacked polynomials (formalized in the Lemma below) is that an evaluation of an unpacked polynomial on $\ZZ_q$-elements can be computed from the evaluation of its packed version.\footnote{Precisely, we mean evaluations on elements of $\Rq$ of degree 0 in $Y$; for simplicity we abuse notation and denote this subset of $\Rq$ by $\ZZ_q$.}

\begin{lemma}\label{lemma:unpackedeval}
Let $\ueval$ be the function that on inputs $a \in \Rq$ and $\vec z \in \ZZ_q^{\nu}$ returns
\[
\ueval(a, \vec z) = \sum_{\vec k \in \bin^{\nu}}  a[\vec k] \cdot \eq(\vec k,\vec z).
\]
Then for any polynomial $\tilde p \in \Rqml[\vec X]$, $\tilde u = \unpack(\tilde p)$, and any $(\vec x, \vec z) \in \ZZ_q^{\ell} \times \ZZ_q^{\nu}$:
$\ueval(\tilde p(\vec x), \vec z) = \tilde u(\vec x, \vec z)$, and $\ueval$ is computable with $O(N)$ operations in $\ZZ_q$.

\end{lemma}
\begin{proof}

Observe that the polynomials $\eq(\vec b, \vec X) \in \Rqml[\vec X]$ are such that $\eq(\vec b, \vec x) \in \ZZ_{q}$ if $\vec x \in \ZZ_q^{\ell}$. This allows to define the coefficients $a[\vec k]$ of $p(\vec x)$ as follows
\begin{eqnarray*}
\tilde p(\vec x) &= & \sum_{\vec b \in \bin^{\ell}} \tilde p(\vec b) \eq(\vec b, \vec x)
	=  \sum_{\vec b \in \bin^{\ell}} \sum_{\vec k \in \bin^{\nu}} \tilde p(\vec b)[\vec k]Y^{\toint(\vec k)} \eq(\vec b, \vec x)\\
	&=& \sum_{\vec k \in \bin^{\nu}} \underbrace{\left( \sum_{\vec b \in \bin^{\ell}}  \tilde p(\vec b)[\vec k] \eq(\vec b, \vec x) \right)}_{= a[\vec k] \in \ZZ_q} Y^{\toint(\vec k)}
	= \sum_{k=0}^{N-1} a[k] Y^{k} .
\end{eqnarray*}
By plugging the above definition of $a[\vec k]$ into the definition of $\ueval$
\[
 \sum_{\vec k \in \bin^{\nu}}  a[\vec k] \cdot \eq(\vec k,\vec z) = \sum_{\vec b \in \bin^{\ell}} \sum_{\vec
 k \in \bin^{\nu}} \tilde p(\vec b)[\vec k] \cdot \eq(\vec b, \vec x) \cdot \eq(\vec k, \vec z),
\]
which is $=\tilde u(\vec x, \vec z)$ by construction of $\unpack$ in Eq.~\eqref{eq:unpack}.
The computability of $\ueval$ in time $\bigO{N}$ follows by an algorithm from \cite{SP:VSBW13} that
computes all the multilinear Lagrange polynomials in time $\bigO{2^{\nu}}$.
\qed
\end{proof}

\headingb{Unpacked Virtual Oracles.}
A useful implication of Lemma~\ref{lemma:unpackedeval} is that we can create virtual oracles (see
Definition \ref{def:virtual-oracles}) of unpacked polynomials and run a PIOP on a virtual oracle
$\ora{g(\tilde u_1, \ldots, \tilde u_k)}$, where each $ \tilde u_i = \unpack(\tilde p_i)$ and $g$ is a
function that admits virtual oracles. %
This holds as long as the PIOP's verifier queries to each unpacked oracle $\ora{\tilde u_i}$ are only
on points of $\ZZ_q$.
This implication follows simply by seeing that for $\unpack$, the associated functions
$ev_{\unpack}, h$ for the virtual oracle (see Definition \ref{def:virtual-oracles}), are $ev_\unpack =
\ueval$ and $h(\vec x, \vec z) = \vec x$.
In Appendix \ref{sec:unpacked-sumcheck}, we give a simple example of how to use a virtual unpacked polynomial to prove a sumcheck
$ y =  \sum_{\vec b \in \bin^{\ell},\vec k \in \bin^{\nu}} \tilde p(\vec b)[\vec k]$ over the $\ZZ_q$ coefficients.

\subsection{PIOP for the Evaluation of Unpacked Polynomials on $\Rq$ points}\label{sec:piopueval}
The result of Lemma \ref{lemma:unpackedeval} shows how to compute the evaluation of an unpacked polynomial on $\ZZ_q$-points, given evaluations of its packed version.
In this section, we address the more general problem of evaluating an unpacked polynomial on points of $\Rq$. While these evaluations cannot be directly reduced to evaluations of the packed polynomial, here we present a protocol that allows proving such evaluation.
Precisely, we present a PIOP $\piopueval$ for the following oracle relation:
\[
\relueval = \{((y_1, \vec x_1, \ldots, y_{\np}, \vec x_{\np}, \ora{\tilde p}); \tilde p) : \forall i \, y_i=\tilde u(\vec x_i) \text{ where } \tilde u = \unpack(\tilde p) \}.
\]

\begin{Boxfig}[h!]{PIOP for $\relueval$}{PIOPueval}{\underline{PIOP $\piopueval$ for $((y_1, \vec x_1, \ldots, y_{\np}, \vec x_{\np}, \ora{\tilde p}); \tilde p) \in \relueval$}}
Let $\ell'=\ell+\nu$, $\vec X = (X_1, \ldots, X_{\ell'})$, $\vec Z = (Z_1, \ldots, Z_\nu)$.\\[-2em]
\begin{enumerate}
\item $\iopV$ sends $\vec\sigma \sample \Esetz^{\nu}, \tau \sample \Esetz$.
\item Let $E(\vec X, \vec Z, T) = \sum_{i=1}^{\np} \sum_{\vec k \in \bin^\nu} \eq(\vec X, \vec x_i)[\vec k] \cdot \eq(\vec Z, \vec k) \cdot T^{i-1} \in \ZZ_q[\vec X, \vec Z, T]$ and $\tilde E_{\sigma,\tau}(\vec X) = E(\vec X, \vec\sigma, \tau)$.
Let $\tilde y(\vec Z,T) = \sum_{\vec k \in \bin^{\nu}}  \eq(\vec k, \vec Z) \sum_{i=1}^{\np} y_i[\vec k] \cdot T^{i-1}$.

$\iopP$ and $\iopV$ run $\piopsumzq$ for $((\tilde y(\vec \sigma, \tau), \ora{f}); f) \in \relsum$ where  $f = \unpack( \tilde p) \cdot \tilde E_{\sigma,\tau}$.

	Given $\piopsumzq.\iopV$'s query to $\ora{f}$ on $\vec r = (\vec r_1, \vec r_2) \in \ZZ_q^{\ell + \nu}$,  the verifier $\piopueval.\iopV$:
	\begin{itemize}
	\item computes $\psi = \sum_{i=1}^{\np}\eq(\vec r,\vec x_i)\tau^{i-1}$, parsed as $\psi = \sum_{\vec k} \psi[\vec k] Y^{\toint(\vec k)}$,

	\item computes $\psi^* = \sum_{\vec k} \psi[\vec k] \eq(\vec k, \vec \sigma) $,

	\item makes a query to $\ora{\tilde p}$ on $\vec r_1$ and receives answer $v=\tilde p(\vec r_1)$,

	\item returns $\ueval(v, \vec r_2) \cdot \psi^*$ to $\piopsumzq.\iopV$.
	\end{itemize}
\end{enumerate}
\end{Boxfig}

The full description of $\piopueval$ is in Figure \ref{fig:PIOPueval}. Here we present the ideas in more
detail and justify its correctness. For simplicity, we focus on the case of evaluating at a single point,
i.e., $\np=1$; the extension to more points is straightforward observing that the aggregation random
challenge $\tau \in \Esetz \subset \ZZ_q$.

The goal is to prove that for $\tilde u = \unpack(\tilde p)$ it holds
$y = \tilde u(\vec x)$, which can be written as follows (where $\ell'=\ell+\nu$). %
\begin{eqnarray*}
\tilde u(\vec x) & = & \sum_{\vec b \in \bin^{\ell'}} \tilde u(\vec b) \eq(\vec b, \vec x)
	=  \sum_{\vec b \in\bin^{\ell'}} \tilde u(\vec b) \sum_{\vec k \in \bin^{\nu}} \underbrace{\eq(\vec b, \vec x)[\vec k]}_{\tilde E(\vec b, \vec k)} Y^{\toint(\vec k)}\\
	& = & \sum_{\vec k \in \bin^{\nu}} \sum_{\vec b \in\bin^{\ell'}} \tilde u(\vec b)  \tilde E(\vec b, \vec k)
	Y^{\toint(\vec k)} = \sum_{\vec k \in \bin^{\nu}} y[\vec k] Y^{\toint(\vec k)}.
\end{eqnarray*}
In particular, the last equality follows from the fact that $\tilde u(\vec b)$ as well as $\tilde E(\vec b, \vec k)$ are in $\ZZ_q$.
Therefore, we can reduce testing $y = \tilde u(\vec x)$ to checking that
for all $\vec k \in \bin^{\nu}$, $y[\vec k] = \sum_{\vec b} \tilde u(\vec b) \tilde E(\vec b, \vec k)$.
These are $2^{\nu}$ sumchecks that we can batch together by sampling a random
$\vec\sigma$ and checking
\[
\tilde y(\vec\sigma) = \sum_{\vec k \in \bin^{\nu}} y[\vec k] \eq(\vec k, \vec \sigma)  =   \sum_{\vec b \in
\bin^{\ell'}} \tilde u(\vec b) \tilde E(\vec b, \vec \sigma).
\]
This is done by running the PIOP $\piopsumzq$ (i.e., the sumcheck PIOP with the verifier's random challenges in $\Esetz \subset \ZZ_q$), which ends with the verifier querying $\tilde u(\vec r)$
and $\tilde E(\vec r, \vec\sigma)$. The query to $\tilde u$ can be computed through a query to $\tilde
p$ using $\ueval$. $\tilde E(\vec r, \vec\sigma)$ can be computed by the verifier as
$\psi^* = \sum_{\vec k} \psi[\vec k] \eq(\vec k, \vec \sigma) $, where
$\psi = \eq(\vec x, \vec r)$, parsed as $\psi = \sum_{\vec k} \psi[\vec k] Y^{\toint(\vec k)}$.
To see correctness:
\begin{eqnarray*}
\tilde E(\vec r, \vec\sigma) & = & \sum_{\vec b, \vec k} \tilde E(\vec b, \vec k) \eq(\vec b, \vec r) \eq(\vec k, \vec \sigma) =  \sum_{\vec k}  \sum_{\vec b} \eq(\vec b, \vec x)[\vec k] \eq(\vec b, \vec r) \eq(\vec k, \vec \sigma) \\
 & = & \sum_{\vec k} \eq(\vec x, \vec r)[\vec k] \cdot \eq(\vec k, \vec \sigma) = \psi^*,
\end{eqnarray*}
where last equality follows from $\eq(\vec x, \vec r) = \sum_{\vec b} \eq(\vec b, \vec x) \eq(\vec b, \vec r) $ and $\eq(\vec b, \vec r) \in \ZZ_q$.

\begin{theorem}\label{thm:piopuevalpacking}
$\piopueval$ is a PIOP for $\relueval$ with perfect completeness, prover's efficiency
$\bigO{\np2^{\ell+\nu}}$, verifier's efficiency $\bigO{\ell + \np \cdot N}$, $3 (\ell+\nu)\log q$ bits of
communication, and $\delta_{ueval}$-knowledge soundness with $\delta_{ueval} = (\np -1 +
2\ell+3\nu)/|\Esetz|$.
\end{theorem}

\begin{remark}[Parallel repetition of $\piopueval$]
If the exceptional set $\Esetz$ is not large enough to make $1/|\Esetz|$ negligible,
we apply $\kappa$-fold parallel repetition, denoted $\piopueval^{(\kappa)}$, that achieves soundness $\delta_{ueval}^{\kappa}$ and RBR-soundness $\delta_{ueval}^{\kappa/(\ell+\nu)}$ (see Theorem \ref{thm:parallelrep}) \cite{STOC:CCHLRR19}.
The efficiency of $\piopueval^{(\kappa)}$ is $\kappa$ times that of $\piopueval$ and its verification requires $\kappa$ queries to the (packed) oracle $\ora{\tilde p}$.
\end{remark}

%% file: new-range-proof-rq.tex
\section{Our PIOP for Range Checks over Packed Polynomials}\label{sec:range}
In this section, we present our new PIOP for range checks that improves over the prior solution from~\cite{C:CCCFGS25}.
More precisely, we construct a PIOP for
\begin{equation}
\label{eq:rangelookup}
\relrange = \{(\ind, \stm; \wit) = (T_{B}, \ora{\tilde v}; \tilde v) : \forall \vec b \in \bin^{\ell} \; \tilde v(\vec
b) \in T_{B} \},
\end{equation}
where $\Rq:=\ZZ_q[Y] / (Y^{N} + 1)$, $\tilde{v} \in \Rqml[X_1,\dots,X_\ell]$, and $T_B$ is the table of size $B^N$ containing all the elements of $\Rq$ with coefficients in $[0, B-1]$.

As shown in~\cite{C:CCCFGS25}, extending ideas of~\cite{EC:SetThaWah24} from finite fields to
$\Rq$, one can reduce a lookup in $T_B$ into $c \cdot N$ lookups in the smaller table $\vec t_{\beta}
= (0, \ldots,\beta - 1)$ for $\beta = B^{1/c}$. This uses a lemma similar to Lemma~\ref{lemma:Rq2Zqrange} below with a slight difference caused by our use of $\unpack$. See Appendix \ref{sec:auxrange} for the proof.

\begin{lemma}\label{lemma:Rq2Zqrange}
Let $B = 2^{b}$, $N = 2^{\nu}$, let $c=2^{\gamma}$ such that $c \mid b$ and $\beta = B^{1/c}$. Let $T_{B}$ be the table defined above, and $\vec t_{\beta} = (0, \ldots,\beta - 1)$.
Then for any $\tilde v \in \Rqml[X_1, \ldots X_{\ell}]$, $(T_{B}, \ora{\tilde v}; \tilde v) \in \relrange$ if and only if there exists \\ $ \tilde \mu \in \Zqml[X_1, \ldots, X_{\ell}, Z_1, \ldots, Z_\nu, Z'_1, \ldots, Z'_{\gamma}] $ such that, for $\tilde u = \unpack(\tilde v)$,
\begin{eqnarray}
 \forall \vec i \in \bin^{\ell+\nu} : && \tilde u(\vec i)  =  \sum_{\vec j \in \bin^{\gamma}} \tilde{\mu}(\vec i, \vec j) \cdot \beta^{\toint(\vec j)} \label{eq:vdecomposition} \\
\forall \vec b \in \bin^{\ell + \nu +\gamma} : && \tilde\mu(\vec b) \in \vec t_{\beta}
\label{eq:mugrange}
\end{eqnarray}
\end{lemma}

Based on this Lemma, one can construct a PIOP $\pioprange$ for proving that $(T_{B}, \ora{\tilde v}; \tilde v)$ is in $\relrange$ by letting the prover send a polynomial $\tilde \mu$ and then showing that (i) $\tilde \mu$ satisfies equation \eqref{eq:mugrange}, i.e., its evaluations are in the smaller table $\vec t_{\beta}$, and (ii) $\tilde \mu$ is correctly related to $\tilde v$ according to equation \eqref{eq:vdecomposition}, i.e, it is a $\beta$-decomposition of  $\unpack(\tilde v)$.
The protocol of \cite{C:CCCFGS25} followed a similar approach by constructing a PIOP for each of these equations;\footnote{Precisely, equation \eqref{eq:vdecomposition} is encoded differently without the $\unpack$ encoding.} however, a large source of inefficiency in their case stems from embedding $\tilde \mu$ in $\Rqml[\vec X, \vec Z, \vec Z']$ which, in spite of its sparsity, is an object that is $c \cdot N$ times bigger than $\tilde v$. In particular, this causes an $\O(N)$ overhead in the entire protocol.

In this paper, we avoid this overhead through the combination of two techniques.
First, we embed $\tilde \mu$ over a different ring $\Hq = \Zq[Y]/h(Y)$ and prove that it satisfies equation \eqref{eq:mugrange} for the table $\vec t_{\beta} \subset \ZZ_q$ via a PIOP over $\Hq$. Crucially, we choose $\Hq$ to be much smaller than $\Rq$, as $d = \deg(h) \ll N$ is a small constant (e.g., $d=2$ or $4$) that ensures the existence of a large enough exceptional set $\Esetd \subset \Hq$ (we explain how we choose $h$ below). With this choice, the PIOP running over $\Hq$ avoids altogether an $\O(N)$ overhead. In Appendix~\ref{sec:newPIOPrange} we also give an optimized variant that takes $\Hq = \FF_p$ for a large prime $p$, which we use in Section~\ref{sec:implementation}.

To complete the picture, we need to connect the range proof about $\tilde \mu$ to the polynomial $\tilde v$, which is the end-goal of $\pioprange$.
To this end, we relate $\tilde\mu \in \Hqml[\vec X, \vec Z, \vec Z']$ with $\tilde v \in \Rqml[\vec X]$ according to equation \eqref{eq:vdecomposition}, and leverage our unpacked evaluation technique to let the verifier query $\tilde u = \unpack(\tilde v)$ without requiring the prover to commit to $\tilde u$.

We remark that our technique is similar to the ring-switching used in \cite{EPRINT:HuaMaoZha25}, although in that work it is primarily used to transform an R1CS statement over $\Rq$ into one over a Galois ring where the entire computation is eventually done, while in our case the change of ring is used (in combination with our unpacking technique) for the specific task of proving ranges. Moreover, the ring-switching technique of \cite{EPRINT:HuaMaoZha25} was later used in \cite{EPRINT:NguRouZha26} to construct lattice-based multilinear polynomial commitments with concrete efficiency in verification.

In what follows we present our protocol in more detail.

\headingb{Our PIOP $\pioprange$.}
Consider the splitting of $\Rq \cong \prod_{i=0}^L \left( \prod_{j=0}^{k-1} R_{i,j} \right)$ described in section \ref{sec:proof-friendly-CKKS}, where each $R_{i,j} = \mathbb{Z}_{p_i}[Y]/(Y^d - \zeta_i^{(2j+1)})$ is isomorphic to the field $\FF_{p_{i}^{d}}$.
We set $\Hq  = \ZZ_q[Y]/h(Y)$ as a ring isomorphic to $\prod_{i=0}^L  R_{i,0}$  by the CRT, i.e., where for all $i$, $h(Y) \bmod p_{i} = (Y^d - \zeta_i)$. %
This choice ensures the existence of an exceptional set  $\Esetd \subset \Hq$ of size $p_{min}^{d}$, where $p_{min}=\min\{p_0, \ldots, p_{L}\}$.

Let $\relgrange$ be an oracle relation over $(\vec t_\beta, \ora{\tilde \mu}; \tilde \mu)$ defined in the same way as $\relrange$ except that its index is restricted to tables $\vec t_{\beta} \subset \ZZ_q$ and the relation is defined over multilinear polynomials $\tilde \mu$ with coefficients in $\Hq$.
It is easy to see that we can rephrase Lemma~\ref{lemma:Rq2Zqrange} by embedding $\tilde\mu$ in $\Hqml[\vec X, \vec Z, \vec Z']$ and replacing the condition of equation \eqref{eq:mugrange} with $(\vec t_\beta, \ora{\tilde \mu}; \tilde \mu)\in\relgrange$.

We construct a PIOP $\pioprange$ for $\relrange$ with table $T_{B}$ by using a PIOP $\piopGqrange$ for $\relgrange$ with the smaller $\vec t_{\beta}$,\footnote{We assume $B$ a power of 2; see \cite{C:CCCFGS25} for the extension to the general case.} combined with a probabilistic test to check equation \eqref{eq:vdecomposition}.
The full description of $\pioprange$ is in Fig.~\ref{fig:pioprange}.

The protocol proceeds as follows. The prover computes the unpacking of $\tilde v$ and its $\beta$-decomposition according to equation \ref{eq:vdecomposition} and sends the oracle $\ora{\tilde\mu}$ to the verifier. Then the goal is to convince the verifier that $(\vec t_\beta, \ora{\tilde \mu}; \tilde \mu)\in\relgrange$ and that $\tilde v$ and $\tilde\mu$ satisfy equation \eqref{eq:vdecomposition}. The former is achieved by executing the PIOP $\piopGqrange$ (step 3 in Fig.~\ref{fig:pioprange}).
For the latter, the idea is to test that the polynomial $\tilde u(\vec X, \vec Z)$ is a linear combination of partial evaluations of $\tilde \mu$, i.e.,  $\tilde u(\vec X, \vec Z)=\sum_{\vec j} \tilde\mu(\vec X, \vec Z, \vec j) \beta^{\toint(\vec j)}$.
Intuitively, this could be tested by evaluating the polynomials on a random point, i.e., querying $\tilde u(\vec r)$ and $\tilde\mu(\vec r, \vec j)$ for all $\vec j$ and testing $\tilde u(\vec r)=\sum_{\vec j} \tilde\mu(\vec r, \vec j) \beta^{\toint(\vec j)}$. However, these polynomials are over different rings with different exceptional sets.
To solve this issue we take the random point $\vec r$ in the exceptional set $\Esetz \subset \Zq$, that is common to both $\Rq$ and $\Hq$. However, since $\Esetz$ is not large enough to make $1/|\Esetz|=p_{min}^{-1} \approx 2^{-\secpar}$ we repeat this test $\kappa$ times. Notably, since this parallel repetition test happens in a single round we can take the smallest $\kappa$ that make $p_{min}^{-\kappa}\approx 2^{-\secpar}$; e.g., for our choice of $45$-bit primes (see Appendix \ref{sec:exp-results}) we take $\kappa=3$.
The last problem to solve is that the test above requires querying a polynomial $\tilde u$ that is larger than $\tilde v$ by a factor of $N$.
We define $\tilde u$ as an unpacked virtual oracle, and using the fact that the random challenges are in $\Esetz \subset \Zq$, we apply our result of Lemma \ref{lemma:unpackedeval}, which reduces queries to $\tilde u = \unpack(\tilde v)$ to queries to the (dense) packed polynomial $\tilde v$.

\begin{Boxfig}[H]{Construction of the PIOP $\pioprange$ from $\piopGqrange$}{pioprange}{\underline{\textbf{PIOP $\pioprange$ for $(T_{B}, \ora{\tilde v}; \tilde v) \in \relrange$}}}
\begin{enumerate}
\item $\iopP$ computes $\tilde u = \unpack(\tilde v) \in \Rqml[\vec X, \vec Z]$ and  $\tilde \mu \in \Gqml[\vec X, \vec Z, \vec Z']$ as the decomposition of $\tilde u$ according to equation \eqref{eq:vdecomposition},
 and sends $\ora{\tilde \mu}$.

$\iopV$ samples $\vec r_1, \ldots, \vec r_\kappa \sample \Esetz^{\ell+\nu}$ and sends them to $\iopP$.

\item $\iopP$ sends $\tilde\mu^*_i(Z'_{1}, \ldots, Z'_{\gamma}) = \tilde \mu(\vec r_i, Z'_{1}, \ldots, Z'_{\gamma})$, for $i=1$ to $\kappa$.

$\iopV$ samples $\vec s \sample \Aq^{\gamma}$ and, for $i=1$ to $\kappa$, queries the oracle $\ora{\tilde\mu}$ on $(\vec r_i, \vec s)$, and the virtual oracle $\ora{\tilde u}=\ora{\unpack(\tilde v)}$ on $\vec r_i$. Then $\iopV$ checks the following:\\
 $\forall i =1,\ldots,\kappa$: $\tilde u(\vec r_i)  =  \sum_{\vec j \in \bin^{\gamma}} \tilde{\mu}^{*}_i(\vec j) \cdot \beta^{\toint(\vec j)}$
and $\tilde\mu(\vec r_i, \vec s)  = \tilde{\mu}^{*}_i(\vec s)$

\item $\iopP$ and $\iopV$ run a PIOP $\piopGqrange$ for $(\vec t_{\beta},  \ora{\tilde \mu};  \tilde \mu) \in \relgrange$.
\end{enumerate}
\end{Boxfig}

The proof of the following theorem appears in Appendix \ref{sec:proof-piopgqrange2pioprange}.

\begin{theorem}\label{thm:piopgqrange2pioprange}
Let $\Aq \subset \Gq$, and $\Esetz \subset \Zq$ be exceptional sets.
Let $\piopGqrange$ be a PIOP for $\relgrange$ with perfect completeness and RBR-knowledge error $\hat\delta_{range}$.
Then $\pioprange$ is a PIOP with perfect completeness and RBR-knowledge error
$\delta_{range} = \max((\frac{\ell+\nu}{|\Esetz|})^{\kappa}, \frac{\kappa \cdot \gamma}{|\Aq|}, \hat\delta_{range})$.
\end{theorem}

To obtain the final protocol $\pioprange$, we need to instantiate the PIOP $\piopGqrange$ for the relation $\relgrange$ where the table in the index is $[0, \beta - 1] \subset \ZZ_q$ and the PIOP works over the ring $\Gq$.
We set this protocol to be the same as that in \cite{C:CCCFGS25} with the only difference that instead of working over $\Rq$ and its exceptional set $\Eset \subset \Rq$, we redefine it over the ring $\Gq$ with exceptional set $\Aq$.
For completeness, we give a full description of $\piopGqrange$ in Appendix \ref{sec:newPIOPrange}, and present an optimized version in Appendix~\ref{sec:pioprangeopt}. We also defer to Appendix \ref{sec:efficiency-pioprangeopt} a discussion on its efficiency, and we give a full efficiency analysis for concrete parameters in Section \ref{sec:implementation}.

%% file: ac_plusplus.tex
\section{Improved Arithmetization and PIOP $\Pi_{AC}$ for CKKS} \label{sec:PIOP_AC}

We construct a PIOP for the relation
$\rel_{AC}$
(Section~\ref{sec:proof-friendly-CKKS}) which, combined with $\pioprange$ from the previous
section, yields a PIOP for proving computations on CKKS ciphertexts, following~\cite{C:CCCFGS25}.
We refer to Appendix \ref{sec:together} for an explanation about which of the
polynomial oracles from this section are in common with the input oracle $\pioprange$, i.e., which
values require a range check.

We improve upon the arithmetization of CKKS in~\cite{C:CCCFGS25} on several fronts.
First, we replace their arithmetic circuit for $\relac$ with a more expressive one which allows to ``copy'' (into a MLE $\Vi[3,copy]$) the results of plaintext-ciphertext products without going through base decomposition and rescaling. We also allow  plaintext inputs from the prover, provided through an oracle MLE $\Vi[3,ptxt]$. This enables commit-and-prove SNARKs where the prover commits to $\Vi[3,ptxt]$.

Second and most importantly, by careful labeling of the wires in this new circuit, we construct MLEs for
the wiring predicates that are sparse in a very structured way, and have significantly fewer variables.
Beyond the obvious improvement of reducing the amount of variables, this structured sparsity
speeds up the sumcheck-based layer consistency equations in different ways:
\begin{itemize}[leftmargin=*]
	\item We provide specialized linear-time prover algorithms that
	improve upon a generic application of the algorithms from~\cite{C:XZZPS19,CCS:ZLWZSX21};
	\item For wiring predicates that are independent of the structure of the HE circuit (i.e. the ones that
	describe base decomposition and rescaling), we provide closed form equations that the verifier can
	easily evaluate on their own;
	\item For those that are not, we provide efficient polynomial commitments in Section~\ref{sec:sparse_PC} that exploit the structured sparsity. There are up to $7$ oracles of this form in our work, depending on the actual circuit structure. For example, if we only wanted to capture  the less  expressive relation  $\rel_{AC}$ in \cite{C:CCCFGS25} without plaintext inputs, we would only have $3$ oracles of this form.
	\item For a HE circuit of depth $D$, the PIOP of \cite{C:CCCFGS25} had $13 + 2(D-1)$ sparse polynomials in the index and $D+1$ dense polynomials in the witness. We describe how to implement our more expressive relation with up to $7$ sparse polynomials (with fewer variables) and a single dense one for the witness.

\end{itemize}

\noindent{\bf Notation:}
Within this section, let $\delta = \ceil{\log_2(D)}$ and let $\mu = \ceil{\log_2(2M)}  + \delta$, where
$D$ is the depth of the HE circuit and $M$ is the number of HE multiplications in its layer with the
most multiplications.
We denote vector of variables as $\vX=(X_0, X_1,\ldots)$ and refer to the leftmost one as the least significant one.
By abuse of notation, we use $d_2$ (resp. $d_0, d_1$) to refer to the set of wires holding the $d_2$ (resp. $d_0, d_1$) output of premultiply, according to the notation in Appendix~\ref{sec:pfckks}.
For a boolean vector $\vx$, $V_{i,S}(\vx)$ denotes the value on wire $\vx$ for the set of wires $S$ in GKR layer $i$ (not to be confused with the CKKS level).

Due to space constraints, the predicates  already introduced in~\cite{C:CCCFGS25} are given in
Appendix~\ref{sec:App_PIOP_AC}.
Table \ref{table:MLE_cheatsheet} gives information about the amount of variables in the different MLEs for the intermediate values of the circuit, as well as what  they are meant to index from their least to their most significant variables.
Some subset of these variables can be thought of as ``metadata''. For example, each predicate has $\delta$ variables which, together, index one of the $D$ layers in the HE circuit.
Table \ref{table:MLE_cheatsheet} does not include
$\Vi[3,copy], \Vi[3,ptxt]$ and $\Vi[in]$, since we do not need closed form equations for the predicates that interface with them.

Notice that the overhead of adding our different forms of metadata is small. For example, without the ``HE level metadata'', predicates in Table \ref{table:MLE_cheatsheet} would need to have as many variables as $\log_2$ of the number of their relevant wires in the whole HE circuit.
Given the metadata, we only need to have as many variables as the \emph{maximum} among such relevant kind of wires on each level.
For example, let $\{d_{2,j}\}_{j=0}^{D-1}$ represent the subset of $d_2$ wires that belong to the $j$-th HE layer. The MLE $\Vi[2, d_2](\vXvar_{[\#(d_{2})]})$ is such that $\#(d_{2}) = \delta + \max(\{\#(d_{2,j})\}_{j=0}^{D-1}) = \mu - 1$.

\begin{table*}[t]

	\centering
	\begin{tabular}{|c|c|c|c|c|c|c|}
		\hline
		&  \multicolumn{2}{|c|}{LSPs} &  \multicolumn{2}{|c|}{Middle positions} &   \multicolumn{2}{|c|}{MSPs} \\
		\hline
		& Bits & Meaning 				& Bits & Meaning 						& Bits & Meaning \\
		\hline
		$\Vi[0]$ & $\Tvar_0, \Tvar_1$& Gate type  		& $\vZvar$ & 	Gate number (for that type)	& - & - \\
		\hline
		$\Vi[1]$ & $\vZvar_{[\delta]}$ & HE level & $\Zvar_\delta$&  key-switch output type: $c'_0$ vs $c'_1$ &  $\vZvar_{(\delta,\mu)}$ & wire on level \\
		\hline
		$\Vi[2,d_{01}]$  & $\vZvar_{[\delta]}$& HE level  &  $\Zvar_\delta$&  pre-mult output type: $d_0$ vs $d_1$ &  $\vZvar_{(\delta,\mu)}$ & wire on level \\
		\hline
		$\Vi[2,d_{2}]$  & $\vZvar_{[\delta]}$& HE level &  $\vZvar_{[\delta,\mu - 1)}$ & wire carrying pre-mult output $d_2$  & - & - \\
		\hline
		$\Vi[2]$ & $\vZvar_0$& $d_{01}$ vs $d_2$ &  \multicolumn{4}{|c|}{As $\Vi[2,d_{01}]$ or $\Vi[2,d_{2}]$}   \\
		\hline
		$\Vi[3,BD]$& $\vXvar_{[\delta]}$& HE level  & $\vZvar_{[\delta,\mu - 1)}$& wire carrying pre-mult output $d_2$ & $\vXvar_{[\delta,2\delta)}$& specific $p_i$   \\
		\hline
		$\Vi[3,quot]$ & $\vYvar_{[\delta]}$& HE level &  $\vYvar_{\delta}$ & mult output type: $c''_0$ vs $c''_1$  & $\vYvar_{(\delta,\mu)}$ & wire on level \\
		\hline
		$\Vi[3,rc_3]$ & $\vXvar_{[\delta]}$& HE level & $\vZvar_{[\delta,\mu)}$ &  $\Zvar_\delta$: $c''_0$ vs $c''_1$ | $\vZvar_{(\delta,\mu)}$: wire on level & $\vXvar_{\delta}$ & rmd vs quot \\
		\hline
	\end{tabular}
	\caption{Variable information for the different intermediate-values MLEs. LSPs = Least significant positions, MSPs = Most significant positions.}

	\label{table:MLE_cheatsheet}
\end{table*}

Let us now explain how we \emph{achieve} the closed form equations for the MLEs and in particular
their structured sparsity and reduced amount of variables. Whereas the first technique originates
from~\cite{CCS:ZLWZSX21}, to the best of our knowledge we are the first ones to exploit the
other two we present.

	\headingb{\cite{CCS:ZLWZSX21} trick:}
	In our layer consistency equations, rather than having a different set of variables for every kind of wiring predicate (as in \cite{C:CCCFGS25}), we use the same set of variables $\vXvar$ for all of them as much as possible. This results in equations of the form $\Vi[out](\vZvar) = \sum_j \sum_{\vx \in \bit^{l}} \tgate[j]'(\vZvar,\vx) \cdot \Vi[i_j](\vx_{[l_j]})$ where $l_j$ is the amount of variables in $\Vi[i_j]$ and $l = \max_j(l_j)$. While it could seem that $\tgate[j]'(\vZvar,\vx)$ is not truncated accordingly, we internally modify that MLE so as to selectively reduce the boolean hypercube, i.e. $\tgate[j]'(\vZvar,\vx) = \teq(\vec{1},\vx_{[l_j,l)}) \cdot \tgate[j](\vZvar,\vx_{[l_j]})$.

	\headingb{Metadata:} Our pieces of ``metadata'' in the wires allow to easily apply the level-appropriate constants when describing the MLEs for the wiring predicates that correspond to the rescaling and base decomposition operations.

	\headingb{$\vXvar$ to $\vZvar$:} Finally, a precise labelling of the wires involved in the expression
	\[\Vi[out](\vZvar) = \sum_j \sum_{\vx \in \bit^{l}} \tgate[j]'(\vZvar,\vx) \cdot \Vi[i_j](\vx_{[l_j]})\]
	results in MLEs $\tgate[j]'(\vZvar,\vx)$ which internally have terms $\teq(\vZvar_{[l'_j]},\vXvar_{[l'_j]})$ of significant length. Whenever such a factor is present, one may hope to drop the corresponding $l'_j$ summation variables altogether and feed $\vZvar_{[l'_j]}$ directly into $\Vi[i_j]$, obtaining expressions that look as follows, where $l' = \max_j(l_j-l'_j)$:
	\begin{equation}\label{eq:XtoZ}
	\Vi[out](\vZvar) = \sum_j \sum_{\vx \in \bit^{l'}} \tgate[j]'(\vZvar,\vx) \cdot
	\Vi[i_j](\vZvar_{[l'_j]},\vx_{[l_j-l'_j]}).
	\end{equation}
	This reduces the amount of $\vXvar$ variables, which are the ones driving the efficiency of the sumcheck protocol. The substitution, however, is not always legitimate: It may only be applied to those variables which the wiring predicate constrains \emph{exclusively} through the equality factor $\teq(\vZvar_{[l'_j]},\vXvar_{[l'_j]})$. If the predicate uses them for anything else, the right hand side of Eq.~(\ref{eq:XtoZ}) increases its degree in those variables and stops being the multilinear polynomial $\Vi[out]$. It still agrees with $\Vi[out]$ on the boolean hypercube, but not at the random point at which the GKR verifier evaluates the layer consistency equation, so an honest prover would be rejected. As we will see in Section~\ref{sec:PIOP_AC_equations}, the substitution is therefore \emph{not} applied to the ``HE level'' metadata variables of the predicates that select level-dependent constants ($\trescon[3], \tbdcon[3], \tevk$), which are kept in $\vXvar$.

\subsection{Layer consistency equations} \label{sec:PIOP_AC_equations}
For a reminder of the wiring predicates in \cite{C:CCCFGS25} see Appendix~\ref{sec:App_PIOP_AC_predicates}. In the following, we adapt those through the application of the three techniques outlined above. We directly provide the MLEs of these new variations.

\headingb{Layer 0:}
Let us start by looking at the MLE for the output layer $\Vi[0]$. We define an extra vector of variables $\vTvar = (\Tvar_0, \Tvar_{1})$ as metadata for output wires. For $\vTvar = (0,0)$ it is the output of a $\prescon$ gate, for $\vTvar = (1,0)$ output of a $\pbdcon$ gate and for $\vTvar = (0, 1)$ it is a $\pcopycon$ gate. The reason to add such variables rather than simply relying on $\vZvar$ to identify a gate is that it is simpler to write the MLEs of each different gate using the $\peq$ predicate, all while being able to group all of them into a single sumcheck for what should be $\Vi[0](\vTvar, \vZvar) = 0$.
Given the presence of $\vTvar$, we can limit the length of the vector $\vZvar$: %
It only has as many variables as the \emph{maximum} among the three kinds of gates: $\vZvar = (\Zvar_0, \ldots, \Zvar_{\max(\mu,\mu - 1,\#(copy)) - 1})$. %

Let $d = \#(V_{3,ptxt})$, let $m = \max(2\delta, d)$ and let $m' = 1 + \max(\mu, \#(in))$.
We apply the $\vXvar \to \vZvar$ optimization to all variables except the $\delta$ ones indexing the HE level of the layer-3 input wire, which we read from $\vXvar_{[\delta]}$.
Eq.~(\ref{eq:GKR0++})  provides a sumcheck-style consistency check for the output layer $\Vi[0]$.
Later on, Equations~(\ref{eq:rescon}), (\ref{eq:bdcon}), (\ref{eq:copycon3copy}) and (\ref{eq:copycon}) describe the MLEs of its wiring predicates. %
\begin{align} \label{eq:GKR0++}
	0 =  \Vi[0](\vTvar, \vZvar) &=  %
	\tbdcon[2](\vTvar) \cdot  \Vi[2, d_2](\vZvar_{[\mu - 1]}) + \trescon[1](\vTvar) \cdot
	\Vi[1](\vZvar_{[\mu]}) \  \\
	&- \tcopycon[3,copy](\vTvar) \cdot \Vi[3,copy](\vZvar_{[\#(copy)]}) \ \nonumber \\
	&+ \sum_{\vx \in \bit^m}  \Big( \tbdcon[3](\vTvar, \vZvar_{[\delta]},\vx) \cdot
	\Vi[3,BD](\vx_{[\delta]}, \vZvar_{[\delta,\mu - 1)}, \vx_{[\delta,2\delta)})  \ \nonumber \\
	&+ \trescon[3](\vTvar,\vZvar_{[\delta]},\vx) \cdot  \Vi[3,rc_3](\vx_{[\delta]}, \vZvar_{[\delta,\mu)}, x_\delta)  \  \nonumber \\
	&+ \sum_{\vy \in \bit^{m'}} \Big( \tcopycon(\vTvar, \vZvar_{[\#(copy)]},\vx,\vy) \cdot f(\vx_{[d]},\vy) \Big)	\Big), \nonumber
\end{align}
where $f(\vXvar_{[d]},\vYvar) =   \Vi[3,ptxt](\vXvar_{[d]}) \cdot g(\vYvar)$ and $g(\vYvar) =  (1 -
\vYvar_{m'-1}) \cdot \Vi[in](\vYvar_{[\#(in)]}) +  \vYvar_{m'-1} \cdot \Vi[3, quot](\vYvar_{[\mu]})$.
One important further improvement that we do just for this layer consistency equation is that we \emph{also truncate $\vZvar$ on the right hand side of Eq.~(\ref{eq:GKR0++})}. This improves efficiency without introducing any new attacks (see Appendix~\ref{sec:App_PIOP_AC_toomany}).

Notice how the $\vXvar \to \vZvar$ optimization is applied to the wire-indexing variables of $\Vi[3,BD]$ and $\Vi[3,rc_3]$ (namely $\vZvar_{[\delta,\mu-1)}$ and $\vZvar_{[\delta,\mu)}$), which the corresponding predicates constrain only through an equality factor, but \emph{not} to their HE-level variables, which are read from $\vXvar_{[\delta]}$. The three predicates that carry no $\vXvar$ at all, $\tbdcon[2], \trescon[1]$ and $\tcopycon[3,copy]$, are trivially compatible with the substitution and their terms are therefore left outside the sum. Every summand of Eq.~(\ref{eq:GKR0++}) is then multilinear in $\vZvar$, so the equation is an identity of polynomials and can be evaluated at the random point at which $\piopac$ invokes it.
Next, we explain the MLEs of each individual wiring predicate.

\paragraph{\underline{Rescaling consistency}}
$\trescon[1](\vTvar)$ is trivial.
The description of $\trescon[3]$ is a bit complicated, since it takes multiple values.
Notice that half of the inputs in layer 3 are quotients (which we will select when $\vXvar_{\delta} = 1$, and simply multiply by the suitable $p_l$) and the other half are remainders (selected when $\vXvar_{\delta} = 0$).
Since the first $\delta$ variables in $\vXvar$ hold the HE level of the input wire, we use them to select whether a particular input wire to the $\vz$-th rescon gate (which holds a quotient value, i.e. with $\vXvar_{\delta} = 1$) has to be multiplied by $-p_l$, or $-p_2$ etc.
Further multiplying $\teq(\tobin(l),\vXvar_{[\delta]})$ with $\teq(\tobin(l),\vZvar_{[\delta]})$ forces the level of the input wire to agree with the level of the rescon gate.

\begin{align} \label{eq:rescon}
	\trescon[1](\vTvar) &= \teq(\vTvar, (0,0)) \\
	\trescon[3](\vTvar,\vZvar_{[\delta]},\vXvar) &= -\teq(\vTvar, \vec{0}) \cdot \teq(\vXvar_{[\delta+1,m)}, \vec{1})\cdot \nonumber \\
	&\quad
	\Big( \sum_{l=0}^{D-1} \teq(\tobin(l),\vZvar_{[\delta]}) \cdot \teq(\tobin(l),\vXvar_{[\delta]}) \cdot \big(1 - \vXvar_{\delta} + \vXvar_{\delta} \,
	p_l \big) \Big). \nonumber
\end{align}
\paragraph{\underline{Base decomposition consistency}}
We define $\tbdcon[3]$ %
similarly to $\trescon[3]$.
\begin{align} \label{eq:bdcon}
	\tbdcon[2](\vTvar) &= \teq(\vTvar, (1,0)) \\
	\tbdcon[3](\vTvar,\vZvar_{[\delta]},\vXvar) &= -\teq(\vTvar, (1,0)) \cdot \teq(\vXvar_{[2\delta,m)}, \vec{1})  \cdot \nonumber \\
	&\quad \Big(\sum_{l=0}^{D-1}   \teq(\tobin(l),\vZvar_{[\delta]}) \cdot \teq(\tobin(l),\vXvar_{[\delta]}) \cdot \nonumber \\
	&\qquad  \big(\sum_{i=0}^{l} \teq(\tobin(i),
	\vXvar_{[\delta,2\delta)} ) \cdot \mathsf{PW}_{\omega_l}(1)[i] \big) \Big). \nonumber
\end{align}
\paragraph{\underline{Copying values consistency}}
The $\tcopycon(\vt,\vz,\vx,\vy)$ MLE selects values of $\vx, \vy$ such that, for a given $\vz$ it holds that $\sum_{\vx, \vy} ptxt[\vx] \cdot ctxt[\vy] = ctxt'[\vz]$, where $ctxt'[\vz]$ is the copied result.
All plaintexts are stored in $\Vi[3,ptxt](\vx_{[d]})$ (for which the verifier has oracle access).
The variable $\vYvar_{m'-1}$ chooses whether $ctxt$ comes from $\Vi[in](\vy_{[\#(in)]})$ (known to the verifier) or $\Vi[3, quot](\vy_{[\mu]})$ (prover's oracle).
$\Vi[3,copy](\vZvar_{[\#(copy)]})$ stores $ctxt'[\vz]$.
Since there is one $ctxt'[\vz]$ copy per check and $\pcopycon(\vt,\vz,\cdot,\cdot)$ is satisfied by as many pairs $(\vx,\vy)$ as the fan-in of that copy gate, the term carrying $\Vi[3,copy]$ has to sit \emph{outside} the sum over $\vx, \vy$, for which we use
\begin{equation} \label{eq:copycon3copy}
	\tcopycon[3,copy](\vTvar) = \teq(\vTvar, (0,1)).
\end{equation}
Note that the choice between $\Vi[in]$ and $\Vi[3,quot]$ is made \emph{inside} the sum, by $\vYvar_{m'-1}$.
See Appendix \ref{sec:App_PIOP_AC} for the precise definition of the $\pcopycon$ predicate.

\headingb{Layer 1:}
In this layer, let $\vZvar = (\Zvar_0, \ldots, \Zvar_{\mu - 1})$. $\Vi[1](\vZvar)$ contains the outputs of key switching.
Eq.~(\ref{eq:GKR1++}) checks consistency between $\Vi[1](\vZvar)$ and the $d_0, d_1$ outputs of pre-multiply (stored in $\Vi[2,d_{01}](\vZvar)$, in layer $2$) mixed with the inner product of the evaluation key and the base decomposition of $d_2$.

\begin{align} \label{eq:GKR1++}
	\Vi[1](\vZvar) = \Vi[2,d_{01}](\vZvar) + \sum_{\vx \in \bit^{2\delta}}  \Big(& \tevk(\vZvar_{[\delta+1]}, \vx)\cdot \Vi[3,BD](\vx_{[\delta]}, \vZvar_{(\delta, \mu)},\vx_{[\delta,2\delta)}) \Big).
\end{align}
Here too the HE-level variables of $\Vi[3,BD]$ are kept in $\vXvar$, since $\tevk$ uses the level to pick the evaluation key digits $\mathsf{evk}_{l,b}[i]$ and hence it is not of the form required by the $\vXvar \to \vZvar$ optimization.
The MLE can be thus written as:
\begin{align*}
	\tevk(\vZvar,\vXvar) &=  \sum_{b=0}^{1}  \teq(\vZvar_{\delta}, b)  \cdot \Big(\sum_{l=0}^{D-1}
	\teq(\tobin(l),\vZvar_{[\delta]}) \cdot \teq(\tobin(l),\vXvar_{[\delta]}) \cdot \\
	& \qquad \big( \sum_{i=0}^{l} \teq(\tobin(i), \vXvar_{[\delta,2\delta)}) \,
	\mathsf{evk}_{l, b}[i] \big) \Big) .
\end{align*}

\headingb{Layer 2:}
Finally, consistency between the values $\Vi[2](\vZvar)$ on layer $2$ and those in layer $3$ can be done through the predicate for inner products $\pXmult[](\vz,\vx,\vy)$
from \cite{CCS:LiuXieZha21}. %
Let $s' = \max(\#(in), \mu, \#(copy))$. In Eq.~(\ref{eq:GKR2++}), we limit the length of the vectors $\vx, \vy$ to $s'$ by having different, specific versions of $\pXmult$ for each kind of product that can occur.
$\pXmult[i,i]$ is an $\pXmult$ between $in$ wires, $\pXmult[q,q]$ between $quot$ wires, $\pXmult[q,c]$ between $quot$ and $copy$ wires, and so on.
In layer $2$, $\Vi[2](\vZvar)$ is such that $\vZvar = (\Zvar_0, \ldots, \Zvar_{\mu})$ and

\begin{align} \label{eq:GKR2++}
	\Vi[2](\vZvar) =& \sum_{\vx, \vy \in \bit^{s'}}
	\Big( \tXmult[i,i](\vZvar, \vx, \vy)\cdot \Vi[in](\vx_{[\#(in)]}) \cdot \Vi[in](\vy_{[\#(in)]}) + \nonumber\\
	&\tXmult[q,q](\vZvar, \vx, \vy)\cdot \Vi[3, quot](\vx_{[\mu]}) \cdot \Vi[3, quot](\vy_{[\mu]}) + \nonumber\\
	&\tXmult[i,q](\vZvar, \vx, \vy)\cdot \Vi[in](\vx_{[\#(in)]}) \cdot \Vi[3, quot](\vy_{[\mu]}) + \nonumber\\
	&\tXmult[i,c](\vZvar, \vx, \vy)\cdot \Vi[in](\vx_{[\#(in)]}) \cdot \Vi[3,copy](\vy_{[\#(copy)]}) + \nonumber\\
	&\tXmult[q,c](\vZvar, \vx, \vy)\cdot \Vi[3, quot](\vx_{[\mu]}) \cdot \Vi[3,copy](\vy_{[\#(copy)]})  + \nonumber\\
	&\tXmult[c,c](\vZvar, \vx, \vy)\cdot \Vi[3,copy](\vx_{[\#(copy)]}) \cdot \Vi[3,copy](\vy_{[\#(copy)]})
	\Big).
\end{align}

Depending on the HE circuit that is captured by this equation, some of the terms of the sum can be
removed if they are known to be zero. %

\ifFullVersion
\subsection{Progressing through the layers} \label{sec:PIOP_AC_progress}
The $\piopac$ PIOP starts by invoking the sumcheck \emph{protocol} on Eq.~(\ref{eq:GKR0++}) and moving towards GKR-layer 3.
We describe such series of reductions. For a set of wires $S_j$, $\Vi[S_j](\vr_j)$ denotes its MLE evaluations for the random sumcheck challenges $\vr_j$.

After the execution of sumcheck for Eq.~(\ref{eq:GKR0++}), and in order to reduce to a claim about $\Vi[1](\vr_1)$ together with a claim about $\Vi[2, d_2](\vr_2)$, the verifier needs to check the final claim from the prover by querying oracles $\ora{\Vi[3,BD]}, \ora{\Vi[3,rc_3]}, \ora{\Vi[3,quot]}, \allowbreak \ora{\Vi[3,ptxt]}, \ora{\Vi[3,copy]}, \ora{\tcopycon}$ and evaluating the closed form equations of $\tbdcon[2], \allowbreak  \tbdcon[3], \allowbreak  \trescon[1], \trescon[3], \tcopycon[3,copy], \Vi[in]$ at the relevant challenge points.

The value $\Vi[1](\vr_1)$ is plugged into Eq.~(\ref{eq:GKR1++}). Completing a sumcheck reduction to $\Vi[2,d_{01}](\vr_{01})$ requires $\iopV$ to query $\ora{\Vi[3,BD]}$ and evaluate $\tevk$ at the new challenges.

Claims  $\Vi[2,d_{01}](\vr_{d_{01}})$ and $\Vi[2, d_2](\vr_{d_2})$ need to be combined into a claim $\Vi[2](\vr_2)$. In order to do this, we follow the approach from \cite{C:CCCFGS25}, which employs sumcheck too. Due to space constraints, we give those details in Appendix~\ref{sec:App_PIOP_AC_progress}.
The reduction requires $\iopV$ to evaluate MLEs $\tcomb[2,d_{01}]$, $\tcomb[2,d_{2}]$ at the relevant challenges.

Finally, $\Vi[2](\vr_2)$ is substituted in Eq.~(\ref{eq:GKR2++}), where for a general circuit $\iopV$ has to evaluate $\Vi[in]$ on their own (twice) and query oracles
$\ora{\pXmult[i,i]}, \allowbreak \ora{\pXmult[q,q]}, \ora{\pXmult[i,q]}, \allowbreak \ora{\pXmult[i,c]}, \ora{\pXmult[q,c]}, \ora{\pXmult[c,c]}$ once
and oracles $\ora{\Vi[3,quot]}, \ora{\Vi[3,copy]}$ twice.
If all checks pass, this concludes $\piopac$.

Notice how the whole protocol requires up to $17$ oracle queries for a generic circuit. In Section \ref{sec:implementation}, since we can simplify Eq.~(\ref{eq:GKR2++}), we require a lot less.

\subsection{Prover algorithms}
Our description in Section~\ref{sec:PIOP_AC_progress} is focused on the verifier.
In Appendix~\ref{sec:GKR_prover}, we provide linear time prover algorithms that follow the techniques in \cite{C:XZZPS19,CCS:ZLWZSX21}. We improve upon them by exploiting our structured sparsity in different forms.
One general philosophy is that we always start to challenge from the least significant variables up to the most significant ones. This has positive impacts due to the fact that the prefixes which enable sparsity are placed in the most significant position.
We provide more efficient algorithms to update the function evaluation tables according to this challenge ordering rule, specialize for our specific wiring predicates and equations, show how to compute the sumcheck for the product of sparse vectors of different length, etc.

\fi

%% file: sparse-poly.tex
\section{Our PIOP for Evaluation of Sparse Polynomials} \label{sec:sparse_PC}
In this section, we present a PIOP, dubbed \spack, for proving the evaluation of a collection of sparse polynomials with coefficients in $\ZZ_q$.
More formally, let us define the indexed relation $\relspeval$ over tuples $(\ind,\stm,\wit)$ such that:
the index $\ind = (\ora{p_0}, \ldots, \ora{p_{\np-1}})$ consists of $\np$ multilinear polynomials in $m$ variables such that each $p_i$ evaluates to a non-zero $\ZZ_q$-value in at most $2^{\ell}$ locations of $\bin^{m}$;
the statement $\stm = (\vec x_0, y_0, \ldots, \vec x_{\np-1}, y_{\np-1})$ consists of $\np$ evaluation points in $\Rq^{m}$, and alleged results $y_i \in \Rq$;
the witness $\wit$ consists of the polynomials $p_0, \ldots, p_{\np-1}$.
Then the relation is
$
\relspeval := \{(\ind,\stm,\wit) : \forall i\in[\np] \; p_i(\vec x_i) = y_i\}.
$

The \spack PIOP is a crucial building block of the PIOP for $\relac$. It enables the prover to convince the verifier about the correct evaluation of the $7$ polynomials
$\tcopycon, \allowbreak \tXmult[i,i], \allowbreak \tXmult[q,q], \allowbreak \tXmult[i,q], \allowbreak
\tXmult[i,c], \allowbreak \tXmult[q,c],\allowbreak  \tXmult[c,c]$ that encode the description of the
circuit (see Section~\ref{sec:PIOP_AC}).
For a circuit of size $|C|$ such polynomials have $m \approx 3\log |C|$ variables and $\approx
2^{\ell}=\bigO{|C|}$ nonzero evaluations, all of them in $\ZZ_q$ and most of them being simply 1 as
they encode wiring predicates.
Naively using a standard polynomial commitment (PC) on these polynomials would incur $\Omega(2^{3\ell})$ proving time, i.e., cubic in the circuit size.
An alternative solution is to use a PC for sparse polynomials
whose state of the art is represented by Spark \cite{C:Setty20} (as generalized in
\cite{EC:SetThaWah24}). Despite being designed to work over finite fields, Spark might be extended to work in
$\Rq$ but unfortunately the large size of $\Rq$, compared to the usual size of finite fields used in
standard SNARKs, reduces its concrete efficiency. We estimated the efficiency of
Spark for parameters $N=2^{13}, \log q\approx135, \ell=15, m=34, \np=2$ that fit one of our neural
network applications (cf. Appendix \ref{sec:spack}) and compared it with the PIOP $\spack$ that we present in this section.

In a nutshell, the Spark-based solution would have a proof size of about 6.5 GiB while $\spack$ yields a 163 MiB proof.
The main reason of this gain is that Spark requires the prover to  generate at least
$a\lceil m/\ell \rceil=6$ commitments to $\Rq$-polynomials of size $2^{\ell}$ and later prove evaluations of $24$ polynomials (6.2 GiB with two batched Basefold proofs over $\Rq$ as in \cite{C:CCCFGS25}). In contrast, our PIOP {\em does not require generating any commitment at proving time} and it only requires the prover to compute a {\em single} evaluation proof for a polynomial commitment computed during preprocessing. Notably, by using our packing techniques, this committed preprocessed polynomial can be of size $2^{\ell + \log m + \log\np + 2 - \log N} = 2^{11}$, much smaller than in Spark. The size of this evaluation proof is 163 MiB with BatchFold. Note that, in terms of computation time, we also eliminate the need to commit to polynomials at proving time, which in Spark would take about 10 minutes of work.
Finally, we also obtain another gain in the size of the information-theoretic part of the proof, which with Spark would be 235 MiB, while in \spack it can be as low as 25 KiB. We provide more details of this comparison in Section~\ref{sec:implementation} and Appendix~\ref{sec:spack}.

\ifFullVersion
\subsection{Overview of the \spack PIOP}
\fi
In this section we explain the main ideas of our \spack PIOP, and refer to Appendix \ref{sec:spack} for its full description, security proof and efficiency analysis.

\paragraph{Sparse encoding}
Let $S=\{s_0,\dots,s_{2^\ell-1}\} \subset \bin^{m}$ be a subset of cardinality $2^{\ell}$ such that $\{\vec k \in \bin^m: p(\vec k)\neq 0\}\subseteq S$.
From now on we assume that $m=2^\mu$ is a power of $2$ (otherwise we can generalize the arguments below by appropriate padding of zeros). Let $\mat S \in \bin^{2^{\ell} \times 2^{\mu}}$ be the matrix with rows indexed by elements in $\bin^\ell$ and whose $\vec b$-th row contains $s_{\toint(\vec b)}\in S\subset\bin^m$.
Then a sparse $p$ can be represented with two multilinear polynomials $\val_{p}$ and $\spidx_{p}$ where $\spidx_p$ is the MLE of the above matrix $\mat S$ and $\val_{p}$ is such that
\[
\forall \vec b \in \bin^{\ell} : \val_{p}(\vec b) = p(\spidx_{p}(\vec b, \vec 0), \ldots, \spidx_{p}(\vec b, \vec
1)).
\]

For $\np$ polynomials $p_0, \ldots, p_{\np-1}$, each assumed to have up to $2^{\ell}$ nonzero evaluations, we can define (larger) polynomials $\val, \spidx$ containing all their sparse encodings such that, denoting $\lnp = \lceil \log \np\rceil$ and setting $p_i := 0$ for $\np \leq i < 2^{\lnp}$,
\[
\forall (\vec b, \vec i) \in \bin^{\ell+\lnp} : \val(\vec b, \vec i) = p_{i}(\spidx(\vec b, \vec i, \vec 0), \ldots,
\spidx(\vec b, \vec i, \vec 1)).
\]

\paragraph{Ideas of the proof.}
To prove $(\forall i\in[\np] ,\ p_i(\vec x_i) = y_i)$, let $\tilde y \in \Rqml[Z_1, \ldots, Z_{\lnp}]$ be the MLE of the vector $(y_0, \ldots, y_{\np-1})$ and let $\tilde x \in \Rqml[Z_1, \ldots, Z_{\lnp+\mu}]$ be the MLE of the vector of inputs $(\vec x_0, \ldots, \vec x_{\np-1})$, i.e., such that $\tilde x(\vec i, \vec j) = (\vec x_i)_j$.
Then we can express the statement to be proven as
\begin{eqnarray}
\tilde y(\vec Z) &=& \sum_{\vec i \in \bin^{\lnp}} p_{i}(\vec x_{i}) \eq(\vec i, \vec Z)
	=  \sum_{\vec i \in \bin^{\lnp}, \vec k \in \bin^{m}} p_{i}(\vec k) \eq(\vec i, \vec Z) \eq(\vec k, \vec x_i) \nonumber \\
	& = &  \sum_{(\vec b, \vec i) \in \bin^{\ell + \lnp}} \val(\vec b, \vec i) \eq(\vec i, \vec Z) \prod_{\vec j \in
	\bin^{\mu}} \eq(\spidx(\vec b, \vec i, \vec j), \tilde x(\vec i,\vec j)). \label{eq:speval-polystatement}
\end{eqnarray}

Using standard techniques, we can test this polynomial equation on a verifier's random challenge $\vec t \sample \Eset^{\lnp}$, reducing the problem to proving
\begin{equation}
\tilde y(\vec t) =  \sum_{(\vec b, \vec i) \in \bin^{\ell + \lnp}} \val(\vec b, \vec i) \eq(\vec i, \vec t)
\prod_{\vec j \in \bin^{\mu}} \eq(\spidx(\vec b, \vec i, \vec j), \tilde x(\vec i,\vec j)). \label{eq:iptree}
\end{equation}

The statement above is a highly structured computation which involves an inner product and $2^{\ell + \lnp}$ trees of multiplications of depth $\mu$, performed in parallel.
Therefore, we present \spack generically as a family of protocols parametrized by a PIOP for proving a
statement like that of Equation \eqref{eq:iptree}.
To instantiate this PIOP we considered two options---Thaler's multiplication protocol \cite{C:Thaler13} and a parallel version of Quark's grand product \cite{EPRINT:SetLee20}.
In \spack, we adopted the former as it avoids creating fresh commitments during the proof. A straightforward use would be expensive due to computing $\mu$ sequential sumchecks and due to the need, in the end, to perform an evaluation proof on the committed polynomials describing the sparse encoding, of size $\np m 2^{\ell}$.
In \spack we leverage the power of our packing technique to reduce these costs.

More formally, for an integer $\beta$, let $\piopiptree$ be a PIOP for the oracle relation
\[
\reliptree  = \{ (v, \ora{f_1}, \ora{f_2}, \ora{E})  :  v =  \sum_{(\vec b, \vec i) \in \bin^{\ell + \lnp}} f_1(\vec
b, \vec i) f_2(\vec i) \prod_{\vec j \in \bin^{\beta}} E(\vec b, \vec i, \vec j)
\},
\]
and in which the verifier makes one query to each oracle polynomial:
$f_1(\vec r_1, \vec r_1')$, $f_2(\vec r_2)$, $E(\vec r, \vec s)$.
It is easy to see how to use $\piopiptree$ to prove Equation \eqref{eq:iptree}
by running it on
$(\tilde y(\vec t), \ora{\val}, \ora{\eq(\vec X, \vec t)}, \ora{E_{\mu}} )$
where, for $\vec Z = (Z_1, \ldots, Z_{\lnp+\mu})$, $E_{\mu}(\vec X, \vec Z) = \eq(\spidx(\vec X, \vec Z), \tilde x(\vec Z))$,
which is multilinear in $\vec X$ and of degree 2 in $\vec Z$.

We improve on this solution through the following techniques.

\headingb{Packing and Preprocessing.}
The polynomials $\val,\spidx$ have all coefficients in $\ZZ_q$. Thus, assuming that $\lnp + \mu + 2 \leq \nu = \log N$, which for our values of $\lnp = 3$ and $\log N \approx 13$ would be satisfied for circuits of size up to $ \approx 2^{2^{\nu - 5}/3} \approx 2^{85}$, we can pack $\val,\spidx$ into a single polynomial $\packpol$ of size $2^{\ell}$.
However, since we have space to pack even more information in $\packpol$, we leverage that the $p_i$ polynomials are part of the index (thus they are not adversarially generated) to create additional polynomials in a deterministic preprocessing that allows us to reduce the depth of the multiplication tree in \eqref{eq:iptree}.

We do this as follows: we observe that the product in Equation \eqref{eq:iptree} can be also written as
$
\prod_{\vec j \in \bin^{\mu-1}} E_{\mu}(\vec b, \vec i, \vec j,0) E_{\mu}(\vec b, \vec i, \vec j,1)
$.
Let $\vec Z = (Z_1, \ldots, Z_{\lnp+\mu-1})$. While running $\piopiptree$ with the polynomial $\hat E_{\mu}(\vec X, \vec Z) = E_{\mu}(\vec X, \vec Z, 0) \cdot E_{\mu}(\vec X, \vec Z, 1)$ would make the tree of smaller depth, the degree of $\hat E_{\mu}$ would double in all variables and proving its evaluation would vanish the gain in depth.

Instead, we use preprocessing to create an additional polynomial that allows us to replace the product of   \eqref{eq:iptree} by
$\prod_{\vec j \in \bin^{\mu-1}} E^{*}_{\mu-1}(\vec b, \vec i, \vec j)$,
where $E^{*}_{\mu-1}$ has degree 1 in $\vec X$ and degree $2$ in $\vec Z$.

To do this, let $\spidx^*$ be the MLE of
$(\spidx(\vec b, \vec i, \vec j,0)\cdot\spidx(\vec b, \vec i, \vec j,1) : \forall \vec b, \vec i, \vec j)$, and
let $\tilde x^{*}$ be the MLE of $\tilde x(\vec i,\vec j,0) \tilde x(\vec i,\vec j,1)$.
Let us define the following polynomials
\begin{eqnarray}
\tilde x^{*}_0(\vec Z) & = & -2 \cdot \tilde x^{*}(\vec Z) +2 \cdot \tilde x(\vec Z,0) + \tilde x(\vec Z,1) -1 \nonumber\\
\tilde x^{*}_1(\vec Z) & = & -2 \cdot \tilde x^{*}(\vec Z) + \tilde x(\vec Z,0) + 2 \cdot \tilde x(\vec Z,1) -1 \nonumber\\
\tilde x^{*}_2(\vec Z) & = & 4 \cdot \tilde x^{*}(\vec Z) - 2 \cdot \tilde x(\vec Z,0) - 2 \cdot \tilde x(\vec Z,1) +1 \nonumber \\
\tilde x^{*}_3(\vec Z) & = &  \tilde x^{*}(\vec Z) - \tilde x(\vec Z,0) - \tilde x(\vec Z,1) +1 \nonumber \\
E_{\mu-1}^{*}(\vec X, \vec Z) & = & \spidx(\vec X, \vec Z,0) \tilde x^{*}_0(\vec Z)
+\spidx(\vec X, \vec Z,1) \tilde x^{*}_1(\vec Z) \label{eq:Epoly}\\
&& \qquad\qquad\qquad\qquad +\spidx^{*}(\vec X, \vec Z) \tilde x^{*}_2(\vec Z) \nonumber
+\tilde x^{*}_3(\vec Z).
\end{eqnarray}

By construction, it holds $\forall \vec b, \vec i, \vec j$:
$E^{*}_{\mu-1}(\vec b, \vec i, \vec j) = E_{\mu}(\vec b, \vec i, \vec j,0)E_{\mu}(\vec b, \vec i, \vec j,1)$.
\noindent
Therefore, in \spack we run $\piopiptree$ with the virtual oracle $\ora{E^{*}_{\mu-1}}$; this ends with a verifier's query $v = E^{*}_{\mu-1}(\vec r, \vec s)$, with $\vec r \in \Eset^{\ell}, \vec s \in \Eset^{\lnp + \mu-1}$.
Interestingly, answering this query requires no additional sumcheck: by Equation \eqref{eq:Epoly}, $E^{*}_{\mu-1}$ is a linear combination of $\spidx(\vec X, \vec Z,0)$, $\spidx(\vec X, \vec Z,1)$ and $\spidx^{*}(\vec X, \vec Z)$ whose coefficients $\tilde x^{*}_{i}(\vec Z)$ depend only on the public inputs. Hence the verifier computes the values $\tilde x^{*}_{i}(\vec s)$ on its own and obtains $v$ from the three queries $\spidx(\vec r, \vec s,0)$, $\spidx(\vec r, \vec s,1)$, and $\spidx^{*}(\vec r, \vec s)$.

In \spack, we create a polynomial $\pol$ of $\ell+\lnp+\mu+2$ variables that puts together $\val, \spidx$ and $\spidx^{*}$, but then we pack $\pol$ into a polynomial $\packpol$ with only $n=(\ell+\lnp+\mu+2 - \nu)$ variables. In particular, the three queries above become evaluations of $\pol$ on points that differ only in two coordinates, and we batch them into a single query to $\pol$ using two random challenges. We refer to Appendix \ref{sec:spack} for the details.

\headingb{Evaluating the (un)packed polynomial.}
The solution sketched above requires querying the unpacked $\ora{\pol}$ on points in $\Rq$, so we cannot use Lemma \ref{lemma:unpackedeval} to compute them from queries to $\ora{\packpol}$.
The last piece of the protocol, which solves this issue, is again a tool of our packing technique, namely our PIOP $\piopueval$ of Section \ref{sec:piopueval} for proving evaluations of unpacked polynomials on $\Rq$ points.

%% file: basefold.tex
\section{BatchFold}\label{sec:basefold}
In this section, we describe our polynomial commitment (PC) scheme to elements in $R_q[\vec X]$,
which  instantiate the PIOP oracles. For efficiency, our goal is to allow for batched openings of the
form $f_m(\vec{z}_m)=y_m$, for $m\in [M]$, at a smaller cost sublinear in $M$. We do this in two steps: we first reduce this question to constructing PC schemes over
\emph{finite fields} with batched openings of polynomials $f_m$ \textit{at a common point}
$\vec{z}$.
Then we instantiate the latter from BaseFold~\cite{C:ZeiCheFis24}.

\subsection{Batching Openings via CRT and HyperPlonk}
Recall that
$R_q\cong \prod_{i=0}^L \prod_{j=0}^{k-1} R_{i,j}$, where each $R_{i,j}$ is a finite field of
size $p_i^{d}$ and $d=N/k$ (in practice, $d=4$). Let $\Phi: R_q\rightarrow
\prod_{i=0}^L
\prod_{j=0}^{k-1} R_{i,j}$ denote the aforementioned isomorphism and $\Phi_{ij}: R_q\rightarrow
R_{i,j}$ its components.
We extend $\Phi$ and $\Phi_{ij}$ to $R_q^\ell$ by applying them coordinate-wise, and to $R_q[\vec
{X}]$ coefficient-wise.

In \cite{C:CCCFGS25}, given polynomial commitments $\PC_{R_{i,j}}$ for each $R_{i,j}[X_1,\dots,X_\ell]$, the authors construct a unified $\PC_{R_q}$ for  $R_q[X_1,\dots,X_\ell]$.
This boils down to committing to each CRT component, i.e.,
 $\PC_{R_{q}}.\Com (f):=(\PC_{R_{i,j}}.\Com (\Phi_{ij}(f)))_{i,j}$, and the evaluation function (for single openings) $\PC_{R_q}.\Eval$ shows $f(\vec{z})=y$ by running each  $\PC_{R_{i,j}}.\Eval$ on the claim $\Phi_{ij}(f)(\Phi_{ij}(\vec{z}))=\Phi_{ij}(y)$.

Our first (elementary) observation is that this can be extended to  more general $\Eval_M$ protocols for batches of $M$ evaluations: to prove $\{f_m(\vec{z}_m)=y_m: m\in [M]\}$, apply each $\PC_{R_{i,j}}.\Eval_M$ on $\{\Phi_{ij}(f_m) (\Phi_{ij}(z_{m}))=\Phi_{ij}(y_m): m\in [M]\}$.

\paragraph{Batching on isomorphic components.}
More interestingly, we can use the observation that for every $i \in [L+1]$, the fields $\{R_{i,j}\}_{j=0}^{k-1}$ are isomorphic to each other.
Therefore, all $\{\PC_{R_{i,j}}\}_{j=0}^{k-1}$ can be captured by a single scheme $\PC_{i}$. Then, if each $\PC_{i}$ is endowed with a $\PC_i.\Eval_{kM}$ (batching $kM$ openings) we can construct $\PC_{R_q}.\Eval_M$ (batching $M$ openings) that only uses each $\PC_i.\Eval_{kM}$ \textit{once}. Indeed, to prove $\{f_m(\vec{z}_m)=y_m: m\in [M]\}$, apply $\PC_i.\Eval_{kM}$ to prove the claim $\{\Phi_{ij}(f) (\Phi_{ij}(\vec{z}_{m}))=\Phi_{ij}(y_m): j\in [k], m\in [M]\}$, for every $i\in [L+1]$. This amounts to
 $L+1$ batched evaluations proofs, rather than $k(L+1)$ (if we were to apply $\PC_i.\Eval_M$ $k$ times) and is beneficial given that one execution  $\PC_i.\Eval_{kM}$ is cheaper than $k$ executions of $\PC_i.\Eval_M$, as we will see.

\paragraph{Reducing to evaluations on the same point.}
We now reduce the task of constructing $\Eval_M$ to creating a protocol $\Evalcommon_M$ that proves batched opening \emph{at a common evaluation point}, i.e. $\Evalcommon_M$ proves claims of the form $\{f_m(\vec z')=y'_m:  m\in [M]\}$ for a common $\vec z'$.
In \figref{fig:PC_hyperplonk_reduction}, we construct $\PC.\Eval_M$ from $\PC.\Evalcommon_M$ by using an idea  from HyperPlonk~\cite{EC:CBBZ23}. The overhead of this protocol is an additional sumcheck over $\{0,1\}^{\ell}$. The soundness error of $\Eval_{M}$ is at most $\frac{2\ell+M-1}{|\FF|}+\mathsf{error}_{\Evalcommon_M}$ where $\mathsf{error}_{\Evalcommon_M}$ is the soundness error of $\Evalcommon_M$, and $\FF$ is the field over which $\PC$ is defined (in our case, for each $\PC_i$, $\FF$ is of size $p_i^d$).

\begin{Boxfig}{Construction of $\PC.\Eval_M$ from $\PC.\Evalcommon_M$}{fig:PC_hyperplonk_reduction}{\underline{Construction of $\PC.\Eval_M$ from $\PC.\Evalcommon_M$}}
\begin{enumerate}
\item The verifier sends $s\sample\FF$ to the prover. For every $\vec{b}\in\{0,1\}^\ell$ the prover defines the value $h_{\vec{b}}:=\sum_{m=0}^{M-1} s^{m} \cdot \eq(\vec{b},\vec{z}_m)\cdot f_m(\vec{b})$. Let $\tilde{h}$ be the multilinear extension of the $h_{\vec{b}}$, i.e. the unique $\tilde h\in \FF^{\leq 1}[X_1,\dots,X_\ell]$ with $\tilde h(\vec{b})=h_{\vec{b}} \ \forall \vec{b}\in\{0,1\}^\ell$.

\item Prover and verifier now engage in a standard sumcheck protocol for the statement $\sum_{\vec b\in \{0,1\}^\ell} \tilde{h}(\vec{b})= S$ for $S:=\sum_{m=0}^{M-1} s^{m}\cdot y_m$ where, when the verifier queries $\tilde h(\vec{r})$ for some $\vec{r}\sample\FF^{\ell}$:

\begin{enumerate}
\item The prover opens $\{q_m=f_m(\vec{r}), m\in [M]\}$ with $\PC.\Evalcommon_M$.
\item If this is correct, the verifier computes $\tilde h(\vec{r})=\sum_{m=0}^{M-1} s^{m} \cdot  \eq(\vec{r},\vec{z}_m)\cdot q_m$
\end{enumerate}

The verifier accepts if the sumcheck (and the $\PC.\Evalcommon_M$ opening) passes.

\end{enumerate}

\end{Boxfig}

\subsection{Batching BaseFold openings at a common point}

\paragraph{BaseFold.}
We summarize the relevant features of BaseFold. A detailed description can be found in Appendix~\ref{app:basefold}.
The commitment to a polynomial $f$ is given by an oracle  $\pi_d$ (in practice, a vector commitment) to an encoding of its coefficient vector with certain linear error-correcting code $C_d$, of dimension $2^\ell$ (the number of coefficients of $f$). The protocol $\PC.\Eval$ for a single opening $f(\vec{z})=y$ consists of a sumcheck for $\sum_{\vec b\in\{0,1\}^{\ell}} f(\vec b)\eq(\vec b,\vec{z})=y$, and a consistency check for the fact that $\pi_d$ is indeed an oracle to an element in $C_d$ and is consistent with the sumcheck. The soundness of the evaluation protocol of $\PC$ depends on the minimum distance of the code $C_d$ and can be found in Appendix~\ref{app:basefold}, where we show an slight improvement with respect to the results in \cite{C:ZeiCheFis24}.

\paragraph{Batching at the same point.}
Our construction for $\PC.\Evalcommon_M$ is given in~\figref{fig:batchfold_same_point_compressed} where we leave the $\PC.\Eval$ of Basefold implicit. In Appendix~\ref{app:basefold},~\figref{fig:batchfold_same_point} we provide the full protocol $\PC.\Evalcommon_M$ including $\PC.\Eval$. Here we point out that we reduce $\PC.\Evalcommon_M$ to applying $\PC.\Eval$ on a linear combination $u=\sum_{m=0}^{M-1} s^{m} f_m$, where queries to $u$ are obtained by querying all $f_m$'s; since other than making these queries and computing the linear combination of the results, the rest of $\PC.\Evalcommon_M$ is independent of $M$ (it is essentially a single application of $\PC.\Eval$), we obtain an advantage over repeating $\PC.\Eval$ $M$ times.

\begin{Boxfig}{Batched opening of $M$ evaluations at common point.}{fig:batchfold_same_point_compressed}{\underline{$\PC.\Evalcommon_M$: Batched opening proof at common point $\vec{z}$}}

\begin{enumerate}
	\item The verifier sends $s\sample\FF$. The prover sets $u(\vec X):=\sum_{m=0}^{M-1} s^{m} f_m (\vec X)$. This implicitly defines $\pi_d:=\pi_{d,u}=\sum_{m=0}^{M-1} s^{m} \pi_{d,f_m}$.
	\item The prover and verifier compute $S=\sum_{m=0}^{M-1} s^{m} y_m$ and engage in $\PC.\Eval$ for the claim:
	``$\pi_{d}$ oracle to $\Enc_d(\vec{u})$ such that $u(\vec{z})=S$''. When the verifier queries $\pi_u$ at a coordinate $\mu$, this is answered with the answers $a_m$ to queries of $\pi_{d,f_m}$ at $\mu$, for all $m\in [M]$. The verifier can then compute the answer as $\sum_{m=0}^{M-1} s^{m} a_m$.

\end{enumerate}

\end{Boxfig}
We have the following soundness  (see Appendix~\ref{app:basefold}, Lemma~\ref{lem:soundness_evalcom_restated}).
\begin{lemma}\label{lem:soundness_evalcom}
	The soundness error of $\PC.\Evalcommon_M$ is at most $\frac{M-1}{|\FF|}+\mathsf{error}_{\PC.\Eval}$
	where $\mathsf{error}_{\PC.\Eval}$ denotes the soundness error of $\PC.\Eval$.
\end{lemma}

As for communication, $\PC.\Evalcommon_M$ has size $\mathcal{O}_{\lambda}(M\ell+\ell^2)$ (for
security parameter  $\lambda$), while naively repeating $M$ times $\PC.\Eval$ would yield
$\mathcal{O}_{\lambda}(M\ell^2)$ communication. See Appendix~\ref{app:basefold} for more precise
statements about proof sizes.

\paragraph{Putting everything together.}
We have shown that the BaseFold commitment $\PC$ with individual openings $\Eval$ can be compiled into a protocol $\PC_{R_q}$ for polynomials in $R_q[\vec X]$. The commitment itself consists in committing to each component $\Phi_{ij}(f)$ with Basefold over $R_{ij}$.
A batched opening $\Eval_{M}$ of $M$ evaluations at arbitrary points in $R_q$ can be obtained from
$L+1$ executions of $\PC.\Evalcommon_{Mk}$ (each in a field $R_{ij}$) and $L+1$ sumchecks needed in the reduction from $\Evalcommon_{Mk}$ to $\Eval_{Mk}$, as shown in the diagram below (where the second arrow is \figref{fig:PC_hyperplonk_reduction} and the first one was explained in \emph{Batching on isomorphic components}):
$$
\boxed{\vphantom{x_{x_x}^{x^x}}\PC_{R_q}.\Eval_M}
\leftarrow
\boxed{\vphantom{x_{x_x}^{x^x}} \{\PC_{i}. \Eval_{Mk}\}_{i\in[L+1]}}
\leftarrow
\boxed{\vphantom{x_{x_x}^{x^x}}\{\textrm \PC_{i}. \Eval_{Mk}^{com} + 1 \textrm{ sumcheck}\}_{i\in [L+1]}}
$$

The soundness error is
$(L+1)(\frac{2\ell+Mk-1}{\min\{p_i\}^d}+\mathsf{error}_{\PC.\Eval})$, where $\mathsf{error}_{\PC.\Eval}$ is an upper bound for the
soundness error of Basefold.
The communication complexity is
$\mathcal{O}_{\lambda}(MkL\ell+L\ell^2)$. In contrast, using $\PC.\Eval$ openings for
individual statements in each of the $Lk$ components would yield
$\mathcal{O}_{\lambda}(MkL\ell^2)$ communication.

%% file: experiments.tex
\section{Implementation and Evaluation}
\label{sec:implementation}
We first implemented our proposals over the verifiable HE framework of~\cite{C:CCCFGS25}\footnote{Available at
\url{https://github.com/vfhe/proof-friendly-CKKS} (commit \texttt{e44d520})}, which we later integrated into the \texttt{vfhe} library~\cite{vfhe-library} for better performance and reproducibility. This full version of the paper considers the updated implementation, and the code reproducing our results is available at~\url{https://github.com/vfhe/Batch-Pack-and-Prove}. Results for the previous implementation are presented in the printed version~\cite{AC:CCCFGS26}.

We follow a similar approach to~\cite{C:CCCFGS25} when analyzing performance. In particular, the performance of their approach is dominated by common proof
operations, such as the sumcheck and polynomial commitment procedures, which they benchmark
over their ``proof-friendly'' CKKS implementation. For fair comparison, we perform a similar analysis, in the same ring, with the same HE parameters.
We run all experiments on an \texttt{r7i.metal-48xl} instance on AWS (Intel Xeon
8488C at 3.2GHz).

\subsection{BatchFold}\label{sec:batchfold}
\begin{figure}[t]
	\centering
	\begin{subfigure}[t]{0.5\columnwidth}
			\centering
			\includegraphics[width=\columnwidth]{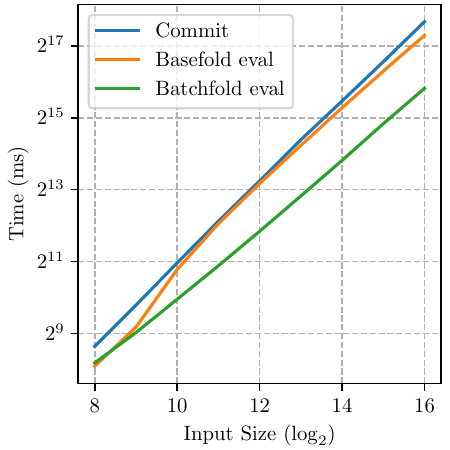}
			\label{fig:results:pc_time}
	\end{subfigure}%
	~
	\begin{subfigure}[t]{0.5\columnwidth}
			\centering
			\includegraphics[width=\columnwidth]{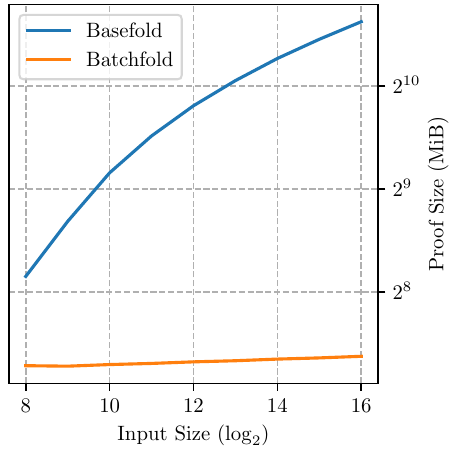}
			\label{fig:results:pc_size}
		\end{subfigure}
	\vspace{-0.6cm}
	\caption{Basefold~\cite{C:ZeiCheFis24} and \textbf{Batchfold} (our work) for
	$\Rq$ with $N = 2^{ 13}$ and 3 RNS components. Time is single-threaded and
	includes the Merkle hashing. We consider only the batching of $N/4$ components in $\Rq$.}
\label{fig:results:pc}
\vspace{-0.3cm}
\end{figure}

Our first improvement over~\cite{C:CCCFGS25} is the change of PC scheme from
Brakedown to Basefold, and in particular to our new BatchFold variant. Basefold commitments are
asymptotically more expensive than Brakedown
(quasilinear instead of linear) but they have logarithmic-time verification (instead of
$\BigO(\sqrt{2^\ell})$). Thanks to our new techniques (notably packing and the use of $\Hq$), many
polynomials we commit to in our protocols are a factor $\BigO(N)$ smaller than in
\cite{C:CCCFGS25}, allowing the prover to bear the extra costs of quasilinear proving time, in
exchange for significantly faster verification. BatchFold improves performance over
Basefold not only for the verifier, but also for the prover computing evaluation proofs, as Figure~\ref{fig:results:pc} shows. These measurements include the
Merkle commitments, where BatchFold also helps, since it hashes significantly less information. Comparisons between plain Basefold and Brakedown can be
found in~\cite{C:ZeiCheFis24}. As additional contribution, we implement an estimator for
computing optimal parameters for BaseFold and BatchFold.

We achieve an up to $2.8\times$ performance improvement upon Basefold for evaluation proofs, while proof size is reduced by up to a factor of $9.5\times$. Our impact to high level protocols could, however, be much greater than this, as our results are only considering the batching of the $N/4 = 2^{11}$ CRT components of $\Rq$ while high level protocols could also batch evaluations of multiple different multivariate polynomials.
Compared to~\cite{C:CCCFGS25}, which takes 8.5 seconds per CRT component of $\Rq$ (i.e., for each $R_{i,j}$) for input size $2^{16}$ using the original implementation of Brakedown, we achieve improvements of up to $302\times$, taking just 57.7 seconds for the same instance.~\cite{C:CCCFGS25} also presents an improved version of Brakedown, but they do not report
proof size and proving time for it. Although potential improvements from the optimized version in~\cite{C:CCCFGS25} should be taken into consideration, we are only measuring single-threaded time, while~\cite{C:CCCFGS25} uses up to 48 threads. Therefore, we expect our results and improvements to hold up, improvements to~\cite{C:CCCFGS25} notwithstanding.

\vspace{-0.3cm}
\subsection{High-level applications}

For the polynomial commitment, we provided a direct comparison between ours and \cite{C:ZeiCheFis24,C:CCCFGS25} based on a generic polynomial; however, demonstrating our broader contributions to verifiable computation requires a less generic benchmark.
In particular, while \cite{C:CCCFGS25} showcases a microbenchmark based on a single CKKS multiplication, we go significantly beyond this and consider small HE applications.
We consider 3 small generic neural network models, each composed of 2 fully-connected hidden
layers using $f(x) = x^2$ as the activation function.
We denote our models by tuples $(n_0,n_1)$, representing the number of neurons in the hidden layers. All models take 30 inputs, and have 2 outputs -- thus being suitable in small classification problems (\emph{e.g.},~\cite{wisconsin,income}). We consider batched evaluations using CKKS's SIMD packing~\cite{AC:CKKS17} (\emph{i.e.}, $N/2 = 2^{12}$ samples are classified at once) and present the detailed architecture in Appendix~\ref{sec:exp-results}. We note that our goal is to demonstrate the feasibility of
our vHE approach, and that machine learning optimization aspects are out of our scope. We compile
these small HE applications into our $\rel_{CKKS}$ relations -- composed of GKR circuits and range
checks -- and compute precisely the number of wires, gates and commitments needed to evaluate
and prove them. From those, we measure the range proofs end to end, and estimate the other components from measurements of their building blocks: we consider the execution times for polynomial commitment operations given in Section~\ref{sec:batchfold}, and we implement and benchmark SumChecks, following \cite{C:CCCFGS25}.

\paragraph{Arithmetic circuit.} In \cite{C:CCCFGS25} the performance analysis focuses on the
range proof protocol $\Pi_{range}$, under the assumption that $\Pi_{AC}$ and the actual HE and
GKR evaluations add negligibly to the overall cost of the protocol. Our estimates confirm
that the GKR evaluation costs are indeed very small, taking 21.9, 5.9, and 3.0 seconds for our three circuits, respectively. On the other hand,
$\Pi_{AC}$ should also prove the evaluations of large sparse polynomials that describe the
circuit, an aspect overlooked by \cite{C:CCCFGS25} and left to be handled through generic
techniques.
In this work, we improve $\Pi_{AC}$ in two ways. First, thanks to our refined arithmetization we
minimize the numbers of polynomials encoding the circuit structure that have to be proven. Second, we propose the \spack PIOP
specialized to prove the evaluation of these sparse polynomials.
Therefore, in our evaluation below we focus on comparing these two components, sparse commitments and range proofs.

\paragraph{Sparse commitments.}
We compare the performance of \spack with that of using Spark \cite{C:Setty20,EC:SetThaWah24}.
Our benchmark are the evaluations of the sparse polynomials arising from the HE circuits corresponding to the three neural networks mentioned earlier.
Table~\ref{tab:res_all} presents our results.
We are able to reduce proof sizes by more than an order of magnitude. Execution time is also improved by
factors of $10$--$23\times$. A large benefit of our construction is that
it does not require any commitments at run time -- all our commitments depend only on the circuit and not on the inputs, enabling them to be
precomputed. Notice that most of the cost of our solution comes from SumChecks, which could be
trivially parallelized. Appendix~\ref{sec:spack} further discusses and details the operations required
by each solution.

We emphasize that here we are not comparing our solution with an application of Spark \cite{C:Setty20,EC:SetThaWah24} to handle the sparse polynomials encoding the circuit wiring in \cite{C:CCCFGS25}.
Such a comparison would likely show even further improvements for us since the arithmetization of \cite{C:CCCFGS25} requires more sparse polynomials (and with more variables) than in our case (see Section \ref{sec:PIOP_AC}).

\begin{table}[h]
	\centering
	\caption{Performance of our \spack and comparison with Spark. Notice that our solution does not
	require execution-time commitments.}
	\label{tab:res_all}
	\begin{tabular}{c|cc|c|cc|cc|cc}
		\hline
		\multirow{3}{*}{\shortstack{Model\\$(n_0, n_1)$}}
		& \multicolumn{5}{c|}{Spark}
		& \multicolumn{4}{c}{\spack} \\
		\cline{2-10}

		& \multicolumn{2}{c}{$\mathtt{SumCheck}$}
		& \multicolumn{1}{c}{$\mathtt{Comm.}$}
		& \multicolumn{2}{c|}{$\mathtt{Eval}$}
		& \multicolumn{2}{c}{$\mathtt{SumCheck}$}
		& \multicolumn{2}{c}{$\mathtt{Eval}$} \\
		\cline{2-10}

		& Size & Time & Time & Size & Time
		& Size & Time & Size & Time \\
		\hline
		$(128, 64)$   & 235.3 MiB & 6m14s & 9m38s & 6.2 GiB & 6m35s & 25.4 KiB &
		1s & 163.0 MiB & 2m6s \\
		$(64, 16)$   & 176.5 MiB & 1m33s & 2m8s & 5.5 GiB & 1m37s &
		17.7 KiB & 1s & 159.4 MiB & 13s  \\
		$(32, 16)$     & 150.3 MiB & 47s & 58s & 5.2 GiB & 48s & 16.7 KiB & 1s & 159.4 MiB
		& 13s \\
		\hline
	\end{tabular}
\end{table}

\paragraph{Range proofs.} Compared to \cite{C:CCCFGS25}, our strategy of packing and working over $\Hq$ enables improvements of
$\BigO(N/d)$ times in both proof size and performance.
We measure execution times of 5m54s, 2m47s, and 1m24s, respectively for each of our three circuits above, with the optimized variant $\hat\Pi^{*}_{range}$ of our range proof that takes $\Hq = \FF_p$ (Fig.~\ref{fig:pioprangeopt}; parameters $\sigma = 2^{11}$, $m=15$, $\kappa=20$, Appendix~\ref{sec:efficiency-pioprangeopt}). Appendix~\ref{app:additional_results} details our results.
It is difficult to precisely estimate the costs for~\cite{C:CCCFGS25}, since their protocol is unlikely to handle the circuits of our benchmark without unrealistic amounts of memory (given the $\BigO(N)$ overhead of their range proof, discussed in Section~\ref{sec:range}). Even if
possible,~\cite{C:CCCFGS25} estimates 25 seconds per multiplication, leading to a
proof
time of at least 1 hour 20 minutes for our largest circuit, which contains 192 multiplications. For the
same circuit, our range proof takes less than 6 minutes. Additionally, their results are for multithreaded
execution with up to 48 threads, whereas we are reporting only single-threaded execution time. Considering this, our wall-clock improvement of more than $13\times$ would
likely
translate to more than $100\times$ times in computational time.

\paragraph{Discussion on concrete verifier costs.}
Our experimental evaluation focuses on the concrete performance of the prover, which is often regarded as the major challenge in verifiable HE with sublinear verification, and leaves the implementation-level analysis of the verifier as future work. The \texttt{vfhe} library provides timings for the verification of our building blocks. For example, a Basefold (resp.\ BatchFold) evaluation proof over $\Rq$ verifies in 0.4--1.7\,s (resp.\ 0.2--0.3\,s) for $2^8$ to $2^{16}$ entries, and the range proof of Table~\ref{tab:res_range_proof} at its largest size, $2^{24}$ digits, takes 1.8\,s, single-threaded. The full verification of our IOPs, however, would require procedures beyond the verification of the building blocks, and extrapolating these timings is difficult.

We nevertheless remark that, asymptotically, our protocol achieves polylogarithmic verification time in the circuit size, improving over the square-root verification complexity of \cite{C:CCCFGS25}. This asymptotic improvement suggests that our approach is fundamentally more scalable and that, for sufficiently large computations, verification costs can be amortized and be faster than trivial solutions, such as recomputing the circuit.
Moreover, recomputation is not always the appropriate baseline in applications of verifiable FHE. In
some scenarios, the verifier may not have access to the prover's inputs and therefore cannot
independently recompute. An example is privacy-preserving ML, where a server holding a
proprietary model evaluates encrypted client queries.
As a future direction, our protocols could be extended to this scenario by using circuit
privacy~\cite{STOC:Gentry09} and a zero-knowledge version of our proof
system. We expect this to work based on standard techniques. For the former, one may either rely on noise flooding~\cite{STOC:Gentry09} or recent proposals, such as Differential-Privacy-based circuit privacy for CKKS~\cite{AC:HMSS25,CiC:KluSan25}, which are concretely more efficient.
Regarding ZK, the conversion can be achieved through  the zero-knowledge
sumcheck of Libra \cite{C:XZZPS19} and PCs with ZK evaluation arguments.

%% file: polycom_def.tex
\subsection{Polynomial Commitments with Batched Evaluations}\label{sec:pcdef}
\begin{definition}[Multilinear Polynomial Commitment]\label{def:mlpolycom}
A polynomial commitment scheme for multilinear polynomials over a ring $\Ring$ is a tuple $\PC = (\Setup, \allowbreak \Com, \allowbreak \VerCom, \allowbreak, \ProveEval, \VerEval)$
\begin{description}
\item[$\Setup(\secparam, \ell) \to (\ck, \vk)$] takes the security parameter and (possibly) an upper bound on the number of variables $\ell$, and outputs a commitment key $\ck$ and a verification key $\vk$.

\item[$\Com(\ck, f) \to (\cm_f, \opn_f)$] takes a multilinear polynomial $f \in \Ring^{(\leq 1)}[X_1, \ldots, X_{\ell}]$, and outputs a commitment $\cm_f$ and an opening $\opn_f$.

\item[$\VerCom(\vk, \cm_f, f, \opn_f) \to b$] checks if $\opn_f$ is a valid opening for the commitment $\cm_f$ to a polynomial $f$, by accepting ($b=1$) or not ($b=0$).

\item[$\langle \ProveEval(\ck, \cm_f, \vec a, y, \ell; f, \opn_f) , \VerEval(\vk,  \cm_f, \vec a, y, \ell) \rangle \to b$] is a protocol in which the prover $\ProveEval$ aims to convince the verifier $\VerEval$ that $y=f(\vec a)$ for the  polynomial $f \in \Ring^{(\leq 1)}[X_1, \ldots, X_{\ell}]$ committed in $\cm_f$.
As a shorthand we write $\Eval$ to refer to $\langle \ProveEval, \VerEval \rangle$.
\end{description}

\noindent{\bf Correctness.}
$\PC$ is correct if for any multilinear $\ell$-variate polynomial $f$
\[
	\Pr \left[
        \begin{array}{c}
		\VerCom(\vk, \cm_f, f, \opn_f) = 1
        \end{array}
        \left \lvert
        \begin{array}{l}
            (\ck, \vk) \gets \Setup(\secparam, \ell) \\
            (\cm_f, \opn_f) \gets \Com(\ck,f)
        \end{array} \right.
        \right] = 1.
\]

\noindent{\bf Binding.}
$\PC$ is computationally binding if for all PPT adversaries $\adv$ the following probability is $\negl$:
\[
	\Pr \left[
        \begin{array}{c}
		\VerCom(\vk, \cm, f, \opn_f) = 1  \\
		\land \VerCom(\vk, \cm, f', \opn'_f) = 1 \\
		\land f \neq f'
        \end{array}
        \left \lvert
        \begin{array}{l}
            (\ck, \vk) \gets \Setup(\secparam, \ell) \\
            (\cm, f, \opn_f, f', \opn'_f) \gets \adv(\ck)
        \end{array} \right.
        \right]
\]

\noindent{\bf Knowledge Soundness.}
We say that $\PC$ is knowledge-sound if $\langle \ProveEval , \VerEval \rangle$ is a complete and knowledge-sound argument of knowledge for the NP relation
$\rel_{\eval} = \{( \stm=(\cm_f, \vec a, y, \ell); \wit = (f, \opn_f) ) \}$ of tuples such that
\[
f \in \Ring^{(\leq 1)}[X_1, \ldots, X_\ell] \land y=f(\vec a) \land \VerCom(\vk, \cm_f, f, \opn_f)=1
\]

\noindent{\bf Succinctness.}
We say that $\PC$ is succinct if the commitments produced by $\Com$ have size sublinear in the size of a polynomial and if $\langle \ProveEval , \VerEval \rangle$ is a succinct argument.

\end{definition}

\subsection{Compiling PIOPs into SNARKs}
Given a multilinear polynomial commitment it is possible to turn a PIOP for a relation $\rel$ into a public-coin interactive argument of knowledge for the same relation. The compilation consists in two steps:
replace every oracle sent by the prover with a commitment to the underlying polynomial, and replace every verifier's query to an oracle $\ora{p}$ at a point $\vec a$ with sending the actual value $p(\vec a)$ and proving its correctness w.r.t. the commitment of $p$ by using the $\langle \ProveEval , \VerEval \rangle$ protocol.
To this standard compilation technique we add the observation that if we start from a PIOP for an oracle relation $\rel$ then we can easily build a (preprocessing) commit-and-prove argument.
Finally, if the PIOP is public-coin so is the resulting argument system and thus it can be further compiled into a non-interactive argument using the Fiat-Shamir transform.

We summarize the compiler in the following theorem whose proof follows from previous work, e.g., \cite{EC:BunFisSze20}.

\begin{theorem}
Let $\PC$ be a multilinear polynomial commitment scheme and let $\iopPi$ be a PIOP for an oracle relation $\rel$ over tuples $(\ind,\stm,\wit)=(\{\ora{p^{{\ind}}_j}\}_j, \{\ora{p^{{\stm}}_i}\}_i, (\{p^{{\ind}}_j\}_j, \{p^{{\stm}}_i\}_i))$ with negligible soundness error. There exists a public-coin interactive argument for the commit-and-prove relation
\begin{align*}
	&\rel' = \left\{(\ind',\stm',\wit') = \right.\\
	& (\{ \cm^\ind_j \}_j, \{ \cm^\stm_i \}_i,
	(  \{p^{\stm}_i, \opn^{\stm}_i\}_i \{\ora{p^{{\ind}}_j}\}_j, \{\ora{p^{{\stm}}_i}\}_i, (\{p^{{\ind}}_j\}_j, \{p^{{\stm}}_i\}_i))) \in \rel \\
	& \bigwedge_{i} \VerCom(\vk, \cm^\stm_i, p^\stm_i, \opn^\stm_i)=1 \big\}
\end{align*}
	where all the commitments $\cm^\ind_j = \Com(\ck, p^{\ind}_j)$ are honestly computed in the preprocessing phase.
\end{theorem}

The efficiency of the resulting argument depends on the efficiency of both $\iopPi$ and $\PC$ as follows.
For  $\iopPi$, let $\mathsf{T}_{\iopP}$ and $\mathsf{T}_{\iopV}$ be its prover and verifier running times, and let $\mathsf{r}, \mathsf{q}, \mathsf{o}, \mathsf{s}$ be, respectively, its round complexity, number of verifier's queries, number of prover's oracles, and total size of the oracles.
For $\PC$, let $\mathsf{T}_{\Com}, \mathsf{T}_{\ProveEval}, \mathsf{T}_{\VerEval}$ be the running times of the corresponding algorithms, and let $|\cm|$ and $|\pi_{\PC}|$ be the size of a commitment and a proof respectively.
\begin{itemize}
\item Proving time is $\mathsf{T}_{\iopP} + \mathsf{T}_{\Com} \cdot \mathsf{s} + \mathsf{T}_{\ProveEval} \cdot \mathsf{q}$
\item Verifier time is $\mathsf{T}_{\iopV} + \mathsf{T}_{\VerEval} \cdot \mathsf{q}$
\item Proof size is $|\cm| \cdot \mathsf{o} + |\pi_{\PC}| \cdot \mathsf{q}$
\end{itemize}

In our work, we use a straightforward extension of this compilation that relies on PCs with {\em batched evaluation proofs} in order to prove all the evaluation proofs involve in the PIOP with single proof. Assume that the proving time (resp. proof size) of batching is a function $\mathsf{T}_{\ProveEval}(\cdot)$ (resp. $\mathsf{S}_{\pi}(\cdot)$) of the batch size.
Then we get the following efficiency:
\begin{itemize}
\item Proving time is $\mathsf{T}_{\iopP} + \mathsf{T}_{\Com} \cdot \mathsf{s} + \mathsf{T}_{\ProveEval} ( \mathsf{q})$
\item Verifier time is $\mathsf{T}_{\iopV} + \mathsf{T}_{\VerEval} \cdot \mathsf{q}$
\item Proof size is $|\cm| \cdot \mathsf{o} + \mathsf{S}_{\pi}( \mathsf{q})$
\end{itemize}

%% file: rbr-soundness.tex
\subsection{Round-by-Round Soundness}\label{sec:rbr-soundness}

We recall the notions of round-by-round (RBR) soundness \cite{STOC:CCHLRR19} and round-by-round knowledge \cite{TCC:ChiManSpo19} for PIOPs.
These notions are useful to precisely argue the knowledge-soundness error of the SNARK obtained after applying the Fiat-Shamir transformation.

Informally, a RBR sound protocol has a well-defined state (accept or
reject) at each round and, if it starts with a false statement, at each round there should be a
negligible probability of turning from a reject to an accept state.

\begin{definition}[Round-by-Round Soundness~\cite{STOC:CCHLRR19}]\label{def:rbr-soundness}
A PIOP $\iopPi$ for an indexed oracle relation $\rel$ satisfies round-by-round (RBR) soundness with error $\delta$ if there exists a state function $\state(\ind, \stm, \tr)$ such that:
\begin{itemize}
\item If $(\ind,\stm)\notin \lang_{\rel}$ then $\state(\ind, \stm, \emptyset) = 0$,
\item For any $(\ind, \stm)$ and any {\em complete} transcript $(\ind, \stm, \tr)$, if $\state(\ind,\stm, \tr)=0$ then $\iopV(\ind, \stm, \tr)$ rejects.
\item If $(\ind, \stm, \tr)$ is a partial transcript such that  $\state(\ind, \stm, \tr) =0$, then for every potential prover's message $\pi$, holds that $\Pr[\state(\ind,\stm,\tr | \pi | c)=1 : c \sample \iopV(\ind,\stm,\tr | \pi)] \leq \delta$.
\end{itemize}
\end{definition}

\begin{definition}[Round-by-Round Knowledge~\cite{TCC:ChiManSpo19}]\label{def:rbrknowledge}
A PIOP $\iopPi$ for an indexed oracle relation $\rel$ satisfies round-by-round (RBR) knowledge with error $\delta$ if there exists a polynomial-time extractor $\ext$ and a state function $\state(\ind, \stm, \tr)$ such that:
\begin{itemize}
\item $\state(\ind, \stm, \emptyset) = 0$,
\item For any $(\ind, \stm)$ and any {\em complete} transcript $(\ind, \stm, \tr)$, if $\state(\ind,\stm, \tr)=0$ then $\iopV(\ind, \stm, \tr)$ rejects.
\item If $(\ind, \stm, \tr)$ is a partial transcript such that  $\state(\ind, \stm, \tr) =0$, if $\Pr[\state(\ind,\stm,\tr | \pi | c)=1 : c \sample \iopV(\ind,\stm,\tr | \pi)]>\delta$, then $(\ind, \stm, \ext(\ind,\stm,\tr | \pi)) \in \rel$.
\end{itemize}
\end{definition}

It is possible to see that an analogous version of Lemma \ref{lemma:sound2ksound} can be stated for RBR soundness and RBR knowledge.
\begin{lemma}\label{lemma:RBRsound2RBRksound}
Consider a $\delta$-RBR-sound PIOP for oracle relations $\rel$ such that for all $(\ind, \stm, \wit) \in \rel$ $\wit$ consists only of polynomials such that the statement contains oracles to these polynomials.
The PIOP has $\delta$-RBR-knowledge and the extractor runs in time $O(|\wit|)$.
\end{lemma}
The proof, in particular the construction of the extractor, is essentially the same as in the proof of Lemma \ref{lemma:sound2ksound} (see \cite{EC:CBBZ23}). The only difference is that we have to define the $\state$ function for round-by-round knowledge, which can be define equal to the $\state$ function of RBR-soundness, except that on empty transcripts it always rejects, regardless of whether the statement is in the language.

\medskip
RBR soundness is particularly useful to understand the (RBR) soundness guarantees of protocols
that need parallel repetitions. This is summarized by the result~\cite[Corollary
5.7]{STOC:CCHLRR19}, restated here for PIOPs.
\begin{theorem}\label{thm:parallelrep}
Suppose that $\iopPi$ is a $2r$-round PIOP with $\delta$-soundness.
Then, its $\kappa$-parallel repetition $\iopPi^{\kappa}$ is $\delta^{\kappa}$-sound and $\delta^{\kappa/r}$-RBR-sound.
\end{theorem}

%% file: sumcheck.tex
\subsection{The Sumcheck Protocol}\label{sec:piop}
Below we describe $\piopsum$ as either an interactive proof or a PIOP. In our protocols, $\iopV$ might not be given (oracle access to) $p(r_1, \ldots, r_{\ell})$ directly, but rather enough information to compute that on its own. For example, for $p = fg +h$, where $\iopV$ knows $f$, it might be given (oracle access to) $g(r_1, \ldots, r_{\ell})$ and $h(r_1, \ldots, r_{\ell})$.
\procb[mode=text,bodylinesep=4pt]{The sum-check protocol/PIOP $\piopsum$ for $\relsum$}{
	$\pcfor j=1$ to $\ell$\\
	\t $\iopP$ computes and sends $g_j(X) \gets \sum_{\vec i \in\bin^{\ell - j}} p(r_{1}, \ldots, r_{j-1}, X, \vec i)$. \\
	\t $\iopV$ sends $r_{j} \sample \Eset$. \\
	$\iopP$ sends $p(r_1, \ldots, r_{\ell})$.     \text{ \ \ \ \ \ \ \ \ \ (In the PIOP version: $\iopV$ queries $\ora p$ on $(r_1, \ldots, r_{\ell})$}).  \\
	$\iopV$ accepts if the following checks are satisfied:\\
	\t $ \forall j \in [\ell] : g_{j-1}(r_{j-1}) = g_{j}(0) + g_{j}(1)$, where $g_{0}(r_0)=y$\\
	\t $g_{\ell}(r_\ell) = p(r_1, \ldots, r_{\ell})$
}

\begin{theorem}\label{thm:sumcheckPIOP}
	Let $\Eset \subseteq \Ring$ be an exceptional set. The PIOP for $\relsum$ described above has perfect completeness and
	$\delta$-knowledge-soundness with $\delta = d \ell / |\Eset|$.
\end{theorem}
\begin{proof}
	It follows from the soundness analysis in \cite{EPRINT:CCKP19} and Lemma \ref{lemma:sound2ksound}. \qed
\end{proof}

\subsubsection{Unpacked Sumcheck}\label{sec:unpacked-sumcheck}
We define the {\em unpacked sumcheck} oracle relation $\relusum$ as the set of tuples $((y, \ora{\tilde p}); \tilde p)$ such that
$ y =  \sum_{\vec b \in \bin^{\ell},\vec k \in \bin^{\nu}} \tilde p(\vec b)[\vec k]$, or equivalently,
\[
\relusum = \{((y, \ora{\tilde p}); \tilde p) : ((y, \ora{\unpack(\tilde p)}); \unpack(\tilde p)) \in \relsum \}.
\]
Given the above reduction of $\relusum$ to $\relsum$ we can construct a PIOP $\piopusum$ for $\relusum$ using the PIOP $\piopsum$ for $\relsum$, changed so that the verifier samples random challenges in an exceptional set $\Esetz \subset \ZZ_q$. Let us call $\piopsumzq$ this variant of $\piopsum$. $\piopusum$ simply runs $\piopsumzq$ on the virtual oracle $\unpack(\tilde p)$ and uses $\ueval$ to simulate queries to the unpacked oracle.

\begin{Boxfig}[h!]{PIOP for $\relusum$}{PIOPusum}{\underline{PIOP $\piopusum$ for $((y, \ora{\tilde p}); \tilde p) \in \relusum$}}
	$\iopP$ and $\iopV$ run the PIOP $\piopsumzq$ for $((y, \ora{\tilde u}); \tilde u) \in \relusum$ with $\tilde u = \unpack( \tilde p)$.

	Given $\piopsumzq.\iopV$'s query to $\ora{\tilde u}$ on $\vec r = (\vec r_1, \vec r_2)$, $\piopusum.\iopV$ makes a query to $\ora{\tilde p}$ on $\vec r_1$ and given its answer $v=\tilde p(\vec r_1)$, returns $\ueval(v, \vec r_2)$ to $\piopsumzq.\iopV$.
\end{Boxfig}

\begin{theorem}
$\piopusum$ is a PIOP with perfect completeness, prover's efficiency $O(2^{\ell+\nu})$, verifier's efficiency $O(\ell + 2^\nu)$, $\delta_{usum}$-knowledge soundness with $\delta_{usum} = (\ell+\nu)/|\Esetz|$, and RBR-knowledge with error $1/|\Esetz|$.
\end{theorem}
The theorem follows by specializing known results for the sumcheck protocol to the case of challenges in the smaller set $\Esetz$.
When $\Esetz$ is not large enough to have $1/|\Esetz|$ negligible, we repeat $\piopusum$ $\kappa (\ell + \nu)$ times in parallel, obtaining RBR-knowledge error $1/|\Esetz|^{\kappa}$ which is negligible for large enough $\kappa$ \cite{STOC:CCHLRR19}.

%% file: pfckks.tex
\subsection{CKKS}
\label{sec:pfckks}
We recall the proof-friendly CKKS construction, as presented in~\cite{C:CCCFGS25}.
Note that the secret and public key generation algorithms, as well as the encryption,
addition and multiplication algorithms are unchanged. For a full description of the ``regular'' CKKS
scheme, we point the reader to~\cite{AC:CKKS17}.
For ease of exposition, we mostly follow the notation of~\cite{C:CCCFGS25}.

Let $q_{0} < q_{1} < \dots < q_{L}$ be the chain of moduli as outlined in the preliminaries and
($\omega_{0}$,
$\omega_{1}$, \dots, $\omega_{L}$) their respective CRT bases, $\chi_\text{key}$ be the secret key
distribution, and $\chi_\text{err}, \chi_\text{enc}$ be the error and encryption distributions, respectively.
\subsubsection{CRT}
Let $l$ denote $l = L - j$, and $\omega_{j} = (p_0, p_1, \dots, p_{l})$ be a
CRT base for $q_j$. Given $a \in R_{q_0}$, the inverse CRT map is defined as follows:

\begin{align}
	\label{eq:CRT}
	\crt^{-1}_{\omega_{j}}(a) \defeq \left( \left[ a
	\right]_{p_0}, \left[ a
	\right]_{p_1}, \dots,
	\left[ a
	\right]_{p_{l}}
	\right) \in R_{q_0}^{l+1}.
\end{align}

Let $Q_{i} = q_0/p_i$ and $\hat{Q}_{i} = \left[(q_0/p_i)^{-1}\right]_{p_i}$
be the CRT recomposition constants. The generator of the ideal defining the embedding
is defined as $z_l \defeq
\sum_{i=0}^{l} Q_i\hat{Q}_i = \crt(\underbrace{1, \dots, 1}_{l+1~\text{times}}, 0, \dots, 0) \in R_{q_0}$.

For each $0 \leq j \leq L$, the CRT recomposition vector for a given $a \in R_{q_0}$ is:

\begin{align}
	\label{eq:PW}
	\pw_{\omega_{l}}(a) \defeq \left(\left[ aQ_{0}\hat{Q}_{0} \right]_{q_0}, \left[
	aQ_{1}\hat{Q}_{1}
	\right]_{q_0}, \dots, \left[ aQ_{l}\hat{Q}_{l} \right]_{q_0} \right) .
\end{align}

\subsubsection{Proof-friendly CKKS}
\begin{description}

	\vspace{0.2cm}
	\item[$\evk \leftarrow \mathsf{KeySwitchGen}(s, s^2)$:]
	For $\sk = (1, s)$,
	for $i = 0, \dots, L$, sample $a_i \xleftarrow{\$} R_{q_0}$, sample $e_i \gets \chi_\text{err}$.
	Compute
	$b_i = - a_i \cdot s + e_i + \pw_{\omega_{L}}(s^2)[i]\pmod{q_0}$. Recall that $l = L - j$, for each level
	$j
	\in \{ 0, \dots, L \}$, compute

	\[
	\evk \defeq (\evk_{j,0}, \evk_{j,1}) \gets \left( (z_l b_i)_{i=0, \dots,
		l}, (z_l a_i)_{i=0, \dots, l} \right) \in \left( R_{q_0}^2 \right)^{l+1}.
	\]

	\vspace{0.2cm}
	\item[$\mu + e' \leftarrow \Dec(\sk, \ct)$:]
	Given a ciphertext $\ct \in R_{q_0}$ for some level $j$, parse $(c_0, c_1) = \ct$. Output the CRT
	recomposition of the first $l = L - j$ components of $\langle\ct, \sk \rangle$, namely
	$\crt_{\omega_{j}}( \langle\ct, \sk \rangle)$.

	\vspace{0.2cm}
	\item[$(d_0, d_1, d_2) \leftarrow \premult(\ct_0, \ct_1, j)$:]
	For some level $j \in \{ 0, \dots, L\}$, compute

	\begin{equation}
		\label{eq:polymult}
		(d_0, d_1, d_2) \defeq (\ct_0[0] \cdot \ct_1[0]
		, \ct_0[0] \cdot \ct_1[1] + \ct_1[0] \cdot \ct_0[1], \ct_0[1] \cdot \ct_1[1]
		),
	\end{equation}
	as an element of $ R_{q_0}^3$. Perform the key switching as follows.

	\item[$\ct' \leftarrow \keyswitch(\evk, \ct=(d_0, d_1, d_2))$:] output $\ct' =(c'_0, c'_1)$ where

	\begin{align}
		\label{eq:keyswitch}
		c'_i \defeq d_i +
		\langle \crt_{\omega_j}^{-1}(d_2), \evk_{j,i} \rangle \in R_{q_0}, \ \ i=0,1.
	\end{align}

	\item[$\ct'' \leftarrow \hescale(\ct', j)$:]
	Let $\ct' = (c'_0, c'_1)$, and $\pres_l^{-1} = \left[q_{j+1}/q_{j}\right]_{q_{j+1} } \cdot z_l \in R_{q_0}$.

	Rescale is done by computing $\ct'' \defeq (c''_0, c''_1) \in R_{q_0}^2$ as

	\begin{align}
		\label{eq:modswitch}
		c''_i \defeq \left( c'_i - \left[c'_i\right]_{p_l}\right) \pres^{-1}_l \in R_{q_0},\ \  i=0,1.
	\end{align}

\end{description}

\subsection{A modular construction of the PIOP $\piopckks$}\label{sec:modular-piopckks}
Belowe recall how we modularly build a PIOP $\piopckks$ for the relation $\rel_{CKKS}$ from PIOPs $\pioprange$ for $\relrange$ and $\piopac$ for $\rel_{AC}$.

\procb[mode=text,bodylinesep=8pt]{\textbf{PIOP $\piopckks$ for $(\ind, \stm; \wit) \in \rel_{CKKS}$}}{
	$\iopP$ and $\iopV$ run $\piopac$ for $(\vC,
	( \ora{\tilde{\layerinputall}}, \ora{\tilde{\layeroutputall}} , \ora{\tilde{\vw}_{2^u}}, \{\ora{\tilde{\vw}_l}\}_{l=0}^{D-2}) ; \layerinputall, \layeroutputall , \vw_{2^u}, \{\vw_l\}_{l=1}^{D-2}) \in \rel_{AC}$ \\
	$\iopP$ and $\iopV$ run $\pioprange$ for $(2^u, \ora{\tilde{\vw}_{2^u}};\vw_{2^u} ) \in \rel_{range}$ and $(q_{L-D+1}/p_{L-D+1}, \ora{\tilde{\layeroutputall}};\layeroutputall ) \in \rel_{range}$ \\
	For $l=0, \dots, D-2$:
	\t$\iopP$ and $\iopV$ run $\pioprange$ for $( q_{L-l}/p_{L-l}, \ora{\tilde{\vw_l}}; \vw_l) \in \rel_{range}$.
}

%% file: apx-packing.tex
\section{Auxiliary Material on PIOP for Unpacked Polynomials}\label{sec:auxpacking}

\subsection{Proof of Theorem \ref{thm:piopuevalpacking}}
\begin{proof}
For correctness, we refer to the explanation in Section~\ref{sec:packing}. Prover efficiency
follows from the runtime of sumcheck, while Verifier efficiency is the sumcheck's verification plus
the time to compute $\psi,\psi^{*}$ and $\ueval$.

For soundness, if the statement is false, i.e., $\exists i : y_i \neq \tilde u(\vec x_i)$,
then
$\tilde y(\vec Z, T) - \sum_{\vec b \in\bin^{\ell'}} \tilde u(\vec b) E(\vec b, \vec Z, T)$ is a nonzero polynomial,
and by generalized Schwartz-Zippel the probability that
$\tilde y(\vec \sigma,\tau) =\sum_{\vec b \in\bin^{\ell'}} \tilde u(\vec b)  E(\vec b, \vec \sigma,\tau)$, for  random $\vec\sigma \sample \Esetz^{\nu},\tau \sample \Esetz$, is $\leq (\nu+\np-1) / |\Esetz|$.
On the other hand, by the soundness of $\piopsumzq$, the probability that the verifier accepts a tuple  $((\tilde y(\vec \sigma), \ora{f}); f) \notin \relsum$,
with $f =\tilde u(\vec X)   E(\vec X, \vec \sigma, \tau)$ of individual degree 2,
is $\leq 2(\ell+\nu) / |\Esetz|$.
\qed
\end{proof}

%% file: apx-range-proof.tex
\section{Auxiliary Material on PIOP for Range Checks}\label{sec:auxrange}

\subsection{Proof of Lemma \ref{lemma:Rq2Zqrange}}

\begin{proof}
Let us index $T_B$ by tuples $(\idx_1, \ldots, \idx_{N}) \in [0,B-1]^{N}$ such that
\[
T_{B}[\idx_1, \ldots, \idx_{N}] = \sum_{k = 1}^{N} \idx_{k} \cdot Y^{k - 1} \in \Rq.
\]
Let $\dig{\beta}{j}(\idx)$
be the function that on input an integer $\idx$, returns the $j$-th digit of its base-$\beta$
decomposition, i.e., $\idx_{j}$ such that $\idx = \sum_{j} \idx_{j} \beta^{j-1}$.
Then we can express each element of $T_{B}$ in terms of elements of $\vec t_{\beta}$ as follows.
\begin{eqnarray*}
T_{B}[\idx_1, \ldots, \idx_{N}] & = & \sum_{k = 1}^{N} \idx_{k} \cdot Y^{k-1}
	=  \sum_{k = 1}^{N} \sum_{j=1}^{c}   \dig{\beta}{j}(\idx_{k}) \cdot \beta^{j-1} \cdot Y^{k-1}  \\
	& = & \sum_{k = 1}^{N} \sum_{j=1}^{c} \vec t_{\beta}[\dig{\beta}{j}(\idx_{k})] \cdot Y^{k-1} \cdot
	\beta^{j-1}.
\end{eqnarray*}
The above equation implies that $\tilde v(\vec i) \in T_{B}$ holds $\forall \vec i \in \bin^{\ell}$ if and only if \linebreak$\exists\tilde \mu\in \Zqml[X_1,\ldots,X_{\ell},Z_1,\dots,Z_\nu,Z'_1,\dots,Z'_\gamma ]$  such that, for $\tilde u = \unpack(\tilde v)$,
\begin{eqnarray}
&& \forall \vec i \in \bin^{\ell}, \vec k\in \bin^\nu : \tilde u(\vec i,\vec k)  =  \sum_{\vec j \in
\bin^{\gamma}} \tilde{\mu}(\vec i, \vec k, \vec j) \cdot \beta^{\toint(\vec j)},
\label{eq:vdecomposition-proof} \\
&& \forall \vec i \in \bin^{\ell}, \vec j \in \bin^{\gamma}, \vec k \in \bin^{\nu}: \tilde{u}(\vec i, \vec k,
\vec j) \in  \vec t_{\beta}. \label{eq:mulookup-proof}
\end{eqnarray}
Concretely, $\tilde{\mu}$ is determined by the values $\tilde{\mu}(\vec i, \vec k, \vec j) =\dig{\beta}{j}(\tilde v(\vec i)[\vec k]),\  \forall \vec i \in \bin^{\ell},\ \vec j \in \bin^{\gamma},\ \vec k \in \bin^{\nu}$.
This completes the proof by definition of $\unpack$ and the equations \eqref{eq:vdecomposition}, \eqref{eq:mugrange}.
\qed
\end{proof}

\subsection{Proof of Theorem \ref{thm:piopgqrange2pioprange}}\label{sec:proof-piopgqrange2pioprange}
\begin{proof}
To prove RBR-soundness, we start by defining a $\state(T_{B}, \ora{\tilde v}, \tr_i)$ function on every (partial) transcript $\tr_{i}$, from round $i=0$ to $i = 2+\mathsf{r}_{range}$, where $\mathsf{r}_{range}$ denotes the number of rounds of the $\piopGqrange$ protocol:

\begin{itemize}
\item For $\tr_0=\emptyset$,  $\state(T_{B}, \ora{\tilde v}, \tr_0)=0$ iff $(T_{B}, \ora{\tilde v}) \notin \lang_{\relrange}$.

\item For $\tr_1 = \ora{\tilde \mu} |  (\vec r_1, \ldots,\vec r_{\kappa})$,  $\state(T_{B}, \ora{\tilde v}, \tr_1)=0$ iff
$(\vec t_{\beta}, \ora{\tilde \mu}; \tilde \mu) \notin \relrange$ or $\exists i \in [\kappa]$: $f(\vec r_i) \neq 0$,
where
\[
f(\vec X, \vec Z) = \tilde u(\vec X, \vec Z)  -  \sum_{\vec j \in \bin^{\gamma}} \tilde{\mu}(\vec X, \vec Z, \vec j) \cdot \beta^{\toint(\vec j)}.
\]

\item For $\tr_2 =  \tr_1 | (\tilde\mu^*_1, \ldots, \tilde\mu^*_{\kappa}) | \vec s$: $\state(T_{B}, \ora{\tilde v}, \tr_2)=0$ iff

$(\vec t_{\beta}, \ora{\tilde \mu}; \tilde \mu) \notin \relrange$ or
$\exists i \in [\kappa]$ such that $\tilde u(\vec r_i)  \neq  \sum_{\vec j \in \bin^{\gamma}} \tilde{\mu}^*_i(\vec j) \cdot \beta^{\toint(\vec j)}$ or
 $\mu^*_i(\vec s) \neq \tilde \mu(\vec r_i, \vec s)$.

\item For $\tr_{2+k} = \tr_2 | \tr_k'$ where, for $1 \leq k \leq \mathsf{r}_{range}$, $\tr'_k$ is the partial transcript of $\piopGqrange$ at the $k$-th round: $\state(T_{B}, \ora{\tilde v}, \tr_{2+k})=0$ iff

$\state'(\vec t_{\beta}, \ora{\tilde \mu}, \tr'_{k})=0$ or
$\exists i \in [\kappa]$ such that $\tilde u(\vec r_i)  \neq  \sum_{\vec j \in \bin^{\gamma}} \tilde{\mu}^*_i(\vec j) \cdot \beta^{\toint(\vec j)}$ or
 $\mu^*_i(\vec s) \neq \tilde \mu(\vec r_i, \vec s)$.

\end{itemize}

Let us now show that this $\state$ function satisfies the three properties of RBR-soundness.

First, it is easy to see that by construction, if $(T_{B}, \ora{\tilde v}; \tilde v)\notin\relrange$ then $\state(T_{B}, \ora{\tilde v}, \tr_0)=0$.

Second, for any {\em complete} transcript $\tr = \tr_{2+\mathsf{r}_{range}} = \tr_2 | \tr'_{\mathsf{r}_{range}}$, by construction we have that if $\state(T_{B}, \ora{\tilde v}, \tr)=0$, then either $\state'(\vec t_{\beta}, \ora{\tilde \mu}, \tr'_{k})=0$, which means that the verifier of $\piopGqrange$ rejects (by the RBR-soundness of $\piopGqrange$), or at least one of the verifier's checks in step 3 fails. Therefore, $\pioprange.\iopV$ rejects.

Third, we show that if $\state$ rejects on a partial transcript of round $i$, for any prover's message of the round $i+1$, we have a concrete (negligible) bound on the probability that $\state$ accepts on the next partial transcript.
We show this property round by round.

\begin{claim}
If $\state(T_{B}, \ora{\tilde v},\tr_0)=0$, then for any prover's message $\ora{\tilde \mu}$ we have that
\[
\Pr_{\vec r_1, \ldots, \vec r_\kappa \sample \Esetz^{\ell+\gamma} }[\state(T_{B}, \ora{\tilde v},\tr_1=(\ora{\tilde \mu} | (\vec r_1, \ldots, \vec r_{\kappa})))=1 ] \leq \left(\frac{\ell+\nu}{|\Esetz|} \right)^{\kappa}.
\]
\end{claim}
By definition, $\state(T_{B}, \ora{\tilde v},\tr_0)=0$ iff $(T_{B}, \ora{\tilde v}) \notin \lang_{\relrange}$ and by Lemma \ref{lemma:Rq2Zqrange} this means that for every $\tilde \mu$ we have that the polynomial $f(\vec X, \vec Z) = \tilde u(\vec X, \vec Z)  -  \sum_{\vec j \in \bin^{\gamma}} \tilde{\mu}(\vec X, \vec Z, \vec j) \cdot \beta^{\toint(\vec j)} \neq 0$ or $(\vec t_{\beta},  \ora{\tilde \mu}) \notin \lang_{\relgrange}$.
Instead, $\state$ accepting on $\tr_1$ means that
$(\vec t_{\beta}, \ora{\tilde \mu}; \tilde \mu) \in \relrange$ and $\forall i \in [\kappa]$: $f(\vec r_i) = 0$.
\begin{eqnarray*}
&&\Pr[\state(T_{B}, \ora{\tilde v},\tr_1)=1 \mid \state(T_{B}, \ora{\tilde v},\tr_0)=0] \\
&& = \Pr[(\vec t_{\beta}, \ora{\tilde \mu}; \tilde \mu) \in \relrange \bigwedge_{ i \in [\kappa]} f(\vec r_i) = 0 \mid f(\vec X, \vec Z) \neq 0 \vee (\vec t_{\beta}, \ora{\tilde \mu}; \tilde \mu) \notin \relrange] \\
&& \leq \Pr[(\bigwedge_{ i \in [\kappa]} f(\vec r_i) = 0 \mid f(\vec X, \vec Z) \neq 0] \\
&& = \prod_{ i \in [\kappa]} \Pr_{\vec r_i \sample \Esetz^{\ell+\gamma}}[( f(\vec r_i) = 0 \mid f(\vec X,
\vec Z) \neq 0 \wedge (\vec t_{\beta}, \ora{\tilde \mu}; \tilde \mu) \in \relrange].
\end{eqnarray*}

By generalized Schwartz-Zippel, $\Pr_{\vec r_i \sample \Esetz}[f(\vec r_i) = 0 \mid f(\vec X, \vec Z) \neq 0] \leq \frac{\ell + \gamma}{|\Esetz|}$, and the claimed bound follows by considering the probability that this occurs for $\kappa$ independent tests.

\begin{claim}
If $\state(T_{B}, \ora{\tilde v},\tr_1)=0$, then for any prover's message $(\tilde\mu^*_1, \ldots, \tilde\mu^*_{\kappa})$ we have that
\[
\Pr_{\vec s \sample \Aq^{\gamma}}[\state(T_{B}, \ora{\tilde v},\tr_2=(\tr_1 | (\tilde\mu^*_1, \ldots, \tilde\mu^*_{\kappa}) | \vec s))=1] \leq \frac{\gamma}{|\Aq|}.
\]
\end{claim}

By definition, $\state(T_{B}, \ora{\tilde v},\tr_1)=0$ implies that either $(\vec t_{\beta}, \ora{\tilde \mu}; \tilde \mu) \notin \relrange$ or $\exists i \in [\kappa]$: $f(\vec r_i) \neq 0$.
On the other hand, if $\state(T_{B}, \ora{\tilde v},\tr_2)=1$ then
$(\vec t_{\beta}, \ora{\tilde \mu}; \tilde \mu) \in \relrange$ and
$\forall i \in [\kappa]$  $\tilde u(\vec r_i)  =  \sum_{\vec j \in \bin^{\gamma}} \tilde{\mu}^*_i(\vec j) \cdot \beta^{\toint(\vec j)}$ and
 $\mu^*_i(\vec s) = \tilde \mu(\vec r_i, \vec s)$.

Consider the specific index $i$ for which $f(\vec r_i) \neq 0$, then we have that
\begin{eqnarray*}
&& \Pr[\state(T_{B}, \ora{\tilde v},\tr_2)=1 \mid \state(T_{B}, \ora{\tilde v},\tr_1)=0] \\
&& \leq \Pr[\bigwedge_{i=1}^{\kappa} \tilde u(\vec r_i)  =  \sum_{\vec j \in \bin^{\gamma}} \tilde{\mu}^*_i(\vec j) \cdot \beta^{\toint(\vec j)} \wedge \mu^*_i(\vec s) = \tilde \mu(\vec r_i, \vec s) \mid \exists i \in [\kappa] \; f(\vec r_i) \neq 0] \\
&& \leq \Pr[\tilde u(\vec r_i)  =  \sum_{\vec j \in \bin^{\gamma}} \tilde{\mu}^*_i(\vec j) \cdot \beta^{\toint(\vec j)} \wedge \mu^*_i(\vec s) = \tilde \mu(\vec r_i, \vec s) \mid f(\vec r_i) \neq 0] \\
&& \leq \Pr[ \mu^*_i(\vec s) = \tilde \mu(\vec r_i, \vec s) \mid \tilde u(\vec r_i)  =  \sum_{\vec j \in \bin^{\gamma}} \tilde{\mu}^*_i(\vec j) \cdot \beta^{\toint(\vec j)} \wedge f(\vec r_i) \neq 0]\\
&& = \Pr[ \mu^*_i(\vec s) = \tilde \mu(\vec r_i, \vec s) \mid  \tilde \mu(\vec r_i, \vec Z') \neq
\tilde{\mu}^*_i(\vec Z') ].
\end{eqnarray*}

To see the last equality, observe that by definition of $f$ we get that
$\sum_{\vec j \in \bin^{\gamma}} \tilde{\mu}(\vec r_i, \vec j) \cdot \beta^{\toint(\vec j)}  \neq \sum_{\vec j \in \bin^{\gamma}} \tilde{\mu}^*_i(\vec j) \cdot \beta^{\toint(\vec j)} $
and since $\beta \neq 0$ and $\beta^{2^{\gamma}-1} < q$ this means that there is a $\vec j$ such that $\mu^*_i(\vec j) \neq \tilde\mu(\vec r_i, \vec j)$, and thus
$g(\vec Z') = \tilde\mu(\vec r_i, \vec Z')  -  \sum_{\vec j \in \bin^{\gamma}} \tilde{\mu}^{*}_i(\vec j) \cdot \eq(\vec Z', \vec j) \neq 0$. However, by generalized Schwartz-Zippel, $\Pr_{\vec s \sample \Aq}[g(\vec s) = 0 \mid g(\vec Z') \neq 0] \leq \frac{\gamma}{|\Aq|}$, which proves the claim.

\begin{claim}
For every $k=0$ to $\mathsf{r}_{range}-1$,
if $\state(T_{B}, \ora{\tilde v},\tr_{2+k})=0$, then for any prover's message $\pi$ and every verifier's random challenge $\rho$ we have that
\[
\Pr_{\rho}[\state(T_{B}, \ora{\tilde v},\tr_{2+k+1}=(\tr_2 | \tr'_{k} | \pi | \rho)=1] \leq \hat\delta_{range}.
\]
\end{claim}

For $k=0$, if $\state(T_{B}, \ora{\tilde v},\tr_{2})=0$ then $(\vec t_{\beta}, \ora{\tilde \mu}; \tilde \mu) \notin \relrange$ and, by the RBR-soundness of $\piopGqrange$, $\state'(\vec t_{\beta}, \ora{\tilde \mu}, \emptyset)=0$.
Let $E$ be the event that $\exists i \in [\kappa]$ such that $\tilde u(\vec r_i)  \neq  \sum_{\vec j \in \bin^{\gamma}} \tilde{\mu}^*_i(\vec j) \cdot \beta^{\toint(\vec j)}$ or
 $\mu^*_i(\vec s) \neq \tilde \mu(\vec r_i, \vec s)$. Then we have

\begin{eqnarray*}
&& \Pr_{\rho}[\state(T_{B}, \ora{\tilde v},\tr_{2+k+1})=1 \mid \state(T_{B}, \ora{\tilde v},\tr_{2+k})=0] \\
&& = \Pr_{\rho}[\state'(\vec t_{\beta}, \ora{\tilde \mu}, \tr'_{k} | \pi | \rho)=1 \wedge \bar E \mid \state'(\vec t_{\beta}, \ora{\tilde \mu}, \tr'_k)=0 \vee E] \\
&& \leq \Pr_{\rho}[\state'(\vec t_{\beta}, \ora{\tilde \mu}, \tr'_{k} | \pi | \rho)=1  \mid \state'(\vec t_{\beta}, \ora{\tilde \mu}, \tr'_k)=0] \leq \hat\delta_{range}.
\end{eqnarray*}
where the last bound follows by the RBR-soundness of $\piopGqrange$.
\qed
\end{proof}

%% file: new-PIOP-range-proof.tex
\newcommand{\pioprangeFF}{\hat\iopPi^{\FF}_{range}}
\newcommand{\pioprangeopt}{\hat\iopPi^{*}_{range}}
\def\prj{\pi}
\def\FFtoInt{\iota}
\def\pmin{p_{\mathsf{min}}}
\def\FFml{\FF^{(\leq 1)}}
\def\GP{\mathsf{GP}}
\def\Fin{\mathsf{V}_6}

\section{Our Instantiation of the PIOP $\piopGqrange$}
\label{sec:newPIOPrange}

In this section, we present an instantiation of the PIOP $\piopGqrange$ for the relation
\[
\relgrange = \{(\ind, \stm; \wit) = (\vec t_\beta, \ora{\tilde \mu}; \tilde \mu) : \forall \vec b \in \bin^{\ell^*} \; \tilde \mu(\vec
b) \in \vec t_\beta \},
\]
where the index $\vec t_\beta = \{0, \ldots, \beta -1\} \subset \ZZ$ and $\tilde \mu \in \Hqml[X_1, \ldots, X_{\ell^*}]$.

A building block of our $\piopGqrange$ is a PIOP $\pioprangeFF$ for the oracle relation $\relgrange$ over a finite field $\FF=\ZZ_p$ for an odd prime number $p$ that is sufficiently larger than $\beta$, following a bound that we specify below.

\headingb{Preliminaries.}
Let us denote $\vec X = (X_1, \ldots, X_{\ell^*})$, let $q = p_{0} \cdot p_{1} \cdots p_{L}$ be the product of $L+1$ distinct prime numbers and let $\pmin$ denote the smallest of them.

Let $p$ be an odd prime number and $I_p = \{-\lfloor p/2 \rfloor, \ldots, \lfloor p/2 \rfloor\} \subset \ZZ$.
For a positive integer $a$, let $\prj_a: \ZZ \to \ZZ_a$ be the canonical quotient map $\prj_a(x) = x \pmod a$.
Let $\prj_{I_{p}}$ be the restriction of $\prj_p$ to $I_p$, which is a bijection,
and let $\FFtoInt = \prj_{I_{p}}^{-1}:\ZZ_{p} \to I_{p}$ be its inverse.

Abusing notation, we extend the definition of these maps to vectors by applying them coordinate-wise and to polynomials by applying them on all their coefficients. For example, given a polynomial $f \in \ZZ[X_1,\ldots,X_n]$ we denote by $\prj_p(f) \in \ZZ_p[X_1,\ldots,X_n]$ the polynomial obtained by applying $\prj_p$ on all the coefficients of $f$.

We rely on the following proposition, which intuitively formalizes the fact that $\ZZ_p$ can emulate computations over signed integers, as long as no overflow occurs.
\begin{proposition}\label{prop:lift_eval}
Let $f \in \ZZ[X_1,\ldots,X_n]$ and $\vec x \in \ZZ^{n}$ such that $\abs{f(\vec x)} < p/2$. Then
$$\FFtoInt( \prj_p(f)(\prj_p(\vec x))) = f(\vec x).$$
\end{proposition}
\begin{proof}
Since $\prj_p$ is a ring homomorphism we have $\prj_p(f(\vec x)) = \prj_p(f)(\prj_p(\vec x))$. Applying $\FFtoInt$ on both sides $\FFtoInt( \prj_p(f)(\prj_p(\vec x))) = \FFtoInt(\prj_p(f(\vec x))) = f(\vec x)$ since $f(\vec x) \in I_p$ and the restriction of $\FFtoInt \circ \prj_p$ to $I_{p}$ acts as the identity function.
\end{proof}

\headingb{Parameters.}
Let $\Sigma = \{0, \ldots, \sigma - 1\} \subseteq \ZZ$ with $\sigma \leq \pmin$, and notice that $\prj_q(\Sigma)$ is a subset of the exceptional set $\Esetz \subset \ZZ_q$.
{}
Let $0 \leq m \leq \ell^*$
be an integer that  represents an efficiency tradeoff parameter,
and $\kappa \geq 1$ be an integer that represents the number of repetitions.

The set of parameters must satisfy the following conditions:
\begin{align}
p & > 2 \beta \cdot (2 \sigma)^{m} \label{eq:params-bound-p}\\
\left(\frac{m}{\sigma}\right)^{\kappa} & \approx 2^{-\secpar} \label{eq:params-bound-soundness}\\
m \cdot 2^{\ell^* - m} & = \mathsf{poly}(\ell^*) \label{eq:params-bound-efficiency}
\end{align}
Intuitively, the first condition ensures correctness of the protocol, the second one ensures soundness, and the third one succinctness.

\headingb{The $\piopGqrange$ protocol.}
We describe the full protocol in Figure \ref{fig:piopGqrange}.

\begin{Boxfig}[H]{Construction of the PIOP $\piopGqrange$ for $\relgrange$ over $\Hq$ from $\pioprangeFF$ for $\relgrange$ over $\ZZ_p$}{piopGqrange}{\underline{\textbf{PIOP $\piopGqrange$ for $(\vec t_\beta, \ora{\tilde \mu}; \tilde \mu) \in \relgrange$}}}
\begin{enumerate}
\item $\iopP$ computes $\tilde\mu' = \prj_p(\tilde\mu'')\in \ZZ_p[\vec X]$ where $\tilde\mu''\in \ZZ^{(\leq 1)}[\vec X]$ is such that, for all $\vec b$, $\tilde\mu''(\vec b) \in \vec t_\beta$ and $\prj_q(\tilde\mu''(\vec b)) = \tilde\mu(\vec b)$, and sends $\ora{\tilde \mu'}$.

$\iopV$ samples $\vec r_0, \ldots, \vec r_{\kappa - 1} \sample \Sigma^{m}$ and sends them to $\iopP$.

\item $\iopP$ sends $(\tilde y_0, \tilde y'_0, \ldots, \tilde y_{\kappa - 1}, \tilde y'_{\kappa - 1})$, where, for $\eta \in [\kappa]$, $\vec{\hat{r}}_{\eta} = \prj_q(\vec r_{\eta})$, $\vec r'_{\eta} = \prj_p(\vec r_{\eta})$ and:
\[
\tilde y_{\eta}(X_{m+1}, \ldots, X_{\ell^*})  = \tilde \mu(\vec{\hat r}_{\eta}, X_{m+1}, \ldots, X_{\ell^*}) \in \Hq[X_{m+1}, \ldots, X_{\ell^*}]
\]\[
\tilde y'_{\eta}(X_{m+1}, \ldots, X_{\ell^*})  = \tilde \mu'(\vec r'_{\eta}, X_{m+1}, \ldots, X_{\ell^*}) \in \ZZ_{p}[X_{m+1}, \ldots, X_{\ell^*}]
\]
$\iopV$ samples random points $\vec{\hat s} \sample \Esetd^{\ell^* - m}$ and $\vec s' \sample \ZZ_p^{\ell^* -m}$.\\
For all $\eta \in [\kappa]$, $\iopV$ queries $\ora{\tilde \mu}$ and $\ora{\tilde \mu'}$ on
	$(\vec{\hat r}_{\eta}, \vec{\hat s}) \in \Hq^{\ell^*}$ and $(\vec r'_{\eta}, \vec s')\in \ZZ_p^{\ell^*}$ respectively, and checks if
\[
\tilde y_{\eta}(\vec{\hat s})  = \tilde \mu(\vec{\hat r}_{\eta}, \vec{\hat s}) \in \Hq \quad \text{ and } \quad
\tilde y'_{\eta}(\vec s')  = \tilde \mu'(\vec r'_{\eta}, \vec s') \in \ZZ_{p}
\]
Then $\iopV$ checks if for all $\eta \in [\kappa]$, $\vec \delta \in \bin^{\ell^* - m}$, and $l \in [0, L]$: {}
{}
\[
\tilde y_{\eta}(\vec \delta) \bmod p_l = \FFtoInt(\tilde y'_{\eta}(\vec \delta)) \bmod p_l
\]

\item $\iopP$ and $\iopV$ run a PIOP $\pioprangeFF$ for $(\vec t'_{\beta},  \ora{\tilde \mu'};  \tilde \mu') \in \relgrange$, where $\vec t_{\beta}' = \prj_p(\vec t_{\beta})$.
\end{enumerate}
\end{Boxfig}

\begin{theorem}
Consider the parameters $q, p, \beta, \sigma, m, \kappa, \ell^*$ satisfying equations \eqref{eq:params-bound-p}, \eqref{eq:params-bound-soundness},\eqref{eq:params-bound-efficiency}.
Let $\Esetd \subset \Hq$ be an exceptional set.
Let $\vec t_\beta = \{0, \ldots, \beta -1\} \subset \ZZ$ and let $\pioprangeFF$ be a PIOP for relation $\relgrange$ over the finite field $\FF=\ZZ_p$ that is perfectly complete and has RBR-knowledge error $\delta'_{range}$.
Then the $\piopGqrange$ protocol of Figure \ref{fig:piopGqrange} is a PIOP for $\relgrange$ with perfect completeness and RBR-knowledge error $\max\left\{\left(\frac{m}{\sigma}\right)^\kappa; \frac{\ell^* - m}{|\Esetd|} + \frac{\ell^* - m }{p}; \delta'_{range}\right\}$.
\end{theorem}
We prove correctness in Section \ref{sec:piopGqrange-correctness} and RBR-knowledge in Section \ref{sec:piopGqrange-security}.
When instantiated with the PIOP $\pioprangeFF$ from Section \ref{sec:PIOPrangeFF}, $p$ should additionally be $>2^{\ell^*}$.

\subsection{Correctness of $\piopGqrange$}
\label{sec:piopGqrange-correctness}
Let $\tilde \mu$ be a polynomial such that $\forall \vec b \in \bin^{\ell^*}$ it holds $\tilde \mu(\vec b) \in \vec t_\beta$. By construction, $\tilde\mu''$ is such that
$\pi_q(\tilde\mu''(\vec b)) = \tilde\mu(\vec b)$, $\tilde \mu(\vec b) \in \vec t_\beta$, and for $\tilde\mu'$ it holds
$\tilde \mu'(\vec b) \in \vec t'_\beta$ $\forall \vec b \in \bin^{\ell^*}$
since
\[\tilde \mu'(\vec b) = \prj_p(\tilde\mu'')(\vec b) = \prj_p(\tilde\mu''(\vec b)) \in \prj_p(\vec t_\beta) = \vec t'_\beta.
\]
Therefore, by its correctness, the protocol $\pioprangeFF$ accepts.

Next, we have to show that the verifier's checks of step 2 are satisfied.

It is straightforward to see that the checks
$\tilde y_{\eta}(\vec{\hat s})  = \tilde \mu(\vec{\hat r}_{\eta}, \vec{\hat s})$ and $\tilde y'_{\eta}(\vec s')  = \tilde \mu'(\vec r'_{\eta}, \vec s')$ are satisfied by construction.

Let us now focus on the last check
$\tilde y_{\eta}(\vec \delta) \bmod p_l = \FFtoInt(\tilde y'_{\eta}(\vec \delta)) \bmod p_l$.

By construction of $\tilde y_{\eta}$, $\tilde y_{\eta}(\vec \delta) = \tilde\mu(\vec{\hat r}_\eta, \vec\delta)$, and since $\vec{\hat r}_{\eta} = \prj_q(\vec r_\eta)$ and $\prj_q(\tilde\mu'') = \tilde\mu$,
$\tilde\mu(\vec{\hat r}_\eta, \vec\delta) = \prj_q(\tilde\mu''(\vec{r}_\eta, \vec\delta))$.
Therefore, $\tilde y_{\eta}(\vec \delta) \bmod p_l = \prj_q(\tilde\mu''(\vec{r}_\eta, \vec\delta)) \bmod p_l = \tilde\mu''(\vec{r}_\eta, \vec\delta) \bmod p_l$.

To conclude the correctness proof, we need to argue that, for every $l$,
$\tilde y_{\eta}(\vec\delta) \bmod p_l = \tilde\mu''(\vec{r}_\eta, \vec\delta) \bmod p_l = \FFtoInt(\tilde y'_{\eta}(\vec\delta)) \bmod p_l$.
The last equality follows by construction of $\tilde y'_{\eta}$ -- by which $\tilde y'_{\eta}(\vec \delta) = \tilde\mu'(\vec r'_\eta, \vec\delta)$ -- and by applying Proposition \ref{prop:lift_eval} along with the bound on the value of $\tilde{\mu}''(\vec r_{\eta}, \vec\delta)$, which we claim below.

\begin{claim}
If $\tilde\mu''(\vec b) \in \vec t_\beta$ for all $\vec b \in \bin^{\ell^*}$ then $\abs{\tilde\mu''(\vec r_{\eta}, \vec\delta)} < 2^{m} \beta \sigma^{m} < p/2$.
\end{claim}
\begin{proof}
By the multi-linearity of the polynomial we have that for all $\vec \delta \in \bin^{\ell^* - m}$:
\[
\tilde\mu''(\vec r_{\eta}, \vec\delta) =  \sum_{\vec \phi \in \bin^{m}} \tilde\mu''(\vec \phi, \vec\delta) \tilde\rho_{\eta}(\vec \phi)
\]
where $\tilde\rho_{\eta}(\vec \phi) = \eq(\vec\phi; \vec r_{\eta}) = \prod_{i=1}^{m} (\phi_{i} r_{\eta,i} + (1-\phi_{i}) (1- r_{\eta,i}))$.
Given that $0 \leq r_{\eta,i} < \sigma$, we have that $-\sigma^{m} < \tilde\rho_{\eta}(\vec \phi) < \sigma^{m}$; thus by combining it with the assumption that $\tilde\mu''(\vec\phi,\vec\delta) < \beta$ we obtain
$\abs{\tilde\mu''(\vec r_{\eta}, \vec\delta)} < 2^{m} \beta \sigma^{m} < p/2$.
\end{proof}

This concludes the correctness proof.

\subsection{Security of $\piopGqrange$}
\label{sec:piopGqrange-security}
We prove the RBR-soundness, from which we get RBR-knowledge by then applying Lemma \ref{lemma:RBRsound2RBRksound}.

To prove RBR-soundness, we start by defining a $\state(\vec t_\beta, \ora{\tilde \mu}, \tr_i)$ function on every (partial) transcript $\tr_{i}$, from round $i=0$ to $i = 2+\mathsf{r}_{range}$, where $\mathsf{r}_{range}$ denotes the number of rounds of the $\pioprangeFF$ protocol:

\begin{itemize}
\item For $\tr_0=\emptyset$,  $\state(\vec t_\beta, \ora{\tilde \mu}, \tr_0)=0$ iff $(\vec t_\beta, \ora{\tilde \mu}) \notin \lang_{\relgrange}$.

 \item For $\tr_1 = \ora{\tilde \mu'} |  (\vec r_0, \ldots,\vec r_{\kappa-1})$,  $\state(\vec t_\beta, \ora{\tilde \mu}, \tr_1)=0$ iff
 $(\vec t'_{\beta}, \ora{\tilde \mu'}; \tilde \mu') \notin \relgrange$ or there exist $ \eta \in [\kappa]$, $\vec\delta \in \bin^{\ell^* - m}$ such that $f_{\vec\delta}(\prj_q(\vec r_\eta)) \neq 0 \bmod q$,
 where $f_{\vec\delta}(\vec X)\in \Hq[\vec X]$ is the polynomial
 \[
 f_{\vec\delta}(\vec X) = \tilde \mu(\vec X, \vec \delta)  -  (\tilde\mu''(\vec X, \vec\delta) \bmod q )
 \]
 and $\tilde\mu''$ is the multilinear polynomial such that $\tilde\mu''(\vec b) = \FFtoInt(\tilde\mu'(\vec b))$ for all $\vec b \in \bin^{\ell^*}$.

\item For $\tr_2 =  \tr_1 | (\tilde y_0, \tilde y'_0, \ldots, \tilde y_{\kappa - 1}, \tilde y'_{\kappa - 1}) | (\vec{\hat s}, \vec s')$: $\state(\vec t_\beta, \ora{\tilde \mu}, \tr_2)=0$ iff
\begin{itemize}
\item $(\vec t'_{\beta}, \ora{\tilde \mu'}; \tilde \mu') \notin \relgrange$
\item or there exist $\eta \in [\kappa]$, $\vec\delta \in \bin^{\ell^* - m}$, $l \in [0,L]$ such that
\[
\tilde y_{\eta}(\vec \delta) \bmod p_l \neq \FFtoInt(\tilde y'_{\eta}(\vec \delta)) \bmod p_l,
\]
\item or $\exists \eta \in [\kappa]$ such that
$\tilde y_{\eta}(\vec{\hat s}) \neq \tilde \mu(\vec{\hat r}_{\eta}, \vec{\hat s}) \vee \tilde y'_{\eta}(\vec s') \neq \tilde \mu'(\vec r'_{\eta}, \vec s')$.

\end{itemize}

\item For $\tr_{2+k} = \tr_2 | \tr_k'$ where, for $1 \leq k \leq \mathsf{r}_{range}$, $\tr'_k$ is the partial transcript of $\pioprangeFF$ at the $k$-th round: $\state(\vec t_\beta, \ora{\tilde \mu}, \tr_{2+k})=0$ iff
\begin{itemize}
\item $\state'(\vec t'_{\beta}, \ora{\tilde \mu'}, \tr'_{k})=0$
\item or there exist $\eta \in [\kappa]$, $\vec\delta \in \bin^{\ell^* - m}$, $l \in [0,L]$ such that
\[
\tilde y_{\eta}(\vec \delta) \bmod p_l \neq \FFtoInt(\tilde y'_{\eta}(\vec \delta)) \bmod p_l,
\]
\item or $\exists \eta \in [\kappa]$ such that
$\tilde y_{\eta}(\vec{\hat s}) \neq \tilde \mu(\vec{\hat r}_{\eta}, \vec{\hat s}) \vee \tilde y'_{\eta}(\vec s') \neq \tilde \mu'(\vec r'_{\eta}, \vec s')$.
\end{itemize}
\end{itemize}

Let us now show that this $\state$ function satisfies the three properties of RBR-soundness.

First, it is easy to see that by construction, if $(\vec t_{\beta}, \ora{\tilde \mu}; \tilde\mu)\notin\relgrange$ then $\state(\vec t_{\beta}, \ora{\tilde \mu}, \tr_0)=0$.

Second, for any {\em complete} transcript $\tr = \tr_{2+\mathsf{r}_{range}} = \tr_2 | \tr'_{\mathsf{r}_{range}}$, by construction we have that if $\state(\vec t_{\beta}, \ora{\tilde \mu}, \tr)=0$, then either $\state'(\vec t'_{\beta}, \ora{\tilde \mu'}, \tr'_{\mathsf{r}_{range}})=0$, which means that the verifier of $\pioprangeFF$ rejects (by the RBR-soundness of $\pioprangeFF$), or at least one of the verifier's checks fails. In any of these cases, $\piopGqrange.\iopV$ rejects.

Third, we show that if $\state$ rejects on a partial transcript of round $i$, for any prover's message of the round $i+1$, we have a concrete (negligible) bound on the probability that $\state$ accepts on the next partial transcript.
We show this property round by round.

\begin{claim}
If $\state(\vec t_{\beta}, \ora{\tilde \mu},\tr_0)=0$, then for any prover's message $\ora{\tilde \mu'}$ we have that
\[
\Pr_{\vec r_0, \ldots, \vec r_{\kappa-1} \sample \Sigma^{m} }[\state(\vec t_{\beta}, \ora{\tilde \mu},\tr_1=(\ora{\tilde \mu'} | (\vec r_0, \ldots, \vec r_{\kappa-1})))=1 ] \leq \left(\frac{m}{\sigma} \right)^{\kappa}.
\]
\end{claim}
\begin{proof}
By definition, $\state(\vec t_{\beta}, \ora{\tilde \mu},\tr_0)=0$ iff $(\vec t_{\beta}, \ora{\tilde \mu}) \notin \lang_{\relgrange}$, that is there exist $\vec\delta^* \in \bin^{\ell^* - m}, \vec\phi^* \in \bin^{m}$ such that $\tilde\mu(\vec\phi^*,\vec\delta^*) \notin \vec t_\beta$.
On the other hand, $\state$ accepting on $\tr_1$ means that:\\
(i) $(\vec t'_{\beta}, \ora{\tilde \mu'}; \tilde \mu') \in \relgrange$, i.e., for all $\vec\delta \in \bin^{\ell^* - m}, \vec\phi \in \bin^{m}$, $\tilde\mu'(\vec\phi,\vec\delta) \in \vec t'_\beta$;\\
(ii) for all $ \eta \in [\kappa]$, $\vec\delta \in \bin^{\ell^* - m}$: $f_{\vec\delta}(\prj_q(\vec r_\eta)) = 0 \bmod q$.

From (i) and the definition of $\tilde\mu''$, we have $\tilde\mu''(\vec\phi, \vec\delta) = \FFtoInt(\tilde\mu'(\prj_p(\vec\phi,\vec\delta))) \in \FFtoInt(\vec t'_\beta)$.
Since $\vec t'_{\beta} = \prj_p(\vec t_\beta)$, $\beta < p/2$ (from  equation \eqref{eq:params-bound-p}), and using $\FFtoInt$'s definition, we have $\FFtoInt(\prj_p(\vec t_\beta)) = \vec t_\beta$, and thus $\tilde\mu''(\vec\phi, \vec\delta) \in \vec t_\beta$ and the polynomial $f_{\vec\delta^*}(\vec X)$ is nonzero.

Therefore,
\begin{eqnarray*}
&&\Pr[\state(\vec t_{\beta}, \ora{\tilde \mu},\tr_1)=1 \mid \state(\vec t_{\beta}, \ora{\tilde \mu},\tr_0)=0] \\
&& = \Pr[(\vec t'_{\beta}, \ora{\tilde \mu'}; \tilde \mu') \in \relgrange \bigwedge_{\substack{ \eta \in [\kappa] \\ \vec\delta \in \bin^{\ell^* - m}}} f_{\vec\delta}(\prj_q(\vec r_\eta)) = 0 \mid (\vec t_{\beta}, \ora{\tilde \mu}; \tilde \mu) \notin \relgrange] \\
&& \leq \Pr[\bigwedge_{\substack{ \eta \in [\kappa] \\ \vec\delta \in \bin^{\ell^* - m}}} f_{\vec\delta}(\prj_q(\vec r_\eta)) = 0 \mid (\vec t'_{\beta}, \ora{\tilde \mu'}; \tilde \mu') \in \relgrange \wedge (\vec t_{\beta}, \ora{\tilde \mu}; \tilde \mu) \notin \relgrange] \\
&& \leq \Pr[\bigwedge_{\eta \in [\kappa]} f_{\vec\delta^*}(\prj_q(\vec r_\eta)) = 0 \mid \exists \vec\delta^* \in \bin^{\ell^* - m}: f_{\vec\delta^*}(\vec X) \neq 0 ] \\
&& = \prod_{ \eta \in [\kappa]} \Pr_{\vec r_\eta \sample \Sigma^{m}}[ f_{\vec\delta^*}(\prj_q(\vec r_\eta))=0 \mid f_{\vec\delta^*}(\vec X) \neq 0].
\end{eqnarray*}

By generalized Schwartz-Zippel \ref{lemma:generalizedSZ}, $\Pr_{\vec r_\eta \sample \Sigma^{m}}[f_{\vec\delta^*}(\prj_q(\vec r_\eta)) = 0 \mid f_{\vec\delta^*}(\vec X) \neq 0] \leq \frac{m}{\sigma}$, and the claimed bound follows by considering the probability that this occurs for $\kappa$ independent samples; by equation \eqref{eq:params-bound-soundness} this bound is $\approx 2^{-\secpar}$.
\qed
\end{proof}

\begin{claim}
If $\state(\vec t_{\beta}, \ora{\tilde \mu},\tr_1)=0$, then for any prover's message $(\tilde y_0, \tilde y'_0, \ldots, \tilde y_{\kappa-1}, \tilde y'_{\kappa-1})$ we have that
\[
\Pr_{\vec{\hat s}, \vec s'}[\state(\vec t_{\beta}, \ora{\tilde \mu},\tr_2=(\tilde y_0, \tilde y'_0, \ldots, \tilde y_{\kappa-1}, \tilde y'_{\kappa-1}) | \vec{\hat s}, \vec s'))=1] \leq  \frac{\ell^* - m}{|\Esetd|} + \frac{\ell^* - m}{p}.
\]
\end{claim}
\begin{proof}
The event $\state(\vec t_{\beta}, \ora{\tilde \mu},\tr_2)=1$ is the intersection of the following events:

\[
(\vec t'_{\beta}, \ora{\tilde \mu'}; \tilde \mu') \in \relgrange
\bigwedge_{\eta \in [\kappa]} \mathsf{V}_{3,\eta} \bigwedge_{\substack{\eta \in [\kappa], l \in [0,L] \\ \vec\delta \in \bin^{\ell^* - m}}} \mathsf{V}_{4,\eta,\vec\delta, l}
\]
where
\[
\mathsf{V}_{3,\eta} :=
(\tilde y_{\eta}(\vec{\hat s}) = \tilde \mu(\vec{\hat r}_{\eta}, \vec{\hat s})) \wedge (\tilde y'_{\eta}(\vec s') = \tilde \mu'(\vec r'_{\eta}, \vec s'))
\]
\[
\mathsf{V}_{4,\eta,\vec\delta,l}  := ( \tilde y_{\eta}(\vec \delta) \bmod p_l = \FFtoInt(\tilde y'_{\eta}(\vec \delta))  \bmod p_l )
\]

Let $\E_2$ be the event ``$\bigwedge_{\eta \in [\kappa]} (\tilde y_{\eta}(X_{m+1}, \ldots, X_{\ell^*})  = \tilde \mu(\vec{\hat r}_{\eta}, X_{m+1}, \ldots, X_{\ell^*})) \wedge
(\tilde y'_{\eta}(X_{m+1}, \ldots, X_{\ell^*}) = \tilde \mu'(\vec r'_{\eta}, X_{m+1}, \ldots, X_{\ell^*}))$''.
By partitioning over $\E_2$, we upper bound
$\Pr[\state(\vec t_{\beta}, \ora{\tilde \mu},\tr_2)=1 \mid \state(\vec t_{\beta}, \ora{\tilde \mu},\tr_1)=0]$ by the sum
 $\Pr[\state(\vec t_{\beta}, \ora{\tilde \mu},\tr_2)=1 \mid  \E_2, \state(\vec t_{\beta}, \ora{\tilde \mu},\tr_1)=0]+\Pr[\state(\vec t_{\beta}, \ora{\tilde \mu},\tr_2)=1  \mid  \bar \E_2, \state(\vec t_{\beta}, \ora{\tilde \mu},\tr_1)=0]$

The second term can be upper bounded as follows (where we use $\vec X$ to compactly denote $X_{m+1}, \ldots, X_{\ell^*}$):
\begin{eqnarray*}
&& \Pr[\state(\vec t_{\beta}, \ora{\tilde \mu},\tr_2)=1  \mid  \bar \E_2, \state(\vec t_{\beta}, \ora{\tilde \mu},\tr_1)=0] \\
&& \leq \Pr[\bigwedge_{\eta \in [\kappa]} \mathsf{V}_{3,\eta} \mid \exists \bar\eta: \tilde y_{\bar\eta}(\vec X)  \neq \tilde \mu(\vec{\hat r}_{\bar\eta}, \vec X) \vee
\tilde y'_{\bar\eta}(\vec X) \neq \tilde \mu'(\vec r'_{\bar\eta}, \vec X) ] \\
&& \leq \Pr[\tilde y_{\bar\eta}(\vec{\hat s}) = \tilde \mu(\vec{\hat r}_{\bar\eta}, \vec{\hat s}) \wedge \tilde y'_{\bar\eta}(\vec s') = \tilde \mu'(\vec r'_{\bar\eta}, \vec s') \mid \exists \bar\eta: \tilde y_{\bar\eta}(\vec X)  \neq \tilde \mu(\vec{\hat r}_{\bar\eta}, \vec X) \vee
\tilde y'_{\bar\eta}(\vec X) \neq \tilde \mu'(\vec r'_{\bar\eta}, \vec X) ] \\
&& \leq \Pr[\tilde y_{\bar\eta}(\vec{\hat s}) = \tilde \mu(\vec{\hat r}_{\bar\eta}, \vec{\hat s}) \mid \tilde y_{\bar\eta}(\vec X)  \neq \tilde \mu(\vec{\hat r}_{\bar\eta}, \vec X)] + \Pr[ \tilde y'_{\bar\eta}(\vec s') = \tilde \mu'(\vec r'_{\bar\eta}, \vec s') \mid \tilde y'_{\bar\eta}(\vec X) \neq \tilde \mu'(\vec r'_{\bar\eta}, \vec X)] \\
&& \leq  \frac{\ell^* - m}{|\Esetd|} + \frac{\ell^* - m}{p}
\end{eqnarray*}
where the latter inequality follows by applying Schwartz-Zippel over $\Hq$ (resp. $\ZZ_p$) with the random $\hat s \sample \Esetd$ (resp. $s' \sample \ZZ_p$).

For the first term, we argue that $\Pr[\state(\vec t_{\beta}, \ora{\tilde \mu},\tr_2)=1  \mid \E_2, \state(\vec t_{\beta}, \ora{\tilde \mu},\tr_1)=0]=0$ as follows.
\begin{eqnarray*}
&& \Pr[\state(\vec t_{\beta}, \ora{\tilde \mu},\tr_2)=1  \mid \E_2, \state(\vec t_{\beta}, \ora{\tilde \mu},\tr_1)=0] \\
&& \leq \Pr\left[ \begin{array}{l}
(\vec t'_{\beta}, \ora{\tilde \mu'}; \tilde \mu') \in \relgrange \\
\bigwedge_{\substack{\eta \in [\kappa],  l \in [0,L] \\ \vec\delta \in \bin^{\ell^* - m}}} \mathsf{V}_{4,\eta,\vec\delta, l}
\end{array}
\; \middle| \;
\E_2 \wedge ( (\vec t'_{\beta}, \ora{\tilde \mu'}; \tilde \mu') \notin \relgrange  \vee \exists \bar\eta, \bar l, \vec{\bar\delta}: f_{\vec{\bar\delta}}(\vec{\hat r}_{\bar\eta}) \neq 0 \bmod p_{\bar l})
 \right] \\
&& \leq \Pr\left[  \bigwedge_{\substack{\eta \in [\kappa],  l \in [0,L] \\ \vec\delta \in \bin^{\ell^* - m}}}  \!\!\!\!\!\!\!\!\!\!  \mathsf{V}_{4,\eta,\vec\delta, l}
\; \middle| \;
\E_2 \wedge  (\vec t'_{\beta}, \ora{\tilde \mu'}; \tilde \mu') \in \relgrange  \wedge \exists \bar\eta, \bar l, \vec{\bar\delta}: f_{\vec{\bar\delta}}(\vec{\hat r}_{\bar\eta}) \neq 0 \bmod p_{\bar l})\right] \\
&& \leq \Pr\left[ \tilde \mu(\vec{\hat r}_{\bar\eta},\vec{\bar\delta}) \bmod p_{\bar l} = \FFtoInt(\tilde\mu'(\vec r'_{\bar\eta},\vec{\bar\delta})) \bmod p_{\bar l}
\; \middle| \;
\begin{array}{l}
(\vec t'_{\beta}, \ora{\tilde \mu'}; \tilde \mu') \in \relgrange  \wedge \\
\exists \bar\eta, \bar l, \vec{\bar\delta}: f_{\vec{\bar\delta}}(\vec r_{\bar\eta}) \neq 0 \bmod p_{\bar l}
\end{array} \right] \\
&& \leq \Pr\left[ \tilde \mu(\vec{\hat r}_{\bar\eta},\vec{\bar\delta}) \bmod p_{\bar l} = \tilde\mu''(\vec r_{\bar\eta},\vec{\bar\delta}) \bmod p_{\bar l}
\; \middle| \;
\begin{array}{l}
(\vec t'_{\beta}, \ora{\tilde \mu'}; \tilde \mu') \in \relgrange  \wedge \\
\exists \bar\eta, \bar l, \vec{\bar\delta}: f_{\vec{\bar\delta}}(\vec r_{\bar\eta}) \neq 0 \bmod p_{\bar l}
\end{array} \right] = 0
\end{eqnarray*}
where the third inequality follows by applying $\E_2$ to the values in the event $\mathsf{V}_{4,\eta,\vec\delta,l}$ and then considering only the event for the tuple $\bar\eta, \bar l, \vec{\bar\delta}$ for which $f_{\vec{\bar\delta}}(\vec{\hat r}_{\bar\eta}) \neq 0 \bmod p_{\bar l})$;
the fourth inequality follows by applying Proposition \ref{prop:lift_eval} and that $|\tilde\mu''(\vec r_{\bar\eta},\vec{\bar\delta})| < p/2$ (which is implied by the fact that $\vec r_\eta \in \Sigma$ and $0 \leq \tilde\mu''(\vec b) < \beta$). %
Finally, the last probability is 0, since we are conditioning on
$f_{\vec{\bar\delta}}(\vec r_{\bar\eta}) \neq 0 \bmod p_{\bar l})$.
\qed
\end{proof}

To prove the last case, recall that, once $\tr_2$ is fixed, so is the event
\[
E := \exists\, \eta \in [\kappa],\, \vec\delta \in \bin^{\ell^* - m},\, l \in [0,L] :
\neg \mathsf{V}_{3,\eta} \vee \neg \mathsf{V}_{4,\eta,\vec\delta,l},
\]
that is, the event that at least one of the verifier's checks of step 2 fails.
We also let $\tr'_0 = \emptyset$.

\begin{claim}
For every $k = 0, \ldots, \mathsf{r}_{range}-1$, if
$\state(\vec t_{\beta}, \ora{\tilde \mu},\tr_{2+k})=0$, then for any prover's message
$\pi$ of the $(k+1)$-th round of $\pioprangeFF$ we have that
\[
\Pr_{\rho}\left[\state(\vec t_{\beta}, \ora{\tilde \mu},\tr_{2+k+1})=1\right] \leq \delta'_{range},
\quad \text{where } \tr_{2+k+1} = \tr_2 | \tr'_{k} | \pi | \rho,
\]
and where the probability is over the $(k+1)$-th verifier's challenge $\rho$ of $\pioprangeFF$.
\end{claim}
\begin{proof}
Since $k+1 \geq 1$, the definition of $\state$ on $\tr_{2+k+1}$ gives
\[
\state(\vec t_{\beta}, \ora{\tilde \mu},\tr_{2+k+1})=1
\iff
\state'(\vec t'_{\beta}, \ora{\tilde \mu'}, \tr'_{k} | \pi | \rho)=1 \;\wedge\; \bar E.
\]
Moreover, $E$ is determined by $\tr_2$, hence it is fixed and independent of $\rho$.
We distinguish two cases.

If $E$ occurs, then $\state(\vec t_{\beta}, \ora{\tilde \mu},\tr_{2+k+1})=1$ never occurs
and the probability of the claim is $0$.

If $E$ does not occur, then the event
$\state(\vec t_{\beta}, \ora{\tilde \mu},\tr_{2+k})=0$ of the precondition implies
$\state'(\vec t'_{\beta}, \ora{\tilde \mu'}, \tr'_{k})=0$: for $k \geq 1$ this is immediate
from the definition of $\state$ on $\tr_{2+k}$, whereas for $k=0$ the definition of $\state$
on $\tr_2$ gives $(\vec t'_{\beta}, \ora{\tilde \mu'}) \notin \lang_{\relgrange}$, which in turn
implies $\state'(\vec t'_{\beta}, \ora{\tilde \mu'}, \tr'_{0})=0$ by the first property of RBR-soundness (Definition \ref{def:rbr-soundness}) applied to $\pioprangeFF$. Therefore, for all $0 \leq k < \mathsf{r}_{range}$,
\begin{eqnarray*}
&& \Pr_{\rho}[\state(\vec t_{\beta}, \ora{\tilde \mu},\tr_{2+k+1})=1] \\
&& = \Pr_{\rho}[\state'(\vec t'_{\beta}, \ora{\tilde \mu'}, \tr'_{k} | \pi | \rho)=1 \wedge \bar E] \\
&& \leq \Pr_{\rho}[\state'(\vec t'_{\beta}, \ora{\tilde \mu'}, \tr'_{k} | \pi | \rho)=1]
\;\leq\; \delta'_{range},
\end{eqnarray*}
where the last inequality follows by the third property of RBR-soundness applied to $\pioprangeFF$ and to the partial
transcript $\tr'_{k}$, on which $\state'$ evaluates to $0$.
\qed
\end{proof}

The final RBR-soundness error is the max of the three per-round bounds, namely $\max\left\{\left(\frac{m}{\sigma}\right)^\kappa; \frac{\ell^* - m}{|\Esetd|} + \frac{\ell^* - m }{p}; \delta'_{range}\right\}$, where the first term is $\approx 2^{-\secpar}$ by equation \eqref{eq:params-bound-soundness}.

\subsection{Efficiency of $\piopGqrange$}
In addition to the cost of running the PIOP $\pioprangeFF$, the efficiency of this protocol is as follows.

\begin{itemize}
	\item The prover sends 1 oracle, $\ora{\tilde\mu'}$, whose size is $2^{\ell^*}$ elements of $\ZZ_p$; in addition the prover sends $\kappa 2^{\ell^* - m}$ elements of $\ZZ_q$ and as many elements of $\ZZ_p$.
	\item The proving time is $O(\kappa \cdot 2^{\ell^*})$, which covers the computation of the partial evaluations of $\tilde \mu$ and $\tilde \mu'$ sent in round 2.

	\item The verifier time is $O(\kappa \cdot 2^{\ell^* - m})$.

	\item The number of verifier's queries is $2 \kappa$, which includes $\kappa$ queries to $\ora{\tilde \mu}$ and $\kappa$ queries to $\ora{\tilde \mu'}$.

\end{itemize}

\headingb{Instantiating the concrete parameters.}
If we target the use of a prime $p$ of 200 bits, and assume $\ell = 20$ and $\beta = 2^{20}$, then we can take for example
$m =  \ell - \lceil \log\ell \rceil = 15$,
$\sigma = 2^{11}$,
$\kappa = 18$.
This gives $\tilde\mu''(\vec r_{\eta},\vec\delta) < 2^{m} \cdot \beta \cdot \sigma^{m} < p/2 \approx 2^{199}$.

Looking specifically at the parameters that fit our use of range proofs for the largest circuit in our experiments, then we have:
$\ell=9$, $\beta=2^{11}$, $\gamma=2$. To fit a 200-bits prime $p$, then we can set the parameters of the PIOP as
$m =  \ell$,
$\sigma = 2^{19}$,
$\kappa = 8$.
These choices satisfy the correctness and soundness bounds since
\[
\tilde\mu''(\vec r_{\eta},\vec\delta) < 2^{m} \cdot \beta \cdot \sigma^{m} =2^{191} < p/2 \approx 2^{199}
\]
\[
\left(\frac{m}{\sigma}\right)^{\kappa} =\left(\frac{9}{2^{19}}\right)^{8} \leq 2^{-120}
\]
and $m=\ell$ clearly ensures the succinctness bound.

\subsection{A PIOP for Range Checks over a Finite Field}
\label{sec:PIOPrangeFF}
The protocol relies on the following lemma from \cite{C:CCCFGS25}, which simplifies Spartan's memory checking technique \cite{C:Setty20} to the case of lookups in $\vec t_{\beta}$.

\begin{lemma}[\cite{C:CCCFGS25}]\label{lemma:lookup2multiset}
	Let $p, \ell^{*}$ be positive integers such that $p > 2^{\ell^{*}}$.
	Let $\FF$ be a field of characteristic $p$, let $\tilde \mu$ be an $\ell^{*}$-variate multilinear polynomial over $\FF$, and let $\tilde t_{\beta}$ be the MLE of a table of size $\beta = 2^{b^*}$ whose elements are all distinct.
	Then we have that $\forall \vec i \in \bin^{\ell^*}$ $\tilde \mu(\vec i) \in \{\tilde t_{\beta}(\vec j) : \vec j \in \bin^{b^*}\}$ if and only if there exist polynomials $\readts \in \FF[X_1, \ldots, X_{\ell^{*}}], \finalts \in \FF[X_1, \ldots, X_{b^*}]$ such that
	$\mathsf{WS} = \mathsf{RS} \cup \mathsf{S}$ where
	\begin{eqnarray*}
		\WS & = & \{(\tilde{t}_{\beta}(\vec j), 0): \vec j \in \bin^{b^*}\} \cup \{(\tilde{\mu}(\vec i),\readts(\vec i)+1): \vec i \in \bin^{\ell^*}\} \\
		\RS & = & \{(\tilde{\mu}(\vec i),\readts(\vec i)): \vec i \in \bin^{\ell^*}\} \\
		\sfS & = &\{(\tilde{t}_{\beta}(\vec j), \finalts(\vec j)): \vec j \in \bin^{b^*}\}.
	\end{eqnarray*}
\end{lemma}

\medskip

Following the lemma, the PIOP prover builds the multilinear polynomials $\readts, \finalts$ and aims to convince the verifier that $\mathsf{WS} = \mathsf{RS} \cup \mathsf{S}$.

To prove this multiset equality, the verifier sends random challenges $\sigma,\tau \sample \FF$, the prover computes $\overline{\WS}_{t}, \overline{\WS}_h, \overline{\RS} , \overline{\sfS}$ as follows
\begin{eqnarray}
\overline{\WS} & = & \overbrace{\prod_{\vec j \in \bin^{b^*}} (\tilde{t}_{\beta}(\vec j) - \tau)}^{\overline{\WS}_{t}}
 \overbrace{\prod_{\vec i \in \bin^{\ell^*}} ( \tilde{\mu}(\vec i) + \sigma (\readts(\vec i)+1) - \tau)}^{\overline{\WS}_{\mu}} \label{eq:WSbar} \\
\overline{\RS} & = & \prod_{\vec i \in \bin^{\ell^*}} (\tilde{\mu}(\vec i) + \sigma\cdot \readts(\vec i)-\tau) \label{eq:RSbar} \\
\overline{\sfS} & = &\prod_{\vec j \in \bin^{b^*}}( \tilde{t}_{\beta}(\vec j) + \sigma\cdot \finalts(\vec j) -
\tau), \label{eq:Sbar}
\end{eqnarray}
and the verifier checks that
$\overline{\WS}_{t} \cdot \overline{\WS}_\mu = \overline{\RS} \cdot \overline{\sfS}$.
The prover must finally convince the verifier that each of the four values is computed correctly; this is done using Quark's grand product.

\begin{lemma}[Grand product Lemma \cite{EPRINT:SetLee20}]\label{lemma:grandprod}
$P = \prod_{\vec i\in\{0,1\}^\ell} v(\vec i)$ if and only if there exists a multilinear polynomial $f$ in $\ell + 1$ variables such that
$f(1, \ldots, 1, 0)  = P$, and $\forall \vec i \in \bin^{\ell}$ it holds
$f(0, \vec i) = v(\vec i)$ and
$f(1, \vec i) =  f(\vec i, 0) \cdot f(\vec i, 1)$.
\end{lemma}

\medskip
In Fig.~\ref{fig:PIOPrangeFF} we present the full specification of the PIOP $\pioprangeFF$ for $\relgrange$.

\begin{Boxfig}[H]{PIOP $\pioprangeFF$, where $\vec t_{\beta} \in \FF^{2^{b^*}}$, $\tilde \mu \in \FFml[X_1, \ldots, X_{\ell^*}]$.}{PIOPrangeFF}{\underline{PIOP $\pioprangeFF$ for $(\vec t_{\beta}, \ora{\tilde \mu};
\tilde\mu) \in \relgrange$}}
	\begin{enumerate}
		\item Let $\tilde \mu \in \FFml[X_1, \ldots, X_{\ell^*}]$.
		$\iopP$ constructs $\readts \in \FFml[X_1, \ldots, X_{\ell^*}]$, $\finalts \in \FFml[X_1, \ldots, X_{b^*}]$ following Lemma \ref{lemma:lookup2multiset} applied to $\tilde \mu$ and $\tilde t_{\beta}$.

		Send $\ora{\readts}, \ora{\finalts}$.

		$\iopV$ replies with random $\sigma, \tau \sample \FF$.

		\item	$\iopP$ constructs multilinear polynomials $f^{(1)}_{\WS_\mu}, f^{(1)}_{\RS}, f^{(1)}_{\WS_t}, f^{(1)}_{\sfS}$ such that\\
		for  $\alpha = \WS_\mu, \RS$:\ \
		$f^{(1)}_{\alpha}(\vec i)  =  f^{(i_1)}_{\alpha}(\vec i', 0) \cdot f^{(i_1)}_{\alpha}(\vec i', 1)\ \ \forall \vec i \in \bin^{\ell^*} $\\
		for $\alpha=\WS_t, \sfS$:\ \
		$f^{(1)}_{\alpha}(\vec j)  =  f^{(j_1)}_{\alpha}(\vec j', 0) \cdot f^{(j_1)}_{\alpha}(\vec j', 1) \ \  \forall \vec j \in \bin^{b^*}$\\
		where for a given $\vec i \in \bin^{m}$, $\vec i'$ denotes $(i_2, \ldots, i_{m})$, and the polynomials $f^{(0)}_{\alpha}$ are
		\begin{center}
		\begin{tabular}{ll}
		$f^{(0)}_{\WS_\mu}  =    \tilde{\mu} + \sigma \cdot ( \readts+1)-\tau, \quad\quad$ &$f^{(0)}_{\RS}   =   \tilde{\mu} + \sigma \cdot \readts-\tau$,\\
		\noalign{\smallskip}
		$f^{(0)}_{\WS_t}   =   \tilde{t}_{\beta}  - \tau$, &$f^{(0)}_{\sfS}  =   \tilde{t}_{\beta} + \sigma\cdot \finalts - \tau$
		\end{tabular}
		\end{center}

		$\iopP$ sends $\ora{{f}^{(1)}_{\WS_\mu}}, \ora{{f}^{(1)}_{\RS}},\ora{{f}^{(1)}_{\WS_t}}, \ora{{f}^{(1)}_{\sfS}}$.

		$\iopV$ replies with random $\chi \sample \FF$, $\vec\rho \sample \FF^{\ell^*}, \vec\xi \sample \FF^{b^*}$.

		\item  $\iopP$ and $\iopV$ run $\piopsum$ for $((0, \ora{\mu^*});  \mu^*) \in \relsum$ and $((0, \ora{t^*});  t^*) \in \relsum$ where $\ora{\mu^*}, \ora{t^*}$ are the virtual oracles: \pccomment{see \ifShortVersion below \else\figref{Functiong} \fi for the definition of $g$} \\
		$\ora{t^*} = g_{\chi,\vec\xi}\left(\ora{\tilde t_{\beta}} - \tau, \ora{\tilde t_{\beta}} + \sigma \cdot \ora{\finalts} - \tau, \ora{f^{(1)}_{\WS_t}}, \ora{f^{(1)}_{\sfS}}\right)$ \\
	 	$\ora{\mu^*} = g_{\chi,\vec\rho}\big(\ora{\tilde \mu} + \sigma \cdot (\ora{\readts }+ 1) - \tau, \ora{\tilde \mu} + \sigma \cdot \ora{\readts} \allowbreak - \tau,\ora{f^{(1)}_{\WS_\mu}}, \ora{f^{(1)}_{\RS}}\big)$.\\

		$\iopV$ queries $\ora{f^{(1)}_{\WS_\mu}}, \ora{f^{(1)}_{\RS}}$ on $(\vec 1, 0) \in \FF^{\ell^*}$, $\ora{f^{(1)}_{\WS_t}}, \ora{f^{(1)}_{\sfS}}$ on $(\vec 1, 0) \in \FF^{b^*}$, and then checks if
		 $f^{(1)}_{\WS_t}(\vec 1, 0) \cdot f^{(1)}_{\WS_\mu}(\vec 1, 0) = f^{(1)}_{\RS}(\vec 1, 0) \cdot f^{(1)}_{\sfS}(\vec 1, 0)$.
	\end{enumerate}
\end{Boxfig}
\begin{Boxfig}{Function $g$ used by PIOP $\pioprangeFF$}{Functiong}{\underline{Function $g_{\chi,\vec\rho}(f^{(0)}_1, f^{(0)}_2, {f^{(1)}_{1}}, {f^{(1)}_{2}} )$}}
	for $k=1,2$, let $f_k(X_0, \vec X) \gets (1-X_0) \cdot f^{(0)}_k(\vec X) + X_0 \cdot {f^{(1)}_k(\vec X)}$\\
	Return $(f_1(1, \vec X) - f_1(\vec X,0) \cdot f_1(\vec X,1) + \chi \cdot ( f_2(1, \vec X) - f_2(\vec X,0) \cdot f_2(\vec X,1))) \cdot \tilde{eq}(\vec\rho, \vec X)$
\end{Boxfig}

Below, we state and prove the security of $\pioprangeFF$. Our proof is inspired to that of \cite[Lemma 3.2]{C:CCCFGS25}; here we specialize it to an instantiation over a finite field and prove round-by-round soundness rather than soundness.

\begin{theorem}\label{thm:smallrangePIOP}
    Let $p$ be a prime number and let us consider the finite field $\FF=\ZZ_p$.
	Let $\ell^*, \beta$ be positive integers such that $\beta = 2^{b^*} < p$ and $2^{\ell^*} < p$.
	Then $\pioprangeFF$ is a PIOP for $\relgrange$, with table $\vec t_{\beta} = \{0, \ldots, \beta - 1\}$, that has perfect completeness and RBR-knowledge error
	$\delta'_{range} = \frac{2^{\ell^{*}} + 2^{b^{*}}}{p}$.
\end{theorem}

\begin{proof}
We show that $\pioprangeFF$ is RBR-sound with error $\delta'_{range} = \frac{2^{\ell^{*}} + 2^{b^{*}}}{p}$.
RBR-knowledge with the same error follows by applying Lemma \ref{lemma:RBRsound2RBRksound},
using the fact that $\relgrange$'s witness is only the polynomial $\tilde\mu$ that is provided as an
oracle in the statement.

\headingb{Notation.}
Given the polynomials $\tilde\mu,\tilde t_\beta, \readts,\finalts$ available after the first round,
let $\WS, \RS, \sfS$ be the multisets of pairs as in the statement of Lemma \ref{lemma:lookup2multiset},
and let $\WS_{t}, \WS_{\mu}$ be the following partition of $\WS$:
\[
\WS_{t}  =  \{(\tilde t_{\beta}(\vec j), 0): \vec j \in \bin^{b^{*}}\} \qquad
\WS_{\mu} = \{(\tilde\mu(\vec i), \readts(\vec i)+1): \vec i \in \bin^{\ell^{*}}\}
\]
For $\alpha \in \{\WS_{t}, \WS_{\mu}, \RS, \sfS\}$ let use define the polynomials
\[
P_\alpha(Z,Z') = \prod_{(a_{0},a_{1}) \in \alpha} (a_{0} + a_{1} \cdot Z - Z') \in \FF[Z,Z'],
\]
and let
\[
P(Z,Z') = P_{\WS_t}(Z,Z') \cdot P_{\WS_\mu}(Z,Z') - P_{\RS}(Z,Z')\cdot P_{\sfS}(Z,Z').
\]
Note that $\deg P \leq 2^{\ell^{*}} + 2^{b^{*}}$ and that the coefficients of $P$ are determined by $\tilde t_{\beta}, \tilde \mu, \readts, \finalts$, hence they are fixed before $\iopV$ sends $\sigma, \tau$.

For $\alpha \in \{\WS_{t}, \WS_{\mu}, \RS, \sfS\}$ let $f_{\alpha}(X_{0}, \vec X) = (1-X_{0})\cdot f^{(0)}_{\alpha}(\vec X) + X_{0} \cdot f^{(1)}_{\alpha}(\vec X)$, where $f^{(0)}_{\alpha}$ is as in \figref{PIOPrangeFF} and $f^{(1)}_{\alpha}$ is the polynomial sent by $\iopP$ as an oracle in round 2.
Let $\bar\alpha = f^{(1)}_{\alpha}(\vec 1, 0) = f_{\alpha}(1, \ldots, 1, 0)$ denote the values that $\iopV$ queries in step 6.
We define two events, both determined by the oracles available after round 2:
\begin{itemize}
\item $\GP$: for every $\alpha$ and every $\vec i$ in the corresponding hypercube ($\bin^{\ell^{*}}$ for $\alpha = \WS_{\mu}, \RS$ and $\bin^{b^{*}}$ for $\alpha = \WS_{t}, \sfS$) it holds $f_{\alpha}(1, \vec i) = f_{\alpha}(\vec i, 0)\cdot f_{\alpha}(\vec i, 1)$, i.e., the grand-product statement is satisfied;
\item $\Fin$: the check of step 6 passes, i.e., $\overline{\WS}_{t} \cdot \overline{\WS}_{\mu} = \overline{\RS} \cdot \overline{\sfS}$.
\end{itemize}
Finally, let $\state_{sum}$ be the state function defined by the RBR-soundness of $\piopsum$, and let us define $\state_{sum}'$ as the state function for the parallel execution of the two sumchecks $\piopsum$ in step 3, which is defined to output $0$ on a partial transcript of the parallel execution iff $\state_{sum}$ outputs 0 on at least one of its two instances.

Since $f_{\alpha}(\vec X, 0) \cdot f_{\alpha}(\vec X, 1)$ has degree $2$ in each variable and $\eq(\vec\rho, \vec X)$ has degree $1$, the virtual oracles $\mu^{*}, t^{*}$ have degree $3$ in each variable.
We use that $\piopsum$, run over $\FF$ on a virtual oracle of degree $d$ in each variable, is RBR-sound with per-round error $d/|\FF|$; this is the round-by-round analogue of Theorem \ref{thm:sumcheckPIOP} and follows from the standard round-by-round analysis of the sum-check protocol \cite{STOC:CCHLRR19}.
Hence the per-round RBR-soundness error of the two instances of $\piopsum$ run in step 3 is $3/p$.
Let us write $\mathsf{r}_{sum} = \max(\ell^{*}, b^{*})$ for the number of rounds of these $\piopsum$ executions in step 3, so that $\mathsf{r}_{range} = \mathsf{r}_{sum} + 2$.

\headingb{The state function.}
We define $\state(\vec t_{\beta}, \ora{\tilde \mu}, \tr_{i})$ on every (partial) transcript, from $i=0$ to $i = 2 + \mathsf{r}_{sum}$, as follows:
\begin{itemize}
\item For $\tr_{0} = \emptyset$: $\state(\vec t_{\beta}, \ora{\tilde \mu}, \tr_{0}) = 0$ iff $(\vec t_{\beta}, \ora{\tilde \mu}) \notin \lang_{\relgrange}$.

\item For $\tr_{1} = \ora{\readts} | \ora{\finalts} | (\sigma, \tau)$: $\state(\vec t_{\beta}, \ora{\tilde \mu}, \tr_{1}) = 0$ iff $P(\sigma,\tau) \neq 0$.

\item For $\tr_{2} = \tr_{1} | \left(\ora{f^{(1)}_{\WS_\mu}}, \ora{f^{(1)}_{\RS}}, \ora{f^{(1)}_{\WS_t}}, \ora{f^{(1)}_{\sfS}}\right) | (\chi, \vec\rho, \vec\xi)$: $\state(\vec t_{\beta}, \ora{\tilde \mu}, \tr_{2}) = 0$ iff $\neg\Fin$, or $((0, \ora{\mu^{*}}); \mu^{*}) \notin \relsum$, or $((0, \ora{t^{*}}); t^{*}) \notin \relsum$.

\item For $\tr_{2+k} = \tr_{2} | \tr''_{k}$, where $\tr''_{k}$ is the partial transcript of $\piopsum$ parallel execution at the $k$-th round, $1 \leq k \leq \mathsf{r}_{sum}$: $\state(\vec t_{\beta}, \ora{\tilde \mu}, \tr_{2+k}) = 0$ iff $\neg\Fin$ or $\state'_{sum}(0, (\ora{\mu^*}, \ora{t^*}),\tr''_{k}) = 0$.
\end{itemize}
We also let $\tr''_{0} = \emptyset$. Note that $\Fin$ is determined by $\tr_{2}$, so it is fixed in all the rounds of step 5.

\medskip
The first property of RBR-soundness holds by construction: if $(\vec t_{\beta}, \ora{\tilde \mu}) \notin \lang_{\relgrange}$ then $\state(\vec t_{\beta}, \ora{\tilde \mu}, \tr_{0}) = 0$.

For the second property, let $\tr = \tr_{2+\mathsf{r}_{sum}}$ be a complete transcript with $\state(\vec t_{\beta}, \ora{\tilde \mu}, \tr) = 0$.
Then either $\neg\Fin$, in which case the check of step 6 fails, or $\state_{sum}(\tr''_{\mathsf{r}_{sum}}) = 0$, in which case the verifier of $\piopsum$ rejects by the second property of RBR-soundness applied to $\piopsum$. In both cases $\pioprangeFF.\iopV$ rejects.

For the third property, we show below that if $\state$ rejects on a partial transcript of round $i$, for any prover's message of round $i+1$, there is a negligible probability that $\state$ turns into accept on the next partial transcript, over the random choice of the round-$(i+1)$ verifier's challenges.

\begin{claim}
If $\state(\vec t_{\beta}, \ora{\tilde \mu},\tr_{0})=0$, then for any prover's message $\ora{\readts}, \ora{\finalts}$,
\[
\Pr_{\sigma, \tau \sample \FF}\left[\state(\vec t_{\beta}, \ora{\tilde \mu},\tr_{1})=1\right] \leq \frac{2^{\ell^{*}} + 2^{b^{*}}}{p}.
\]
\end{claim}
\begin{proof}
By definition $\state(\vec t_{\beta}, \ora{\tilde \mu},\tr_{0})=0$ there exists $\vec i^{*} \in \bin^{\ell^{*}}$ such that $\tilde\mu(\vec i^{*}) \notin \vec t_{\beta}$.
Since $\beta = 2^{b^{*}} < p$, the elements of $\vec t_{\beta} = \{0, \ldots, \beta-1\}$ are pairwise distinct in $\FF$, and since $2^{\ell^{*}} < p$ we can apply Lemma \ref{lemma:lookup2multiset} to $\tilde\mu$ and $\tilde t_{\beta}$.
This gives us that $\WS \neq \RS \cup \sfS$ as multisets, {\em for every} choice of the polynomials $\readts, \finalts$ sent by $\iopP$ in round 1.
By \cite[Lemmas 7.2--7.4]{C:Setty20}, we have that the polynomial $P(Z,Z') \neq 0 \in \FF[Z,Z']$.

Since $\state(\vec t_{\beta}, \ora{\tilde \mu},\tr_{1})=1$ iff $P(\sigma,\tau) = 0$,
the claim follows by applying the Schwartz-Zippell lemma to $P$, which is of total degree at most $2^{\ell^{*}} + 2^{b^{*}}$.
\qed
\end{proof}

Next we analyze the transition from $\tr_{1}$ to $\tr_{2}$.

\begin{claim}
If $\state(\vec t_{\beta}, \ora{\tilde \mu},\tr_{1})=0$, then for any prover's message $\ora{f^{(1)}_{\WS_\mu}}, \ora{f^{(1)}_{\RS}}, \ora{f^{(1)}_{\WS_t}}, \ora{f^{(1)}_{\sfS}}$,
\[
\Pr_{\chi, \vec\rho, \vec\xi}\left[\state(\vec t_{\beta}, \ora{\tilde \mu},\tr_{2})=1\right] \leq \frac{\max(\ell^{*}, b^{*}) + 1}{p}.
\]
\end{claim}
\begin{proof}
Recall that $\state(\vec t_{\beta}, \ora{\tilde \mu},\tr_{1})=0$ means $P(\sigma,\tau) \neq 0$, and that whether $\GP$ occurs is determined only by the prover's message, {\em independently} of $\chi, \vec\rho, \vec\xi$.
We distinguish the two cases.

If $\GP$ occurs, then for every $\alpha$ the polynomial $f_{\alpha}$ satisfies the assumption of Lemma \ref{lemma:grandprod} (with polynomial $v = f^{(0)}_{\alpha}$). Indeed, $f_{\alpha}(0, \vec X) = f^{(0)}_{\alpha}(\vec X)$ by construction and therefore the value
$\bar\alpha = f_{\alpha}(1, \ldots, 1, 0) = \prod_{\vec i} f^{(0)}_{\alpha}(\vec i)$.
By the definition of the polynomials $f^{(0)}_{\alpha}$ we have $\prod_{\vec i} f^{(0)}_{\alpha}(\vec i) = P_\alpha(\sigma,\tau)$.
Hence $\bar\alpha = P_\alpha(\sigma,\tau)$ for all four $\alpha$, and thus $\Fin$ would give $P(\sigma,\tau) = 0$, contradicting $\state(\vec t_{\beta}, \ora{\tilde \mu},\tr_{1})=0$. Therefore $\neg\Fin$ and $\state(\vec t_{\beta}, \ora{\tilde \mu},\tr_{2})=1$ with probability $0$.

If $\GP$ does not occur, there are a multiset $\alpha^{*}$ and a point $\vec i^*$ of the corresponding hypercube for which the grand-product statement fails.
Belowe we analyze the case $\alpha^{*} \in \{\WS_{\mu}, \RS\}$; the case $\alpha^{*} \in \{\WS_{t}, \sfS\}$ is symmetric, following by considering $\vec\xi$ and $b^{*}$ in place of $\vec\rho$ and $\ell^{*}$ respectively.

For $\vec i \in \bin^{\ell^{*}}$ let
\[
A(\vec i) = f_{\WS_{\mu}}(1, \vec i) - f_{\WS_{\mu}}(\vec i, 0)\cdot f_{\WS_{\mu}}(\vec i,1),
\qquad
B(\vec i) = f_{\RS}(1, \vec i) - f_{\RS}(\vec i, 0)\cdot f_{\RS}(\vec i,1),
\]
and let $\tilde G$ be the multilinear polynomial such that $\tilde G(\vec i) = A(\vec i) + \chi \cdot B(\vec i)$ for all $\vec i$.

By the definition of $g$ in \figref{Functiong} and since $\sum_{\vec i \in \bin^{\ell^{*}}} h(\vec i)\cdot\eq(\vec\rho, \vec i) = \tilde h(\vec\rho)$ for the MLE $\tilde h$ of any $h: \bin^{\ell^{*}} \to \FF$, we have
\[
\sum_{\vec i \in \bin^{\ell^{*}}} \mu^{*}(\vec i) = \widetilde{G}(\vec\rho).
\]

By the assumption that $\GP$ does not occur, there is an $\vec i^*$ such that $A(\vec i^{*}) \neq 0$ or $B(\vec i^{*}) \neq 0$. Hence, $A(\vec i^{*}) + X \cdot B(\vec i^{*})$ is a non-zero polynomial in $X$ of degree at most $1$, and thus $A(\vec i^{*}) + \chi \cdot B(\vec i^{*}) \neq 0$ except with probability $\leq 1/p$ over $\chi \sample \FF$.
Therefore, $\widetilde G$ is a non-zero multilinear polynomial in $\ell^{*}$ variables, and by Lemma \ref{lemma:generalizedSZ} $\widetilde G(\vec\rho) = 0$ with probability at most $\ell^{*}/p$ over $\vec\rho \sample \FF^{\ell^{*}}$.
Since $\state(\vec t_{\beta}, \ora{\tilde \mu},\tr_{2})=1$ implies in particular $((0, \ora{\mu^{*}}); \mu^{*}) \in \relsum$, i.e., $\widetilde G(\vec\rho) = 0$, in this case of $\alpha^{*} \in \{\WS_{\mu}, \RS\}$ we can bound
$\Pr[\state(\vec t_{\beta}, \ora{\tilde \mu},\tr_{2})=1] \leq (\ell^{*}+1)/p$.
In the symmetric case of $\alpha^{*} \in \{\WS_{t}, \sfS\}$ we instead can bound
$\Pr[\state(\vec t_{\beta}, \ora{\tilde \mu},\tr_{2})=1] \leq (b^{*}+1)/p$.
This proves the bound of the claim.
\qed
\end{proof}

Finally, we analyze the transition from $\tr_2$ to $\tr_{2+k}$.

\begin{claim}
For every $k = 0, \ldots, \mathsf{r}_{sum}-1$, if $\state(\vec t_{\beta}, \ora{\tilde \mu},\tr_{2+k})=0$, then for any prover's message $\pi$ of the $(k+1)$-th round of step 5,
\[
\Pr_{\rho}\left[\state(\vec t_{\beta}, \ora{\tilde \mu},\tr_{2+k+1})=1\right] \leq \frac{3}{p},
\]
where $\tr_{2+k+1} = \tr_{2} | \tr''_{k} | \pi | \rho$ and the probability is over the $(k+1)$-th verifier's challenge $\rho$ of step 5.
\end{claim}
\begin{proof}
Since $k+1 \geq 1$, by definition $\state(\vec t_{\beta}, \ora{\tilde \mu},\tr_{2+k+1})=1$ iff $\Fin$ and $\state_{sum}(\tr''_{k} | \pi | \rho)=1$.
As observed above, $\Fin$ is determined by $\tr_{2}$, hence it is fixed and independent of $\rho$.

If $\Fin$ does not occur, then $\state(\vec t_{\beta}, \ora{\tilde \mu},\tr_{2+k+1})=1$ never occurs and the probability of the claim is $0$.

If $\Fin$ occurs, then $\state(\vec t_{\beta}, \ora{\tilde \mu},\tr_{2+k})=0$ implies
$\state'_{sum}(0, (\ora{\mu^*}, \ora{t^*}),\tr''_{k}) = 0$.
For $k \geq 1$ this is immediate from the definition of $\state$ on $\tr_{2+k}$, whereas for $k=0$ the definition of $\state$ on $\tr_{2}$ gives that at least one of $((0, \ora{\mu^{*}}); \mu^{*}), ((0, \ora{t^{*}}); t^{*})$ is not in $\relsum$, which implies
$\state'_{sum}(0, (\ora{\mu^*}, \ora{t^*}),\tr''_{0}) = \state'_{sum}(0, (\ora{\mu^*}, \ora{t^*}),\emptyset)=0$
by the first property of RBR-soundness applied to the parallel execution of $\piopsum$.
The claim then follows by the third property of RBR-soundness applied to the instance of $\piopsum$ on which $\state_{sum}$ evaluates to $0$, whose virtual oracle has degree $3$ in each variable.
\qed
\end{proof}

The RBR-soundness error of $\pioprangeFF$ is the maximum of the three per-round bounds established by the claims above, namely
\[
\max\left\{\frac{2^{\ell^{*}} + 2^{b^{*}}}{p}; \frac{\max(\ell^{*}, b^{*})+1}{p}; \frac{3}{p}\right\} = \frac{2^{\ell^{*}} + 2^{b^{*}}}{p}.
\]
\qed
\end{proof}

\subsection{Our Optimized PIOP $\pioprangeopt$}
\label{sec:pioprangeopt}
In this section, we present a PIOP protocol, called $\pioprangeopt$, for the relation $\relrange$ of equation \eqref{eq:rangelookup}, which is more optimized than the PIOP $\pioprange$ obtained by combining the PIOP $\pioprange$ of Section \ref{sec:range} (Figure \ref{fig:pioprange}) with the PIOPs $\piopGqrange$ and $\pioprangeFF$ described in the previous sections.

For this protocol, we reuse the same preliminaries and building blocks of Section \ref{sec:newPIOPrange}, except that we use the following bound instead of that of equation \eqref{eq:params-bound-soundness}:
\begin{align}
\left(\frac{\ell + \nu}{\sigma}\right)^{\kappa} & \approx 2^{-\secpar} \label{eq:params-bound-soundness-opt}
\end{align}

We present the protocol in Figure \ref{fig:pioprangeopt}.

\begin{Boxfig}[htbp]{Construction of the PIOP $\pioprangeopt$ from $\pioprangeFF$ over $\ZZ_p$}{pioprangeopt}{\underline{\textbf{PIOP $\pioprangeopt$ for $(T_{B}, \ora{\tilde v}; \tilde v) \in \relrange$}}}
\begin{enumerate}
\item $\iopP$ computes $\tilde u = \unpack(\tilde v) \in \Rqml[\vec X, \vec Z]$ and  $\tilde \mu \in \ZZ^{(\leq 1)}[\vec X, \vec Z, \vec Z']$ as the decomposition of $\tilde u$ according to equation \eqref{eq:vdecomposition}, computes $\tilde\mu' = \prj_p(\tilde\mu)\in \ZZ_p[\vec X, \vec Z, \vec Z']$ and sends $\ora{\tilde \mu'}$.

$\iopV$ samples $\vec r_0, \ldots, \vec r_{\kappa - 1} \sample \Sigma^{\ell + \nu}$ and sends them to $\iopP$.

\item $\iopP$ sends $(\tilde\psi'_0, \ldots, \tilde\psi'_{\kappa - 1})$, where, for $\eta \in [\kappa]$, $\vec r'_{\eta} = \prj_p(r_{\eta,1}, \ldots, r_{\eta,m})$ and:
\[
\tilde\psi'_{\eta}(X_{m+1}, \ldots X_\ell, \vec Z, \vec Z')  = \tilde \mu'(\vec r'_{\eta},X_{m+1}, \ldots X_\ell, \vec Z, \vec Z')
\]
$\iopV$ samples $\vec s' \sample \FF_p^{\ell -m + \nu + \gamma}$.

For all $\eta \in [\kappa]$, queries $\ora{\tilde \mu'}$ on $(\vec r'_{\eta}, \vec s')\in \FF_p^{\ell+\nu+\gamma}$ and the virtual oracle $\ora{\tilde u}$ on $\vec{\hat r}_\eta = \prj_q(\vec r_{\eta}) \in \Zq^{\ell+\nu}$.
Then $\iopV$ checks if:
\[
 \forall \eta \in [\kappa]: \tilde\psi'_{\eta}(\vec s')  = \tilde \mu'(\vec r'_{\eta}, \vec s') \in \ZZ_{p}
\]
and $\forall l \in [0, L], \eta \in [\kappa]$
\[
\tilde u(\vec{\hat r}_{\eta}) \bmod p_l = \!\!\!\!\!\!\!\!\!\! \sum_{\substack{\vec\delta\in\bin^{\ell+\nu -m}\\ \vec k \in \bin^{\gamma}}} \!\!\!\!\!\!\!\!\!\! \FFtoInt(\tilde\psi'_{\eta}(\vec \delta, \vec k)) \cdot \eq(\vec\delta; (r_{\eta,m+1}, \ldots, r_{\eta,\ell+\nu})) \cdot \beta^{\toint(\vec k)} \bmod p_l
\]

\item $\iopP$ and $\iopV$ run a PIOP $\pioprangeFF$ for $(\vec t'_{\beta},  \ora{\tilde \mu'};  \tilde \mu') \in \relgrange$, where $\vec t_{\beta}' = \prj_p(\vec t_{\beta})$.
\end{enumerate}
\end{Boxfig}

The following theorem states the correctness and security of $\pioprangeopt$.

\begin{theorem}\label{thm:pioprangeopt}
Let $B = 2^{b}$, $N = 2^{\nu}$, $c = 2^{\gamma}$ be such that $c \mid b$ and $\beta = B^{1/c}$, and let $T_{B}$ and $\vec t_{\beta} = \{0, \ldots, \beta - 1\} \subset \ZZ$ be the tables of Lemma \ref{lemma:Rq2Zqrange}.
Consider the parameters $q, p, \beta, \sigma, m, \kappa$ satisfying equations \eqref{eq:params-bound-p}, \eqref{eq:params-bound-soundness-opt} and \eqref{eq:params-bound-efficiency} for $\ell^{*} = \ell + \nu + \gamma$.
Let $\pioprangeFF$ be a PIOP for the relation $\relgrange$ over the finite field $\FF=\ZZ_p$ that is perfectly complete and has RBR-knowledge error $\delta'_{range}$.
Then the $\pioprangeopt$ protocol of Figure \ref{fig:pioprangeopt} is a PIOP for $\relrange$ with perfect completeness and RBR-knowledge error
$\max\left\{\left(\frac{\ell+\nu}{\sigma}\right)^{\kappa}; \frac{\ell+\nu+\gamma -m}{p}; \delta'_{range}\right\}$.
\end{theorem}
When instantiated with the PIOP $\pioprangeFF$ from Section \ref{sec:PIOPrangeFF}, $p$ should additionally be $> 2^{\ell + \nu + \gamma}$.

Throughout the two proofs below we write $\vec r_{\eta} = (\vec r^{\leq m}_{\eta}, \vec r^{>m}_{\eta})$, where $\vec r^{\leq m}_{\eta} = (r_{\eta,1}, \ldots, r_{\eta,m})$ and $\vec r^{>m}_{\eta} = (r_{\eta,m+1}, \ldots, r_{\eta,\ell+\nu})$, so that $\vec r'_{\eta} = \prj_p(\vec r^{\leq m}_{\eta})$ and $\vec{\hat r}_{\eta} = \prj_q(\vec r_{\eta})$.

\headingb{Correctness of $\pioprangeopt$.}
Let $\tilde v$ be a polynomial such that $\forall \vec b \in \bin^{\ell}$ it holds $\tilde v(\vec b) \in T_{B}$.
By Lemma \ref{lemma:Rq2Zqrange}, the polynomial $\tilde \mu \in \ZZ^{(\leq 1)}[\vec X, \vec Z, \vec Z']$ computed by $\iopP$ in step 1 satisfies equation \eqref{eq:vdecomposition} and $\tilde\mu(\vec b) \in \vec t_{\beta}$ for all $\vec b \in \bin^{\ell+\nu+\gamma}$.
By construction, for $\tilde\mu' = \prj_p(\tilde\mu)$ it holds
$\tilde \mu'(\vec b) \in \vec t'_\beta$ $\forall \vec b \in \bin^{\ell+\nu+\gamma}$
since
\[\tilde \mu'(\vec b) = \prj_p(\tilde\mu)(\vec b) = \prj_p(\tilde\mu(\vec b)) \in \prj_p(\vec t_\beta) = \vec t'_\beta.
\]
Therefore, by its correctness, the protocol $\pioprangeFF$ accepts.

Next, we have to show that the verifier's checks of step 2 are satisfied.

It is straightforward to see that the checks $\tilde\psi'_{\eta}(\vec s')  = \tilde \mu'(\vec r'_{\eta}, \vec s')$ are satisfied by construction.

Let us now focus on the last check.
By construction of $\tilde\psi'_{\eta}$ we have $\tilde\psi'_{\eta}(\vec\delta, \vec k) = \tilde \mu'(\vec r'_{\eta}, \vec\delta, \vec k)$ and, by applying Proposition \ref{prop:lift_eval} to $\tilde\mu$ on the point $(\vec r^{\leq m}_{\eta}, \vec\delta, \vec k) \in \ZZ^{\ell+\nu+\gamma}$ along with the bound on the value of $\tilde\mu(\vec r^{\leq m}_{\eta}, \vec\delta, \vec k)$, which we claim later,
\[
\FFtoInt(\tilde\psi'_{\eta}(\vec\delta, \vec k)) = \FFtoInt\left(\prj_p(\tilde\mu)\left(\prj_p(\vec r^{\leq m}_{\eta}, \vec\delta, \vec k)\right)\right) = \tilde\mu(\vec r^{\leq m}_{\eta}, \vec\delta, \vec k).
\]
Therefore, by the multilinearity of $\tilde\mu$, for the right-hand side of the last check we obtain the following equality over the integers:
\begin{align*}
&\sum_{\substack{\vec\delta\in\bin^{\ell+\nu -m}\\ \vec k \in \bin^{\gamma}}} \!\!\!\!\!\! \FFtoInt(\tilde\psi'_{\eta}(\vec \delta, \vec k)) \cdot \eq(\vec\delta; \vec r^{>m}_{\eta}) \cdot \beta^{\toint(\vec k)} \\
& \quad =  \sum_{\vec k \in \bin^{\gamma}} \beta^{\toint(\vec k)} \!\!\!\!\!\! \sum_{\vec\delta\in\bin^{\ell+\nu-m}} \!\!\!\!\!\! \tilde\mu(\vec r^{\leq m}_{\eta}, \vec\delta, \vec k) \cdot \eq(\vec\delta; \vec r^{>m}_{\eta})
 = \sum_{\vec k \in \bin^{\gamma}} \tilde\mu(\vec r_{\eta}, \vec k) \cdot \beta^{\toint(\vec k)}.
\end{align*}
Finally, both sides of equation \eqref{eq:vdecomposition} are multilinear in $(\vec X, \vec Z)$ and agree on $\bin^{\ell+\nu}$, so that equation, once reduced modulo $q$, holds as an identity of polynomials, i.e.
$\prj_q\left(\sum_{\vec k \in \bin^{\gamma}} \tilde\mu(\vec X, \vec Z, \vec k) \cdot \beta^{\toint(\vec k)}\right) = \tilde u(\vec X, \vec Z)$.
Since $\prj_q$ is a ring homomorphism, evaluating on $\vec r_{\eta}$ gives
$\tilde u(\vec{\hat r}_{\eta}) = \prj_q\left(\sum_{\vec k \in \bin^{\gamma}} \tilde\mu(\vec r_{\eta}, \vec k) \cdot \beta^{\toint(\vec k)}\right)$,
and thus the last check is satisfied for all $\eta \in [\kappa]$ and $l \in [0,L]$.

\begin{claim}
If $\tilde\mu(\vec b) \in \vec t_\beta$ for all $\vec b \in \bin^{\ell+\nu+\gamma}$ then, for all $\eta \in [\kappa]$, $\vec\delta \in \bin^{\ell+\nu-m}$ and $\vec k \in \bin^{\gamma}$, $\abs{\tilde\mu(\vec r^{\leq m}_{\eta}, \vec\delta, \vec k)} < 2^{m} \beta \sigma^{m} < p/2$.
\end{claim}
\begin{proof}
By the multi-linearity of the polynomial we have that
\[
\tilde\mu(\vec r^{\leq m}_{\eta}, \vec\delta, \vec k) =  \sum_{\vec \phi \in \bin^{m}} \tilde\mu(\vec \phi, \vec\delta, \vec k) \, \tilde\rho_{\eta}(\vec \phi)
\]
where $\tilde\rho_{\eta}(\vec \phi) = \eq(\vec\phi; \vec r^{\leq m}_{\eta}) = \prod_{i=1}^{m} (\phi_{i} r_{\eta,i} + (1-\phi_{i}) (1- r_{\eta,i}))$.
Given that $0 \leq r_{\eta,i} < \sigma$, we have that $-\sigma^{m} < \tilde\rho_{\eta}(\vec \phi) < \sigma^{m}$; thus by combining it with the assumption that $\tilde\mu(\vec\phi,\vec\delta,\vec k) \in \vec t_{\beta}$, i.e. $0 \leq \tilde\mu(\vec\phi,\vec\delta,\vec k) \leq \beta -1$, we obtain
$\abs{\tilde\mu(\vec r^{\leq m}_{\eta}, \vec\delta, \vec k)} < 2^{m} \beta \sigma^{m}$, which is smaller than $p/2$ by equation \eqref{eq:params-bound-p}.
\qed
\end{proof}

This concludes the correctness proof.

\headingb{Security of $\pioprangeopt$.}
We prove the RBR-soundness, from which we get RBR-knowledge by then applying Lemma \ref{lemma:RBRsound2RBRksound}.

As in Section \ref{sec:piopGqrange-security}, let $\tilde\mu''$ be the multilinear polynomial in $\ZZ^{(\leq 1)}[\vec X, \vec Z, \vec Z']$ such that $\tilde\mu''(\vec b) = \FFtoInt(\tilde\mu'(\vec b))$ for all $\vec b \in \bin^{\ell+\nu+\gamma}$; note that $\tilde\mu'' = \tilde\mu$ when $\iopP$ is honest.
We define the $\state(T_{B}, \ora{\tilde v}, \tr_i)$ function on every (partial) transcript $\tr_{i}$, from round $i=0$ to $i = 2+\mathsf{r}_{range}$, where $\mathsf{r}_{range}$ denotes the number of rounds of the $\pioprangeFF$ protocol:

\begin{itemize}
\item For $\tr_0=\emptyset$, $\state(T_{B}, \ora{\tilde v}, \tr_0)=0$ iff $(T_{B}, \ora{\tilde v}) \notin \lang_{\relrange}$.

\item For $\tr_1 = \ora{\tilde \mu'} |  (\vec r_0, \ldots,\vec r_{\kappa-1})$, $\state(T_{B}, \ora{\tilde v}, \tr_1)=0$ iff
$(\vec t'_{\beta}, \ora{\tilde \mu'}) \notin \lang_{\relgrange}$ or there exists $\eta \in [\kappa]$ such that $F(\vec{\hat r}_{\eta}) \neq 0$, where $F(\vec X, \vec Z) \in \ZZ_q[\vec X, \vec Z]$ is the multilinear polynomial
\[
F(\vec X, \vec Z) = \tilde u(\vec X, \vec Z) - \prj_q\left(\sum_{\vec k \in \bin^{\gamma}} \tilde\mu''(\vec X, \vec Z, \vec k) \cdot \beta^{\toint(\vec k)}\right).
\]

\item For $\tr_2 = \tr_1 | (\tilde\psi'_0, \ldots, \tilde\psi'_{\kappa-1}) | \vec s'$: $\state(T_{B}, \ora{\tilde v}, \tr_2)=0$ iff
\begin{itemize}
\item $(\vec t'_{\beta}, \ora{\tilde \mu'}) \notin \lang_{\relgrange}$
\item or there exists $\eta \in [\kappa]$ such that $\neg \mathsf{V}_{1,\eta}$, or there exist $\eta \in [\kappa]$ and $l \in [0,L]$ such that $\neg \mathsf{V}_{2,\eta,l}$, where
\[
\mathsf{V}_{1,\eta} := \left(\tilde\psi'_{\eta}(\vec s')  = \tilde \mu'(\vec r'_{\eta}, \vec s')\right)
\]
\[
\mathsf{V}_{2,\eta,l} := \left(\tilde u(\vec{\hat r}_{\eta}) \bmod p_l = \!\!\!\!\!\! \sum_{\substack{\vec\delta\in\bin^{\ell+\nu -m}\\ \vec k \in \bin^{\gamma}}} \!\!\!\!\!\! \FFtoInt(\tilde\psi'_{\eta}(\vec \delta, \vec k)) \cdot \eq(\vec\delta; \vec r^{>m}_{\eta}) \cdot \beta^{\toint(\vec k)} \bmod p_l \right)
\]
are the two checks of step 2.
\end{itemize}

\item For $\tr_{2+k} = \tr_2 | \tr_k'$ where, for $1 \leq k \leq \mathsf{r}_{range}$, $\tr'_k$ is the partial transcript of $\pioprangeFF$ at the $k$-th round: $\state(T_{B}, \ora{\tilde v}, \tr_{2+k})=0$ iff
$\state'(\vec t'_{\beta}, \ora{\tilde \mu'}, \tr'_{k})=0$, or there exists $\eta \in [\kappa]$ such that $\neg \mathsf{V}_{1,\eta}$, or there exist $\eta \in [\kappa]$ and $l \in [0,L]$ such that $\neg \mathsf{V}_{2,\eta,l}$.
\end{itemize}

Let us now show that this $\state$ function satisfies the three properties of RBR-soundness.

First, it is easy to see that by construction, if $(T_{B}, \ora{\tilde v})\notin\lang_{\relrange}$ then $\state(T_{B}, \ora{\tilde v}, \tr_0)=0$.

Second, for any {\em complete} transcript $\tr = \tr_{2+\mathsf{r}_{range}} = \tr_2 | \tr'_{\mathsf{r}_{range}}$, by construction we have that if $\state(T_{B}, \ora{\tilde v}, \tr)=0$, then either $\state'(\vec t'_{\beta}, \ora{\tilde \mu'}, \tr'_{\mathsf{r}_{range}})=0$, which means that the verifier of $\pioprangeFF$ rejects by the second property of RBR-soundness (Definition \ref{def:rbr-soundness}) applied to $\pioprangeFF$, or at least one of the verifier's checks of step 2 fails. In any of these cases, $\pioprangeopt.\iopV$ rejects.

Third, we show that if $\state$ rejects on a partial transcript of round $i$, for any prover's message of the round $i+1$, we have a concrete (negligible) bound on the probability that $\state$ accepts on the next partial transcript.
We show this property round by round.

\begin{claim}
If $\state(T_{B}, \ora{\tilde v},\tr_0)=0$, then for any prover's message $\ora{\tilde \mu'}$ we have that
\[
\Pr_{\vec r_0, \ldots, \vec r_{\kappa-1} \sample \Sigma^{\ell+\nu} }[\state(T_{B}, \ora{\tilde v},\tr_1=(\ora{\tilde \mu'} | (\vec r_0, \ldots, \vec r_{\kappa-1})))=1 ] \leq \left(\frac{\ell+\nu}{\sigma} \right)^{\kappa}.
\]
\end{claim}
\begin{proof}
By definition, $\state(T_{B}, \ora{\tilde v},\tr_1)=1$ means that:\\
(i) $(\vec t'_{\beta}, \ora{\tilde \mu'}) \in \lang_{\relgrange}$, i.e., $\tilde\mu'(\vec b) \in \vec t'_\beta$ for all $\vec b \in \bin^{\ell+\nu+\gamma}$;\\
(ii) for all $\eta \in [\kappa]$: $F(\vec{\hat r}_{\eta}) = 0$.

From (i) and the definition of $\tilde\mu''$ we have $\tilde\mu''(\vec b) = \FFtoInt(\tilde\mu'(\vec b)) \in \FFtoInt(\vec t'_\beta)$; since $\vec t'_{\beta} = \prj_p(\vec t_\beta)$ and $\beta < p/2$ (from equation \eqref{eq:params-bound-p}), using $\FFtoInt$'s definition we have $\FFtoInt(\prj_p(\vec t_\beta)) = \vec t_\beta$, and thus $\tilde\mu''(\vec b) \in \vec t_\beta$ for all $\vec b \in \bin^{\ell+\nu+\gamma}$.
In particular, the multilinear polynomial $\prj_q(\tilde\mu'') \in \Zqml[\vec X, \vec Z, \vec Z']$ satisfies equation \eqref{eq:mugrange}.
On the other hand, $\state(T_{B}, \ora{\tilde v},\tr_0)=0$ means $(T_{B}, \ora{\tilde v}) \notin \lang_{\relrange}$, and therefore, by Lemma \ref{lemma:Rq2Zqrange}, $\prj_q(\tilde\mu'')$ must violate equation \eqref{eq:vdecomposition}, i.e., there exists $\vec i^{*} \in \bin^{\ell+\nu}$ such that
$\tilde u(\vec i^{*}) \neq \prj_q\left(\sum_{\vec k \in \bin^{\gamma}} \tilde\mu''(\vec i^{*}, \vec k)\cdot \beta^{\toint(\vec k)}\right)$, i.e. $F(\vec i^{*}) \neq 0$ and thus the polynomial $F(\vec X, \vec Z)$ is nonzero.

Notice that $F$ is determined by $\tilde v$ and by the oracle $\ora{\tilde \mu'}$, hence it is fixed before $\iopV$ samples $\vec r_0, \ldots, \vec r_{\kappa-1}$, and that these are sampled independently of each other. Therefore,
\begin{eqnarray*}
&&\Pr[\state(T_{B}, \ora{\tilde v},\tr_1)=1 \mid \state(T_{B}, \ora{\tilde v},\tr_0)=0] \\
&& \leq \Pr\left[\bigwedge_{\eta \in [\kappa]} F(\prj_q(\vec r_\eta)) = 0 \; \middle| \; F(\vec X, \vec Z) \neq 0 \right] \\
&& = \prod_{ \eta \in [\kappa]} \Pr_{\vec r_\eta \sample \Sigma^{\ell+\nu}}\left[ F(\prj_q(\vec r_\eta))=0 \mid F(\vec X, \vec Z) \neq 0\right].
\end{eqnarray*}
Since $F$ is multilinear in $\ell + \nu$ variables, hence of total degree at most $\ell+\nu$, and since $\prj_q(\Sigma)$ is a subset of the exceptional set $\Esetz$ of size $\sigma$, by generalized Schwartz-Zippel (Lemma \ref{lemma:generalizedSZ}) we get $\Pr_{\vec r_\eta \sample \Sigma^{\ell+\nu}}[F(\prj_q(\vec r_\eta)) = 0 \mid F(\vec X, \vec Z) \neq 0] \leq \frac{\ell+\nu}{\sigma}$, and the claimed bound follows by considering the probability that this occurs for $\kappa$ independent samples; by equation \eqref{eq:params-bound-soundness-opt} this bound is $\approx 2^{-\secpar}$.
\qed
\end{proof}

\begin{claim}
If $\state(T_{B}, \ora{\tilde v},\tr_1)=0$, then for any prover's message $(\tilde\psi'_0, \ldots, \tilde\psi'_{\kappa-1})$ we have that
\[
\Pr_{\vec s'}[\state(T_{B}, \ora{\tilde v},\tr_2=(\tr_1 | (\tilde\psi'_0, \ldots, \tilde\psi'_{\kappa-1}) | \vec s'))=1] \leq  \frac{\ell - m + \nu + \gamma}{p}.
\]
\end{claim}
\begin{proof}
Let $\E_2$ be the event ``$\bigwedge_{\eta \in [\kappa]} \left(\tilde\psi'_{\eta}(X_{m+1}, \ldots, X_{\ell}, \vec Z, \vec Z') = \tilde \mu'(\vec r'_{\eta}, X_{m+1}, \ldots, X_{\ell}, \vec Z, \vec Z')\right)$'', i.e. the event that the polynomials sent by $\iopP$ are the correct partial evaluations of $\tilde\mu'$.
Notice that $\E_2$ is determined by the prover's message, hence it is fixed and independent of $\vec s'$; the same holds for the event $(\vec t'_{\beta}, \ora{\tilde \mu'}) \in \lang_{\relgrange}$, which is determined by the oracle sent in step 1. We distinguish the two cases.

If $\E_2$ does not occur, let $\bar\eta$ be an index for which the two polynomials differ. Since $\state(T_{B}, \ora{\tilde v},\tr_2)=1$ requires in particular $\mathsf{V}_{1,\bar\eta}$, we have
\begin{eqnarray*}
&& \Pr[\state(T_{B}, \ora{\tilde v},\tr_2)=1  \mid  \bar \E_2, \state(T_{B}, \ora{\tilde v},\tr_1)=0] \\
&& \leq \Pr\left[\tilde\psi'_{\bar\eta}(\vec s')  = \tilde \mu'(\vec r'_{\bar\eta}, \vec s') \; \middle| \; \tilde\psi'_{\bar\eta} \neq \tilde \mu'(\vec r'_{\bar\eta}, \cdot) \right]
\;\leq\; \frac{\ell - m + \nu + \gamma}{p},
\end{eqnarray*}
where the last inequality follows by the Schwartz-Zippel lemma over $\ZZ_p$, applied to the non-zero multilinear polynomial $\tilde\psi'_{\bar\eta} - \tilde \mu'(\vec r'_{\bar\eta}, \cdot)$ in $\ell - m + \nu + \gamma$ variables, of total degree at most $\ell - m + \nu + \gamma$, with the random $\vec s' \sample \ZZ_p^{\ell-m+\nu+\gamma}$.

If $\E_2$ occurs, we argue that $\Pr[\state(T_{B}, \ora{\tilde v},\tr_2)=1  \mid \E_2, \state(T_{B}, \ora{\tilde v},\tr_1)=0] = 0$.
We may assume $(\vec t'_{\beta}, \ora{\tilde \mu'}) \in \lang_{\relgrange}$, as otherwise $\state(T_{B}, \ora{\tilde v},\tr_2)=0$ by definition.
As in the proof of the previous claim, this gives $\tilde\mu''(\vec b) \in \vec t_\beta$ for all $\vec b \in \bin^{\ell+\nu+\gamma}$, so that the claim of the correctness proof, applied to $\tilde\mu''$ in place of $\tilde\mu$, gives $\abs{\tilde\mu''(\vec r^{\leq m}_{\eta}, \vec\delta, \vec k)} < p/2$. Hence, by Proposition \ref{prop:lift_eval} and $\E_2$,
\[
\FFtoInt(\tilde\psi'_{\eta}(\vec \delta, \vec k)) = \FFtoInt\left(\tilde\mu'(\vec r'_{\eta}, \vec\delta, \vec k)\right) = \tilde\mu''(\vec r^{\leq m}_{\eta}, \vec\delta, \vec k)
\qquad \forall \eta \in [\kappa], \vec\delta \in \bin^{\ell+\nu-m}, \vec k \in \bin^{\gamma}.
\]
By the same computation as in the correctness proof --- which only uses the multilinearity of $\tilde\mu''$ --- the right-hand side of $\mathsf{V}_{2,\eta,l}$ equals $\prj_q\left(\sum_{\vec k \in \bin^{\gamma}} \tilde\mu''(\vec r_{\eta}, \vec k)\cdot\beta^{\toint(\vec k)}\right) \bmod p_l$, and therefore
\[
\bigwedge_{l \in [0,L]} \mathsf{V}_{2,\eta,l}
\iff
\tilde u(\vec{\hat r}_{\eta}) = \prj_q\left(\sum_{\vec k \in \bin^{\gamma}} \tilde\mu''(\vec r_{\eta}, \vec k)\cdot\beta^{\toint(\vec k)}\right)
\iff
F(\vec{\hat r}_{\eta}) = 0,
\]
where the first equivalence holds since $p_0 \cdots p_L = q$ and thus, by the CRT, two elements of $\ZZ_q$ are equal if and only if they are equal modulo $p_l$ for all $l \in [0,L]$.
Since $\state(T_{B}, \ora{\tilde v},\tr_1)=0$ and we assumed $(\vec t'_{\beta}, \ora{\tilde \mu'}) \in \lang_{\relgrange}$, there exists $\bar\eta \in [\kappa]$ with $F(\vec{\hat r}_{\bar\eta}) \neq 0$, hence $\neg\mathsf{V}_{2,\bar\eta,\bar l}$ for some $\bar l \in [0,L]$ and $\state(T_{B}, \ora{\tilde v},\tr_2)=0$.
\qed
\end{proof}

To prove the last case, recall that, once $\tr_2$ is fixed, so is the event
\[
E := \exists\, \eta \in [\kappa],\, l \in [0,L] : \neg \mathsf{V}_{1,\eta} \vee \neg \mathsf{V}_{2,\eta,l},
\]
that is, the event that at least one of the verifier's checks of step 2 fails.
We also let $\tr'_0 = \emptyset$.

\begin{claim}
For every $k = 0, \ldots, \mathsf{r}_{range}-1$, if
$\state(T_{B}, \ora{\tilde v},\tr_{2+k})=0$, then for any prover's message
$\pi$ of the $(k+1)$-th round of $\pioprangeFF$ we have that
\[
\Pr_{\rho}\left[\state(T_{B}, \ora{\tilde v},\tr_{2+k+1})=1\right] \leq \delta'_{range},
\quad \text{where } \tr_{2+k+1} = \tr_2 | \tr'_{k} | \pi | \rho,
\]
and where the probability is over the $(k+1)$-th verifier's challenge $\rho$ of $\pioprangeFF$.
\end{claim}
\begin{proof}
Since $k+1 \geq 1$, the definition of $\state$ on $\tr_{2+k+1}$ gives
\[
\state(T_{B}, \ora{\tilde v},\tr_{2+k+1})=1
\iff
\state'(\vec t'_{\beta}, \ora{\tilde \mu'}, \tr'_{k} | \pi | \rho)=1 \;\wedge\; \bar E.
\]
Moreover, $E$ is determined by $\tr_2$, hence it is fixed and independent of $\rho$.
We distinguish two cases.

If $E$ occurs, then $\state(T_{B}, \ora{\tilde v},\tr_{2+k+1})=1$ never occurs and the probability of the claim is $0$.

If $E$ does not occur, then the event $\state(T_{B}, \ora{\tilde v},\tr_{2+k})=0$ of the precondition implies $\state'(\vec t'_{\beta}, \ora{\tilde \mu'}, \tr'_{k})=0$: for $k \geq 1$ this is immediate from the definition of $\state$ on $\tr_{2+k}$, whereas for $k=0$ the definition of $\state$ on $\tr_2$ gives $(\vec t'_{\beta}, \ora{\tilde \mu'}) \notin \lang_{\relgrange}$, which in turn implies $\state'(\vec t'_{\beta}, \ora{\tilde \mu'}, \tr'_{0})=0$ by the first property of RBR-soundness (Definition \ref{def:rbr-soundness}) applied to $\pioprangeFF$. Therefore, for all $0 \leq k < \mathsf{r}_{range}$,
\begin{eqnarray*}
&& \Pr_{\rho}[\state(T_{B}, \ora{\tilde v},\tr_{2+k+1})=1] \\
&& = \Pr_{\rho}[\state'(\vec t'_{\beta}, \ora{\tilde \mu'}, \tr'_{k} | \pi | \rho)=1 \wedge \bar E] \\
&& \leq \Pr_{\rho}[\state'(\vec t'_{\beta}, \ora{\tilde \mu'}, \tr'_{k} | \pi | \rho)=1]
\;\leq\; \delta'_{range},
\end{eqnarray*}
where the last inequality follows by the third property of RBR-soundness applied to $\pioprangeFF$ and to the partial transcript $\tr'_{k}$, on which $\state'$ evaluates to $0$.
\qed
\end{proof}

The final RBR-soundness error is the max of the three per-round bounds, namely $\max\left\{\left(\frac{\ell+\nu}{\sigma}\right)^{\kappa}; \frac{\ell - m + \nu + \gamma}{p}; \delta'_{range}\right\}$, where the first term is $\approx 2^{-\secpar}$ by equation \eqref{eq:params-bound-soundness-opt}.

\subsubsection{Efficiency of $\pioprangeopt$}
\label{sec:efficiency-pioprangeopt}
In this section, we detail the efficiency of the PIOP $\pioprangeopt$ when instantiated with the PIOP $\pioprangeFF$ (instantiated with $\ell^{*} = \ell + \nu + \gamma$) described in the previous section.

The PIOP has the following efficiency profile:
\begin{itemize}
\item the number of rounds is $\mathsf{r} = \max(\ell^*, b/c) + 4$, as $\pioprangeFF$ involves two rounds plus the  parallel execution of two sum-checks of size $2^{\ell^*}$ and $2^{b/c}$, and then there are two rounds in $\pioprangeopt$ itself;
\item the prover sends $\mathsf{o} = 7$ oracles that are multilinear polynomials over $\ZZ_p$: $3$ of size $2^{b/c}$ and $4$ of size $2^{\ell^*}$; in addition, the prover sends the polynomials $(\tilde\psi'_0, \ldots, \tilde\psi'_{\kappa - 1})$ that are $\kappa \cdot 2^{\ell^* - m}$ elements of $\ZZ_p$;
\item the proving time is $\mathsf{T}_{\iopP} = O(\kappa \cdot 2^{\ell^*} + 2^{b/c})$ since this is the time needed to build the $7$ oracle polynomials and since the two sum-checks can be run in linear time as they involve a product of multilinear polynomials, plus considering the cost of computing the $\kappa$ partial evaluations of $\tilde \mu'$;
\item the verifier time is $\mathsf{T}_{\iopV} = O(\ell^* + b/c + \kappa \cdot 2^{\ell^* - m})$ since $O(\ell^* + b/c)$ is the  verification time of the two sum-check executions and since $O(\kappa \cdot 2^{\ell^* - m})$ is the time to perform the last verification checks in the round 2 of $\pioprangeopt$; we recall that this term is $\mathsf{poly}(\ell + \nu+\gamma)$ by equation \eqref{eq:params-bound-efficiency}.
Additionally, as we show below $\iopV$ has to compute the evaluation of $\tilde t'_{\beta}$ on $(\vec r^*_t, 0), (\vec r^*_t, 1)$, which can be done in time $O(b/c)$ since
$
\tilde t'_{\beta}(X_1, \ldots, X_{b/c}) =  \sum_{k=1}^{b/c} X_k \cdot 2^{k-1}
$.

\item the number of verifier's queries is $\mathsf{q} = 15 + 2\kappa$, as justified below.
\end{itemize}

Assume that in $\pioprangeFF$ the sum-check on $\mu^*$ (resp. $t^*$) ends with a query of $\iopV$ of the virtual oracle on $\vec r_\mu$ (resp. $\vec r_t$).
Below, for a vector $\vec x$ we denote $\vec x[1,\ldots,m]$ its prefix of length $m$, and for $\vec x = (x_1, \ldots, x_m)$, we denote by $\vec x^{*}$ the vector $(x_2, \ldots, x_m)$.

The protocol requires the following queries to the 7 oracles over $\ZZ_p$:
\begin{itemize}
\item $\ora{\tilde \mu'}$ on $(\vec r'_0, \vec s'), \ldots, (\vec r'_{\kappa -1}, \vec s')$, $(\vec r^*_\mu, 0), (\vec r^*_\mu, 1)$;
\item $\ora\readts$ on $(\vec r^*_\mu, 0), (\vec r^*_\mu, 1)$;
\item $\ora\finalts$ on  $(\vec r^*_t, 0), (\vec r^*_t, 1)$;
\item $\ora{f^{(1)}_{\WS_\mu}}$ on $\vec r_\mu, (\vec r^*_\mu, 0), (\vec r^*_\mu, 1), (\vec 1, 0)$
\item $\ora{f^{(1)}_{\RS}}$ on $\vec r_\mu, (\vec r^*_\mu, 0), (\vec r^*_\mu, 1), (\vec 1, 0)$
\item $\ora{f^{(1)}_{\WS_t}}$ on $\vec r_t, (\vec r^*_t, 0), (\vec r^*_t, 1), (\vec 1, 0)$
\item $\ora{f^{(1)}_{\sfS}}$ on $\vec r_t, (\vec r^*_t, 0), (\vec r^*_t, 1), (\vec 1, 0)$
\end{itemize}
In addition, in the protocol $\pioprangeopt$, the verifier queries $\ora{\unpack(\tilde v)}$ on $\vec{\hat r}_1, \ldots, \vec{\hat r}_{\kappa}$, which turn into querying the packed polynomial $\ora{\tilde v}$ on $\vec{\hat r}_1[1, \ldots, \ell], \ldots, \vec{\hat r}_\kappa[1, \ldots, \ell]$.

To minimize the number of evaluation proofs, it is possible to apply a simple batching optimization to all those evaluations of the form $(\vec r^{*}, 0), (\vec r^{*}, 1)$. In this case, the prover first sends the results $y$ and $y'$ of the evaluations in  $(\vec r^{*}, 0)$ and $ (\vec r^{*}, 1)$ respectively, the verifier chooses a random point $\theta$, and queries the polynomial on $(\vec r^{*}, \theta)$. Then, assuming the result of this query is $y''$, the verifier checks that $y'' = (1 - \theta) y + \theta y'$.

By using this optimization, we obtain in total $\mathsf{q}=15+2\kappa$ queries, as mentioned earlier.

\headingb{Instantiating the concrete parameters.}
By equations \eqref{eq:params-bound-p} and \eqref{eq:params-bound-soundness-opt}, the parameters must satisfy
\[
\log p > 1 + \log \beta + m \cdot (1 + \log \sigma)
\qquad \text{ and } \qquad
\kappa \geq \frac{\secpar}{\log \sigma - \log (\ell+\nu)},
\]
while the dominant term of the proof size is $\kappa \cdot 2^{\ell + \nu + \gamma - m}$ elements of $\ZZ_p$.
Differently from $\piopGqrange$, where the challenges $\vec r_{\eta}$ are used to fix $m \leq \ell^{*}$ variables of $\tilde\mu$, here they are sampled in $\Sigma^{\ell+\nu}$ and thus $m$ can be taken as large as $\ell + \nu$.
This is what makes the proof size independent of $N$: for $m \leq \ell$ we would have $2^{\ell+\nu+\gamma-m} \geq 2^{\nu + \gamma} = c \cdot N$, i.e., a proof size linear in $N$, whereas taking $m > \ell$ removes this factor.

Let us consider the parameters used in the range proofs for the largest circuit in our experiments, that is $\ell = 9$, $\beta = 2^{11}$, $\gamma = 2$ and $N = 2^{13}$ (hence $\nu = 13$ and $\ell + \nu = 22$), and let us target $\secpar = 128$ and a prime $p$ of $208$ bits.
Then we can take for example
$m = 15$,
$\sigma = 2^{11}$,
$\kappa = 20$,
which gives
\[
2 \beta \cdot (2\sigma)^{m} = 2^{192} < p,
\qquad
\left(\frac{\ell+\nu}{\sigma}\right)^{\kappa} = \left(\frac{22}{2^{11}}\right)^{20} \leq 2^{-130},
\]
and a proof size of $\kappa \cdot 2^{\ell+\nu+\gamma-m} = 20 \cdot 2^{9} = 10240$ elements of $\ZZ_p$, i.e., about $260$ KB.
This is the choice we use in our experiments (Table~\ref{tab:res_range_proof}).

Alternatively, by lowering $\sigma$ we can push $m$ up to its maximum $\ell + \nu$ at the price of more repetitions. Taking
$m = \ell + \nu = 22$,
$\sigma = 2^{7}$,
$\kappa = 51$,
we get
\[
2 \beta \cdot (2\sigma)^{m} = 2^{188} < p,
\qquad
\left(\frac{\ell+\nu}{\sigma}\right)^{\kappa} = \left(\frac{22}{2^{7}}\right)^{51} \leq 2^{-129},
\]
and a proof size of $51 \cdot 2^{2} = 204$ elements of $\ZZ_p$, i.e., about $5$ KB.
The price of this second choice is a larger $\kappa$, which affects the prover's running time, that grows as $O(\kappa \cdot 2^{\ell+\nu+\gamma})$, and the batch size of the two evaluation proofs.

%% file: app_GKR.tex
\section{Additional material for Section~\ref{sec:PIOP_AC}} \label{sec:App_PIOP_AC}

\subsection{The $\pcopycon$ wiring predicate in Equation (\ref{eq:GKR0++})}
Throughout this subsection $\vz$ denotes the $\#(copy)$-bit index of the copy gate, i.e.\ the block $\vZvar_{[\#(copy)]}$ that Eq.~(\ref{eq:GKR0++}) feeds to the predicate.
Let $s := \max(\mu, \#(in))$, so that $m' = s+1$ and $\vy = (\vy_{[s]}, \Yvar_{s})$, where the most significant variable $\Yvar_{s} = \Yvar_{m'-1}$ is the one that $g$ uses to choose between $\Vi[in]$ and $\Vi[3,quot]$ (see Section~\ref{sec:PIOP_AC_equations}).
$\pcopycon(\vt,\vz,\vx, \vy)$ always evaluates to zero, except in the two following cases when it evaluates to $1$:
\begin{equation} \label{eq:copycon}
	\begin{cases}
		\Big( (\Yvar_{m'-1} = 0) \land (\vx_{[d,m)} = \vec{1}) \land (\vy_{[\#(in),s)} = \vec{1}) \land (\vt = (0,1))    \\
		\land (\Vi[3,ptxt](\vx_{[d]})  \text{ multiplies } \Vi[in](\vy_{[\#(in)]}) 		\text{ in the } (\vz)\text{-th check}) \Big)\\
		\text{ or }  \\
		\Big( (\Yvar_{m'-1} = 1) \land (\vx_{[d,m)} = \vec{1}) \land (\vy_{[\mu,s)} = \vec{1}) \land (\vt = (0,1))\\
		\land (\Vi[3,ptxt](\vx_{[d]})  \text{ multiplies } \Vi[3, quot](\vy_{[\mu]}) \text{ in the } (\vz)\text{-th check}) \Big)
	\end{cases}
\end{equation}
$\tcopycon(\vt,\vz,\vx, \vy)$ can be described as:

$\tcopycon(\vTvar, \vZvar,\vXvar, \vYvar) = \teq(\vTvar,(0,1)) \cdot \tcopycon'(\vZvar,\vXvar,
\vYvar)$,

where $\tcopycon'(\vZvar,\vXvar, \vYvar)$ can be described exactly as in Eq.~(\ref{eq:copycon}) by
removing the condition $\vt = (0,1)$.
Whenever a party needs to evaluate $\tcopycon(\vTvar, \vZvar,\vXvar, \vYvar)$ at a specific point,
they can do it from the evaluation of $\tcopycon'(\vZvar,\vXvar, \vYvar)$.
In particular, the committed polynomial is $\tcopycon'$, which has $\#(copy) + m + m'$ variables.

\subsection{Wiring predicates from \cite{C:CCCFGS25}} \label{sec:App_PIOP_AC_predicates}
Let us provide the description of the wiring predicates introduced in \cite{C:CCCFGS25}.

\begin{align*}
	\prescon[1](\vz,\vx)&=
	\begin{cases}
		1, & \text{if wire}\ \vx  \text{ holds the } \vz \text{-th } c' \text{ value.}  \\
		0, & \text{otherwise.}\
	\end{cases}
	\\
	\prescon[3](\vz,\vx)&=
	\begin{cases}
		-1, & \text{if wire } \vx  \text{ holds the } \vz \text{-th remainder.}  \\
		-\pres_l, & \text{if } \vx  \text{ belongs to the } l \text{-th HE layer and holds the } \vz \text{-th quotient.}  \\
		0, & \text{otherwise.}
	\end{cases}
	\\
	\pbdcon[2](\vz,\vx)&=
	\begin{cases}
		1, & \text{if wire}\ \vx  \text{ holds the } \vz \text{-th } d_2 \text{ value.}  \\
		0, & \text{otherwise.}\
	\end{cases}
	\\
	\pbdcon[3](\vz,\vx)&=
	\begin{cases}
		-\mathsf{PW}_{\omega_l}(1)[i], & \text{if } \vx  \text{ belongs to the } l \text{-th HE layer and holds}\\
		& \text{the } i\text{-th decomposition of the } \vz \text{-th } d_2 \text{ value.}  \\
		0, & \text{otherwise.}\
	\end{cases}
\end{align*}

\begin{equation*}
	\pevk[](\vz,\vx)=
	\begin{cases}
		\mathsf{evk}_{l, b}[i],  & \text{if wire } \vx  \text{ belongs to the } l \text{-th HE layer and}\\
		& \text{holds the } i\text{-th decomposition of the }  d_2 \text{ value}  \\
		& \text{corresponding to the } \vz \text{-th } c'[b] \text{, where } b \in \bit.\\
		0, & \text{otherwise.}
	\end{cases}
\end{equation*}

The following predicate was used in \cite{C:CCCFGS25} but originates from \cite{CCS:LiuXieZha21}

\begin{equation} \label{eq:Xmult}
	\pXmult[](\vz,\vx,\vy)=
	\begin{cases}
		1, & \text{if }\ V(\vx) \cdot V(\vy)  \text{ is added to } V_2(\vz).  \\
		0, & \text{otherwise.}\
	\end{cases}
\end{equation}

In our work, we have introduced several changes so as to improve efficiency.
First of all, we have added the metadata variables $\vTvar$ that distinguish among the different kinds of gate.
Second, we have drastically reduced the amount of $\vXvar$ variables by converting them into $\vZvar$ variables. Notice how, in our case, $\prescon[1](\vz,\vx)$ and $\pbdcon[2](\vz,\vx)$ have simplified practically to the point of disappearance. We elaborate on this next, in Appendix~\ref{sec:App_PIOP_AC_toomany}.
This conversion is not applied everywhere. The predicates $\prescon[3], \pbdcon[3]$ and $\pevk[]$ above do not merely test $\vz$ against $\vx$: they read the HE layer $l$ of the wire $\vx$ in order to apply the level-dependent constants $\pres_l$, $\mathsf{PW}_{\omega_l}(1)[i]$ and $\mathsf{evk}_{l,b}[i]$. The $\delta$ variables encoding that layer must therefore stay in $\vXvar$. What our version changes is only that the layer is matched against $\vZvar_{[\delta]}$ and used to select the constant by a single fused factor $\teq(\tobin(l),\vZvar_{[\delta]}) \teq(\tobin(l),\vXvar_{[\delta]})$, keeping the MLE multilinear in both blocks.

The evolution of the predicates in \cite{C:CCCFGS25} to the ones we propose is implicit by the  definitions above, Appendix~\ref{sec:App_PIOP_AC_toomany} and the detailed description of our MLEs in Section~\ref{sec:PIOP_AC_equations}.

\subsection{Too many gates in Equation (\ref{eq:GKR0++})?} \label{sec:App_PIOP_AC_toomany}
Notice how $\trescon[1](\vTvar) = \teq(\vTvar, (0,0))$ does not include any $\vZvar$ variables and hence $\forall \vz \in \bit^{\max(\mu,\#(copy))}$, $\trescon[1](0,0) \cdot  \Vi[1](\vz_{[\mu]}) = \Vi[1](\vz_{[\mu]})$.
This would seem to have the effect of having ``too many $\trescon[]$ gates''. There are two different cases to consider here, on which we elaborate next:

\begin{itemize}
	\item 	First we have the excess gates that result from the difference between the actual number of $\trescon[]$ gates and the next power of two following that number, i.e. $2^{\mu}$ (all possible values for $\vz_{[\mu]}$).
	Let us label those exceeding gates as $\hat{E} \subseteq \bit^{\mu}$. Whereas an honest prover would set $\Vi[1](\hat{\vz}) = 0 \ \forall \hat{\vz} \in \hat{E}$, a malicious prover could choose to put some non-zero values in there. The same discussion applies verbatim to $\tcopycon[3,copy](\vTvar) \cdot \Vi[3,copy](\vZvar_{[\#(copy)]})$ and to the exceeding copy gates in $\bit^{\#(copy)}$, since $\tcopycon[3,copy]$ carries no $\vZvar$ either.
	Such malicious prover would then have two options: Either these values respect the constraints given in Equations~(\ref{eq:GKR0++}), (\ref{eq:GKR1++}) and their dependencies, or they do not.
	In the latter case, the soundness of sumcheck will detect this and abort. In the former case, this would ultimately do no harm, since the $\tXmult$ predicates from Equation~(\ref{eq:GKR2++}) are such that $\tXmult[](\hat{\vz},\vx,\vy) = 0 \ \forall (\hat{\vz},\vx,\vy)  \in (\hat{E} \times \bit^{s'} \times \bit^{s'})$. This is by the very definition of $\hat{E}$.

	\item 	Second, there are the potentially exceeding $\max(\mu,\#(copy)) - \mu$ variables in $\vZvar$. These extra values mean repeating the same check that many times, which is harmless both in terms of soundness and efficiency.
	More concretely, $\forall \vz'_{extra} \in \bit^{\max(\mu,\#(copy)) - \mu}$ the ``$\trescon$ part'' (i.e. $\vTvar = (0,0)$) of Equations~(\ref{eq:GKR0++}) simplifies to
	\begin{align*}
		\Vi[0](0,0, \vz_{[\mu]}, \vz'_{extra}) &= \Vi[1](\vz_{[\mu]}) \\
		& + \sum_{\vx \in \bit^m} \trescon[3](0,0,\vz_{[\delta]},\vx) \cdot  \Vi[3,rc_3](\vx_{[\delta]}, \vz_{[\delta,\mu)},x_\delta) \\
		&= \Vi[1](\vz_{[\mu]}) + \sum_{b=0}^{1} \trescon[3](0,0,\vz_{[\delta]},\vx^{(b)}) \cdot  \Vi[3,rc_3](\vz_{[\mu]}, b),
	\end{align*}
	where $\vx^{(b)} := (\vz_{[\delta]}, b, \vec{1})$. The second equality uses that, for boolean $\vz$, the fused factor $\teq(\tobin(l),\vz_{[\delta]}) \teq(\tobin(l),\vx_{[\delta]})$ inside $\trescon[3]$ leaves a single non-zero value of $\vx_{[\delta]}$, namely $\vx_{[\delta]} = \vz_{[\delta]}$.
	Notice how the right hand side is independent of $\vz'_{extra} \in \bit^{\max(\mu,\#(copy)) - \mu}$.
\end{itemize}

\ifFullVersion
\subsection{Additional material for Section \ref{sec:PIOP_AC_progress}} \label{sec:App_PIOP_AC_progress}

\subsubsection{\textit{Recombining into Equation~(\ref{eq:GKR2++})}}
Here we continue with the explanation of how to regroup values coming from Equation~(\ref{eq:GKR0++}) and Equation~(\ref{eq:GKR1++}) so as to execute sumcheck on Equation~(\ref{eq:GKR2++}). This essentially follows \cite{C:CCCFGS25}.

For a set $S$ among $\{d_{01}, d_2\}$, define predicates $\pcomb[2,S](\vYvar,\vZvar)$ as follows:
\begin{equation*}
	\pcomb[2,S](\vy,\vz)=
	\begin{cases}
		1, & \text{if wire}\ \vy  \text{ in } V_{2,S} \text{ is the } \vz \text{-th wire in } V_2.  \\
		0, & \text{otherwise.}\
	\end{cases}
\end{equation*}
$\iopV$ samples $s_{01}, s_{2} \sample \Eset$ so as to randomly combine $\iopP$'s claims.
Now, by running the sum-check protocol on Equation~(\ref{eq:V2comb++}), $\iopP$ ends by claiming an evaluation $\Vi[2](\vr_2)$ (for a random sum-check challenge $\vr_2 \sample \Eset^{\mu + 1}$, since $\#(V_2) =\mu + 1$) which can be substituted into Equation~(\ref{eq:GKR2++}) so as to advance towards the input layer.
\begin{align} \label{eq:V2comb++}
	&s_{01} \Vi[2,d_{01}](\vr_{d_{01}}) + s_2 \Vi[2, d_2](\vr_2)   = \nonumber \\
	&	s_{01} \big( \sum_{\vz \in \bit^{\mu + 1}} \tcomb[2,d_{01}](\vr_{d_{01}},\vz) \Vi[2](\vz)  \big)  + 	s_{2} \big( \sum_{\vz \in \bit^{\mu + 1}} \tcomb[2,d_{2}](\vr_{2},\vz) \Vi[2](\vz) \big)    \nonumber \\
	& = \sum_{\vz \in \bit^{\mu + 1}} \Vi[2](\vz) \Big( 	s_{01} \tcomb[2,d_{01}](\vr_{d_{01}},\vz)  + 	s_{2} \tcomb[2,d_{2}](\vr_{2},\vz) \Big).
\end{align}
For simplicity, the most significant variable of $\Vi[2]$ chooses between $\Vi[2,d_{01}]$ and $\Vi[2,d_{2}]$, i.e.\footnote{It is arbitrary whether to have $\Vi[2](\vz_{[\mu - 1]},1,1) = \Vi[2,d_{2}](\vz_{[\mu - 1]})$ and $\Vi[2](\vz_{[\mu - 1]},0,1) = 0$ or viceversa.} $\Vi[2](\vz_{[\mu]}, 0) = \Vi[2,d_{01}](\vz_{[\mu]})$ and $\Vi[2](\vz_{[\mu - 1]}, 1, 1) = \Vi[2,d_{2}](\vz_{[\mu - 1]})$.
Let us define the corresponding predicates. Notice that $\vXvar$ is of length $\mu$ and $\vYvar$ is of length $\mu - 1$.

\begin{align*}
	\tcomb[2,d_{01}](\vXvar,\vZvar)& = (1 - \vZvar_{\mu})	\cdot \teq(\vXvar, \vZvar_{[0]})\\
	\tcomb[2,d_2](\vYvar,\vZvar)& =	 \vZvar_{\mu}	\cdot \vZvar_{\mu - 1} \cdot \teq(\vYvar, \vZvar_{[1]})
\end{align*}

While we could have defined $\Vi[2]$ directly as we just have described and use it to replace calls to  $\Vi[2,d_{01}]$ and $\Vi[2,d_{2}]$ in Equations~(\ref{eq:GKR0++}) and (\ref{eq:GKR1++}), that would have made our protocol more expensive.
Had we done so, we would have required a standard two-to-one reduction in Equation~(\ref{eq:GKR2++}), which requires twice the amount of queries to the $\tXmult$ predicates in there. Since those encode the ``actual HE circuit'' and we implement them using polynomial commitments, we would rather do the simpler reassignment from Equation~(\ref{eq:V2comb++}).
\fi

\input{algorithms}

\input{global-piop}\label{sec:App_PIOP_AC_orac_comb}

\subsection{Rearranging into the final oracles and final queries}
After executing the sumcheck protocol on Equation~(\ref{eq:GKR2++}), we still need to match the syntax of $\piopac$ given in $\piopckks$.
In \cite{C:CCCFGS25}, claimed values
were reduced to claims about $D+1$ oracles:
$\ora{\Vi[out]}$ for the claimed ciphertext outputs,
$\ora{\Vi[aux,2^u]}$ for all non-deterministic inputs from the prover that need to be proven smaller than $2^u$
and finally, for $j = 1, \ldots, D-1$, $\ora{\Vi[aux,j]}$ for non-deterministic inputs from the prover that need to be proven smaller than $q/q_{L-j-1}$ %

Rather than following the strategy from \cite{C:CCCFGS25}, in our work we exploit existing relations between the oracles that were already employed during $\piopac$ to produce the ones we need.
This avoids the cost of one sumcheck to do recombinations and reassignments \cite[Eq (23)]{C:CCCFGS25}, together with the computation of the relevant reassigning predicates by both the prover and the verifier. Moreover, it also requires less queries. %

\begin{enumerate}
	\item Given the oracle $\ora{\Vi[3,rc_3]}$, we have implicit access to $\ora{\Vi[3, quot]}$,  so we don't need to split it. Given a query $\Vi[3, quot](\vx_{[quot]})$, translate it to $\Vi[3, rc_3](\vx_{[quot]},1)$.
	Following the same idea, we can also produce oracles $\ora{\Vi[aux, j]}_{j=1}^{D-1}$: Given a query $\Vi[aux, j](\vx_{[\#(aux, j)]})$, it can be answered with $\Vi[3, rc_3](\tobin(j),\vx_{[\#(aux, j)]}, 1)$.

	\item Going in the opposite direction, we can also avoid using sumcheck to recombine $\ora{\Vi[3,BD]}, \ora{\Vi[3,rc_3]}$ into a single oracle $\ora{\Vi[aux, 2^u]}$.
	Instead, we create a virtual oracle that includes the two smaller ones.  All values in $\ora{\Vi[3,BD]}$ would belong to $\ora{\Vi[aux, 2^u]}$, whereas for $\ora{\Vi[3,rc_3]}$ the legal queries are those of the form $\Vi[3, rc_3](0, \vx_{[\mu]})$.
	Let $\vZvar = (\Zvar_0, \ldots, \Zvar_{\mu + \delta - 1})$. The following MLE shows us how to query the virtual oracle (we are assuming $\delta \geq 1$):
	\[\Vi[aux, 2^u](\vZvar) = (1 - \vZvar_{\mu + \delta - 1}) \cdot \Vi[3,BD](\vZvar_{[\mu + \delta - 1]})  + \vZvar_{\mu + \delta - 1} \cdot \Vi[3, rc_3](\vZvar_{[\mu]},0)  \]

	\item Finally, the $\Vi[out]$ oracle can also be easily implemented by querying directly $\ora{\Vi[3,rc_3]}$ or $\ora{\Vi[3, copy]}$ at the relevant wire that has to be revealed.
	\item Hence, merely by implementing  $\ora{\Vi[3, copy]}, \ora{\Vi[3,rc_3]}$ and $\ora{\Vi[3,BD]}$ through a polynomial commitment, we can verify all claimed values  $\Vi[3, quot](\vr_{quot}), \allowbreak \Vi[3, quot](\vr'_{quot}), \allowbreak \Vi[3,BD](\vr_{BD})$, $\Vi[3,rc_3](\vr_{rc_3})$, $\allowbreak \Vi[3, copy](\vr_{copy})$ and $\Vi[3, copy](\vr'_{copy})$ as well as we can perform the necessary range checks.
	Moreover, we apply a final optimization: we implement the three oracles as a \emph{single} PC with $2 + \max(\#(copy), \mu +1, \mu + \delta - 1)$ variables.
	This can be done using the same idea as when we defined $\ora{\Vi[aux, 2^u](\vZvar)}$
\end{enumerate}

%% file: algorithms.tex
\subsection{Prover algorithms} \label{sec:GKR_prover}
Here we list the necessary algorithms for a linear time prover when executing sumcheck on algorithms, adapted from Libra \cite{C:XZZPS19}.
We take verbatim from there the following algorithms and their dependencies: \Sumcheck~(Algorithm 2), \SumcheckProduct~(Algorithm 3) and  \Precomp~(given as a procedure within Algorithm 4).
Notice that these algorithms represent polynomials as their evaluation at the set $\{0,1,2\}$, which has to be an exceptional set in $\Ring_q$ in order to establish an isomorphism with their coefficient form.
This is always the case for our choice of parameters, but otherwise $\{0,1,2\}$ can be replaced with any other exceptional set of the same size.

In all our algorithms, we start challenging from the least to the most significant variables. This has positive efficiency impacts due to the sparsity provided by prefixes in the MLEs of the wiring predicates (which are placed at their most significant position). Some of our algorithms exploit that kind of sparsity, as well as the possibility of multiplying vectors of different length. %

Algorithm~\ref{Alg:Prover_V0} is one of the central pieces of our efficient proofs for $\relac$.
It requires several subroutines, given in Algorithms \ref{Alg:InitCopyY} to \ref{Alg:SumcheckX}.
Throughout, we abbreviate the following partially evaluated factors:
\begin{align}
	\widehat{V}_{pt}(\vx) &:= \Vi[3,ptxt](\vx_{[d]}) \cdot g(\chall{Y}), \nonumber \\
	\widehat{V}_{BD}(\vx) &:= \Vi[3,BD](\vx_{[\delta]},\chall{Z_{[\delta,\mu-1)}},\vx_{[\delta,2\delta)}), \nonumber \\
	\widehat{V}_{rc}(\vx) &:= \Vi[3,rc_3](\vx_{[\delta]},\chall{Z_{[\delta,\mu)}},x_\delta), \label{eq:ProvAbbv}
\end{align}
whose look-up tables we write $\booktable[\vXvar]{\widehat{V}_{pt}} := g(\chall{Y}) \cdot \booktable[\vXvar]{V_{3,ptxt}}$, $\booktable[\vXvar]{\widehat{V}_{BD}} := \booktable[\vXvar]{\Vi[3,BD]}$ and $\booktable[\vXvar]{\widehat{V}_{rc}} := \booktable[\vXvar]{\Vi[3,rc_3]}$.
Moreover, inside Algorithms~\ref{Alg:SumcheckX} and \ref{Alg:Prover_V0}, once $\vTvar, \vZvar$ and $\vYvar$ are all fixed to their challenges, we write $\widehat{\pcopycon}(\vx), \widehat{\pbdcon[3]}(\vx)$ and $\widehat{\prescon[3]}(\vx)$ for the resulting functions of $\vx$ alone (i.e. $\widehat{\pbdcon[3]}(\vx):= \tbdcon[3](\chall{T},\chall{Z_{[\delta]}},\vx)$,  $\widehat{\prescon[3]}(\vx):= \trescon[3](\chall{T},\chall{Z_{[\delta]}},\vx)$, $\widehat{\pcopycon}(\vx) := \tcopycon(\chall{T},\chall{Z_{[\#(copy)]}},\vx,\chall{Y})$).
The rounds that challenge the $\vXvar$ variables are collected in Algorithm~\ref{Alg:SumcheckX}, which Algorithm~\ref{Alg:Prover_V0} calls once its look-up tables are in place.
To simplify the writing, Algorithm~\ref{Alg:Prover_V0}'s description we use the $\SumcheckProduct$ algorithm in lines \ref{line:bdcon3} and \ref{line:rescon3}. In practice and in our implementation, those are replaced with a more efficient $\SparseSumcheckProduct$ version which appears in Algorithm~\ref{Alg:SumcheckProdSparse}. %

\begin{algorithm}
	\caption{Algorithm $\InitCopyY(\tcopycon,\booktable[\vXvar]{V_{3,ptxt}},\chall{T},\chall{Z})$}
	\label{Alg:InitCopyY}
	\SetAlgoLined
	\SetKwInOut{Input}{Input}
	\SetKwInOut{Output}{Output}
	\LinesNumbered
	\Input{ Sparse multi-linear $\tcopycon(\vTvar,\vZvar,\vXvar,\vYvar)$. Multi-linear $V_{3,ptxt}(\vXvar_{[d]})$ and its initial look-up table.
		Random challenges $\chall{T} \in \Eset^2, \chall{Z} \in \Eset^{\#(copy)}$. }
	\Output{Look-up table $\booktable[\vYvar]{f_1}$.}
	$\booktable{\chall{T}} \leftarrow \Precomp(\chall{T})$.	\\
	$\booktable{\chall{Z}} \leftarrow \Precomp(\chall{Z})$.\\
	$\forall \vy \in \bit^{m'}$, set $\booktable[\vYvar]{f_1}[\vy] = 0$.\\
	For every $(\vt, \vz,\vx,\vy) \in \bit^{2+\#(copy)+m+m'}$ such that $\tcopycon(\vt,\vz,\vx,\vy) \neq 0$, do:\\
	$\booktable[\vYvar]{f_1}[\vy] \leftarrow \booktable[\vYvar]{f_1}[\vy] + \booktable{\chall{T}}[\vt] \cdot \booktable{\chall{Z}}[\vz]\cdot \tcopycon(\vt,\vz,\vx,\vy) \cdot \booktable[\vXvar]{V_{3,ptxt}}[\vx_{[d]}].$\\
	\Return{$\booktable[\vYvar]{f_1}$}
\end{algorithm}

\begin{algorithm}
	\caption{Algorithm $\InitCopyX(\tcopycon,\chall{T},\chall{Z},\chall{Y})$}
	\label{Alg:InitCopyX}
	\SetAlgoLined
	\SetKwInOut{Input}{Input}
	\SetKwInOut{Output}{Output}
	\LinesNumbered
	\Input{ Sparse multi-linear $\tcopycon(\vTvar,\vZvar,\vXvar,\vYvar)$.
		Random challenges $\chall{T} \in \Eset^2, \chall{Z} \in \Eset^{\#(copy)}, \chall{Y} \in \Eset^{m'}$. }
	\Output{Look-up table $\booktable[\vXvar]{f_3}$.}
	$\booktable{\chall{T}}[\vt] \leftarrow \Precomp(\chall{T})$.	\\
	$\booktable{\chall{Z}}[\vz] \leftarrow \Precomp(\chall{Z})$.\\
	$\booktable{\chall{Y}}[\vy] \leftarrow \Precomp(\chall{Y})$.	\\
	$\forall \vx \in \bit^{m}$, set $\booktable[\vXvar]{f_3}[\vx] = 0$.\\
	For every $(\vt, \vz,\vx,\vy) \in \bit^{2+\#(copy)+m+m'}$ such that $\tcopycon(\vt,\vz,\vx,\vy) \neq 0$, do:\\
	$\booktable[\vXvar]{f_3}[\vx] \leftarrow \booktable[\vXvar]{f_3}[\vx] + \booktable{\chall{T}}[\vt] \cdot \booktable{\chall{Z}}[\vz]\cdot \booktable{\chall{Y}}[\vy] \cdot \tcopycon(\vt,\vz,\vx,\vy).$\\
	\Return{$\booktable[\vXvar]{f_3}$}
\end{algorithm}

\begin{algorithm}
	\caption{Algorithm $\InitX(f,\chall{T},\chall{Z})$}
	\label{Alg:InitX}
	\SetAlgoLined
	\SetKwInOut{Input}{Input}
	\SetKwInOut{Output}{Output}
	\LinesNumbered
	\Input{ Multi-linear $f(\vTvar,\vZvar,\vXvar)$.
		Random challenges $\chall{T} \in \Eset^2$ and $\chall{Z}$, the block of $\vZvar$-challenges that $f$ depends on. }
	\Output{Look-up table $\booktable[\vXvar]{f}$.}
	$\booktable{\chall{T}}[\vt] \leftarrow \Precomp(\chall{T})$.	\\
	$\booktable{\chall{Z}}[\vz] \leftarrow \Precomp(\chall{Z})$.	\\
	$\forall \vx \in \bit^{m}$, set $\booktable[\vXvar]{f}[\vx]= 0$.\\
	For every $(\vt, \vz,\vx)$ in the domain of $f$ such that $f(\vt,\vz,\vx) \neq 0$, do:\\
	$\booktable[\vXvar]{f}[\vx] \leftarrow \booktable[\vXvar]{f}[\vx] + \booktable{\chall{T}}[\vt] \cdot \booktable{\chall{Z}}[\vz]\cdot f(\vt,\vz,\vx).$\\
	\Return{$\booktable[\vXvar]{f}$}
\end{algorithm}

We invoke $\InitX$ also on the intermediate-value MLEs $\Vi[3,BD]$ and $\Vi[3,rc_3]$, which carry no $\vTvar$ variables and for which only \emph{part} of their arguments comes from $\vZvar$, their HE-level block being read from $\vXvar_{[\delta]}$. In every call we therefore pass exactly the block of challenges that the argument depends on: $\InitX(\Vi[3,BD], \chall{Z_{[\delta,\mu-1)}})$ returns a table indexed by the $2\delta$ variables $\vXvar_{[2\delta]}$ (level and $p_i$), and $\InitX(\Vi[3,rc_3], \chall{Z_{[\delta,\mu)}})$ one indexed by the $\delta+1$ variables $\vXvar_{[\delta+1]}$ (level and the rmd-vs-quot bit). Likewise $\tbdcon[3], \trescon[3]$ only depend on $\vZvar_{[\delta]}$ and $\tevk$ on $\vZvar_{[\delta+1]}$.

The tables $\booktable[\vXvar]{V_{3,ptxt}}$, $\booktable[\vYvar]{g}$ and $\booktable[\vZvar]{\Vi[2]}$ are not produced by $\InitX$: the polynomials they tabulate depend neither on $\vTvar$ nor on any block of $\vZvar$ that is fixed to a challenge, so they are simply the initial bookkeeping tables of $\Vi[3,ptxt]$, $g$ and $\Vi[2]$, which the prover assembles from values it already holds.

\begin{algorithm}
	\caption{Algorithm $\SumcheckX(\widehat{\pcopycon},\booktable[\vXvar]{f_3}, \widehat{V}_{pt},\booktable[\vXvar]{\Vi[3,ptxt]} \cdot g(\chall{Y}),
		\widehat{\pbdcon[3]},\booktable[\vXvar]{\tbdcon[3]}, \widehat{V}_{BD},\booktable[\vXvar]{\Vi[3,BD]},
		\widehat{\prescon[3]},\booktable[\vXvar]{\trescon[3]}, \widehat{V}_{rc},\booktable[\vXvar]{\Vi[3,rc_3]})$}
	\label{Alg:SumcheckX}
	\SetAlgoLined
	\SetKwInOut{Input}{Input}
	\SetKwInOut{Output}{Output}
	\LinesNumbered
	\Input{ The look-up tables $\booktable[\vXvar]{f_3}, \booktable[\vXvar]{\tbdcon[3]}, \booktable[\vXvar]{\trescon[3]}$ of the three wiring predicates and $\booktable[\vXvar]{\Vi[3,ptxt]} \cdot g(\chall{Y}), \booktable[\vXvar]{\Vi[3,BD]}, \booktable[\vXvar]{\Vi[3,rc_3]}$ of the three factors they are paired with, as built in Algorithm~\ref{Alg:Prover_V0}. }
	\Output{$\va_{[m',m+ m']}$, which are the prover sumcheck messages for the $\vXvar$ variables in Equation~(\ref{eq:GKR0++})}
	\textit{Comment: The three products are added up inside a \emph{single} sumcheck, so their rounds must be interleaved rather than run one after the other: the message of round $i$ is the sum of the three round-$i$ contributions, and only once it has been sent is $r_{\Xvar_i}$ available and the six look-up tables advanced by one variable. We write $\SumcheckProduct^{(i)}$ for the $i$-th round of $\SumcheckProduct$, that is, for the three elements $(a_{i,0},a_{i,1},a_{i,2})$ it outputs in that round. Interleaving at the level of rounds is also what lets each product keep its own number of variables for the second term, namely $d$ for $\Vi[3,ptxt]$, $2\delta$ for $\Vi[3,BD]$ and $\delta+1$ for $\Vi[3,rc_3]$.}\\
		\For{$i \gets 0$ \KwTo $m-1$}{
			The next three steps are executed in parallel:\\
			$\va'_{m'+i} \leftarrow \SumcheckProduct^{(i)}(\widehat{\pcopycon}(\vx),\booktable[\vXvar]{f_3}, \widehat{V}_{pt}(\vx),\booktable[\vXvar]{\Vi[3,ptxt]} \cdot g(\chall{Y}))$\\
			$\va''_{m'+i} \leftarrow \SumcheckProduct^{(i)}(\widehat{\pbdcon[3]}(\vx),\booktable[\vXvar]{\tbdcon[3]}, \widehat{V}_{BD}(\vx),\booktable[\vXvar]{\Vi[3,BD]})$ \label{line:bdcon3} \\
			$\va'''_{m'+i} \leftarrow \SumcheckProduct^{(i)}(\widehat{\prescon[3]}(\vx),\booktable[\vXvar]{\trescon[3]}, \widehat{V}_{rc}(\vx),\booktable[\vXvar]{\Vi[3,rc_3]})$ \label{line:rescon3} \\
			$\va_{m'+i}  \leftarrow \va'_{m'+i} + \va''_{m'+i} + \va'''_{m'+i}$ \label{line:interleave} \\
			Send $\va_{m'+i}$, receive $r_{\Xvar_i} \in \Eset$ and advance the six look-up tables with it, as $\FuncEval$ and $\SparseFuncEval$ do.\\
		}
		\Return{$\va_{[m',m+ m']}$}
\end{algorithm}

\begin{algorithm}
	\caption{$\Sumcheck$ Equation~(\ref{eq:GKR0++})}
	\label{Alg:Prover_V0}
	\SetAlgoLined
	\SetKwInOut{Input}{Input}
	\SetKwInOut{Output}{Output}
	\LinesNumbered
		\Output{$\va_{[m+ m']}$, which are the prover sumcheck messages for Equation~(\ref{eq:GKR0++})}
		\textit{Comment: We challenge first on the $\vYvar$ variables, then, in Algorithm~\ref{Alg:SumcheckX}, on the $\vXvar$ ones.
			Notice the abuse of notation in lines \ref{line:BDtable} and \ref{line:rc3table}, since $\Vi[3,BD]$ and $\Vi[3,rc_3]$ have no $\vTvar$ variables and only their wire-indexing block comes from $\vZvar$.
			The abbreviations $\widehat{V}_{BD}(\vx)$ and $\widehat{V}_{rc}(\vx)$ are the ones introduced in Equation~(\ref{eq:ProvAbbv}), and we drop the fixed arguments $\chall{T},\chall{Z_{[\delta]}}$ from $\tbdcon[3]$ and $\trescon[3]$.
		 }\\
		$\booktable[\vXvar]{\tbdcon[3]} \leftarrow \InitX(\tbdcon[3],\chall{T},\chall{Z_{[\delta]}})$\\
		$\booktable[\vXvar]{\trescon[3]} \leftarrow \InitX(\trescon[3],\chall{T},\chall{Z_{[\delta]}})$\\
		$\booktable[\vXvar]{\Vi[3,BD]} \leftarrow \InitX(\Vi[3,BD],\chall{Z_{[\delta,\mu-1)}})$ \label{line:BDtable} \\
		$\booktable[\vXvar]{\Vi[3,rc_3]} \leftarrow \InitX(\Vi[3,rc_3],\chall{Z_{[\delta,\mu)}})$ \label{line:rc3table} \\
		{}
		$\booktable[\vYvar]{f_1} \leftarrow \InitCopyY(\tcopycon,\booktable[\vXvar]{V_{3,ptxt}},\chall{T},\chall{Z_{[\#(copy)]}})$\\
		The next two steps are executed in parallel:\\
		$\va'_{[m']} \leftarrow \SumcheckProduct(\sum_{\vx \in \bit^m}\tcopycon(\chall{T},\chall{Z_{[\#(copy)]}},\vx,\vy) \cdot \Vi[3,ptxt](\vx_{[d]}),\booktable[\vYvar]{f_1},g(\vy),\booktable[\vYvar]{g},\chall{Y})$\\
		$\va_{[m']} \leftarrow \va'_{[m']} +
		 \vec{1} \cdot \Big(\sum_{\vx \in \bit^m} \big(\widehat{\pbdcon[3]}(\vx) \cdot  \widehat{V}_{BD}(\vx) 	+ \widehat{\prescon[3]}(\vx) \cdot  \widehat{V}_{rc}(\vx)\big)\Big) $  \label{line:x-terms}	\\
		$\booktable[\vXvar]{f_3} \leftarrow \InitCopyX(\tcopycon,\chall{T},\chall{Z_{[\#(copy)]}},\chall{Y})$\\
		$\va_{[m',m+ m']} \leftarrow \SumcheckX(
		\widehat{\pcopycon},\booktable[\vXvar]{f_3}, \widehat{V}_{pt},\booktable[\vXvar]{\Vi[3,ptxt]} \cdot g(\chall{Y}), \allowbreak
		\widehat{\pbdcon[3]},\booktable[\vXvar]{\tbdcon[3]}, \widehat{V}_{BD},\booktable[\vXvar]{\Vi[3,BD]},
		\widehat{\prescon[3]},\booktable[\vXvar]{\trescon[3]}, \widehat{V}_{rc},\booktable[\vXvar]{\Vi[3,rc_3]}
		)   $\\
		$\va_{[m+ m']} \leftarrow \va_{[m+ m']} + \tbdcon[2](\chall{T}) \cdot  \Vi[2, d_2](\chall{Z_{[\mu - 1]}}) + \trescon[1](\chall{T}) \cdot \Vi[1](\chall{Z_{[\mu]}}) - \tcopycon[3,copy](\chall{T}) \cdot \Vi[3,copy](\chall{Z_{[\#(copy)]}})$\\
		\Return{$\va_{[m+ m']}$}
\end{algorithm}

\begin{algorithm}
	\caption{$\Sumcheck$ Equation~(\ref{eq:GKR1++})}
	\label{Alg:Prover_V1}
	\SetAlgoLined
	\SetKwInOut{Input}{Input}
	\SetKwInOut{Output}{Output}
	\LinesNumbered
	\Output{$\va_{[2\delta]}$, which are the prover sumcheck messages for Equation~(\ref{eq:GKR1++}).}
	\textit{Comment: The challenges $\chall{Z}$ of this equation are those output by the sumcheck of Eq.~(\ref{eq:GKR0++}) for $\Vi[1]$. Still, the table of $\Vi[3,BD]$ must be rebuilt from scratch, since the one from line \ref{line:BDtable} of Algorithm~\ref{Alg:Prover_V0} is conditioned on a different block of $\vZvar$.}\\
	$\booktable[\vXvar]{\tevk} \leftarrow \InitX(\tevk,\chall{Z_{[\delta+1]}})$\\
	$\booktable[\vXvar]{\Vi[3,BD]} \leftarrow \InitX(\Vi[3,BD],\chall{Z_{(\delta,\mu)}})$\\
	$\va'_{[2\delta]} \leftarrow \SumcheckProduct(\tevk(\chall{Z_{[\delta+1]}},\vx),\booktable[\vXvar]{\tevk}, \widehat{V}_{BD}(\vx),\booktable[\vXvar]{\Vi[3,BD]},\chall{X})$\\
	$\va_{[2\delta]}  \leftarrow \va'_{[2\delta]} + \Vi[2,d_{01}](\chall{Z})  $\\
	\Return{$\va_{[2\delta]}$}
\end{algorithm}

\begin{algorithm}
	\caption{$\Sumcheck$ Equation~(\ref{eq:V2comb++})}
	\label{Alg:Prover_V2comb}
	\SetAlgoLined
	\SetKwInOut{Input}{Input}
	\SetKwInOut{Output}{Output}
	\LinesNumbered
	\Output{$\va_{[\mu+1]}$, which are the prover sumcheck messages for Equation~(\ref{eq:V2comb++}).}
	$\booktable[\vZvar]{\tcomb[2,d_{01}]} \leftarrow \InitX(\tcomb[2,d_{01}],\vr_{d_{01}})$\\
	$\booktable[\vZvar]{\tcomb[2,d_{2}]} \leftarrow \InitX(\tcomb[2,d_{2}],\vr_{d_2})$\\
	$\va_{[\mu+1]} \leftarrow \SumcheckProduct(s_{01} \tcomb[2,d_{01}](\vr_{d_{01}},\vz)  + 	s_{2} \tcomb[2,d_{2}](\vr_{d_2},\vz), s_{01} \booktable[\vZvar]{\tcomb[2,d_{01}]} + s_{2} \booktable[\vZvar]{\tcomb[2,d_{2}]}, \Vi[2](\vz),\booktable[\vZvar]{\Vi[2]},\chall{Z})$\\
	\Return{$\va_{[\mu+1]}$}
\end{algorithm}

For Equation~(\ref{eq:GKR2++}), follow essentially \texttt{SumCheckGKR} from Libra \cite{C:XZZPS19}.

\subsubsection{Exploiting sparsity}
\paragraph{Function Evaluations for sparse polynomials}
There are different MLEs that we use in the algorithms above which are very sparse.
In particular, this is the case for $\tbdcon[3]$ and $\trescon[3]$. The former only has $D(D+1)/2$ non-zero evaluations in $\vXvar$, despite that vector being much longer and the latter, only $2D$.
Moreover, these predicates are sparse in a very structured way, where all non-zero entries share the all-one vector as a common prefix: the padding factors $\teq(\vXvar_{[2\delta,m)},\vec{1})$ and $\teq(\vXvar_{[\delta+1,m)},\vec{1})$ pin the $L = m - 2\delta$, resp.\ $L = m-\delta-1$, most significant positions.
Algorithm~\ref{Alg:SparseFuncEval} shows how to compute the sumcheck messages for this functions more efficiently. Whenever $\FuncEval$ would be called upon these predicates, replace it with $\SparseFuncEval$.
Notice that whenever Algorithm~\ref{Alg:SparseFuncEval} is called upon $f \in \{\tbdcon[3], \trescon[3]\}$, they have already become a function only in $\vXvar$ variables, due to the previous execution of $\InitX(f,\chall{T},\chall{Z_{[\delta]}})$.

\begin{algorithm}
	\caption{Algorithm $\SparseFuncEval(f(\Xvar_0, \ldots, \Xvar_{m-1}),\booktable{f},L,\chall{\vXvar})$}
	\label{Alg:SparseFuncEval}
	\SetAlgoLined
	\SetKwInOut{Input}{Input}
	\SetKwInOut{Output}{Output}
	\LinesNumbered
	\Input{ Multi-linear $f$ and its initial bookkeeping table $\booktable{f}$.
		Prefix length $L$, i.e. length of the all-one vector $\vec{1}$.
		Random challenges $\vr \in \Eset^?$. }
	\Output{All function evaluations for sumcheck.
		}
	\For{$i \gets 0$ \KwTo $m-1$}{
		\If{$i < m - L$}{
			\ForEach{$\vec{b} \in \{0,1\}^{m - i - 1}$}{
				\For{$t \gets 0$ \KwTo $2$}{
					$f(r_0, \dots, r_{i-1}, t, \vb_{[m-i-1]}) \gets \booktable{f}[\vec{0}_{[i]}, 0, \vb_{[m-i-1]}] \cdot (1 - t) + \booktable{f}[\vec{0}_{[i]}, 1, \vb_{[m-i-1]}] \cdot t$\;
				}
				$\booktable{f}[\vec{0}_{[i+1]},\vb_{[m-i-1]}] \gets \booktable{f}[\vec{0}_{[i]}, 0, \vb_{[m-i-1]}] \cdot (1 - r_{i}) + \booktable{f}[\vec{0}_{[i]}, 1, \vb_{[m-i-1]}] \cdot r_{i}$\;
		}}
		\ElseIf{$i \geq m-L$}{
	\For{$t \gets 0$ \KwTo $2$}{
		$f(\vec{r}_{[i]}, t, \vec{1}_{[m-i-1]}) \gets \booktable{f}[\vec{0}_{[i]}, \vec{1}_{[m-i]}] \cdot t$\  %
	}
	$\booktable{f}[\vec{0}_{[i+1]}, \vec{1}_{[m-i-1]}] \gets \booktable{f}[\vec{0}_{[i]}, \vec{1}_{[m-i]}] \cdot r_{i}$\;
	}
	}
	Let $\mathcal{F}$ contain all function evaluations $f(\cdot)$ computed above.
	\Return $\mathcal{F}$
\end{algorithm}

\paragraph{SumcheckProduct with different length and sparsity}
Using Algorithm~\ref{Alg:SparseFuncEval}, we construct a more efficient, sparse version of $\SumcheckProduct$ which we denote $\SparseSumcheckProduct$ and provide in Algorithm~\ref{Alg:SumcheckProdSparse}.
This is a replacement for lines \ref{line:bdcon3} and \ref{line:rescon3} in Algorithm~\ref{Alg:SumcheckX}.

\begin{algorithm}
	\caption{Algorithm $\SparseSumcheckProduct(f,\booktable[\vXvar]{f}, g,\booktable[\vXvar]{g}, \vr)$}
	\label{Alg:SumcheckProdSparse}
	\SetAlgoLined
	\SetKwInOut{Input}{Input}
	\SetKwInOut{Output}{Output}
	\LinesNumbered
	\Input{ Multi-linear $f(\Xvar_0, \ldots, \Xvar_{L-1})$, $g(\vXvar_{[L']})$ where $L' < L$.
		$f$ is a sparse wiring predicate where non-zero values share the prefix in $\vXvar$.
		Random challenges $\vr \in \Eset^L$. }
	\Output{sumcheck messages for $\sum_\vx f(\vx) \cdot g(\vx_{[L']})$ Each message $a_{i,t}$ consists of 3 elements.
	}
	$\mathcal{F} \leftarrow \SparseFuncEval(f, \booktable[\vXvar]{f}, \chall{X})$.	\\
	$\mathcal{G} \leftarrow \FuncEval(g, \booktable[\vXvar]{g}, \chall{X_{[L']}})$.	\\
	\For{$i \gets 0$ \KwTo $L-1$}{
		\For{$t = 0, 1, 2$}{
			\If{$i < L'$}{$a_{i,t} = \sum_{\vb_{[i+1,L)} \in \bit^{L-i-1}} f(r_0, \ldots, r_{i-1}, t, \vb) \cdot g(r_0, \ldots, r_{i-1}, t, \vb_{[i+1,L')})$}
			\ElseIf{$L' \leq i < L$}{$a_{i,t} = \sum_{\vb_{[i+1,L)} \in \bit^{L-i-1}} f(r_0, \ldots, r_{i-1}, t, \vb) \cdot g(r_0, \ldots, r_{L'-1})$}
			}
		 }
	\Return{$\va = (a_0, \ldots, a_{L-1})$}
\end{algorithm}

%% file: global-piop.tex
\section{Putting everything together: The $\piopckks$ PIOP} \label{sec:together}
We are finally in a position to define our global PIOP $\piopckks$.
In \cite{C:CCCFGS25}, the final claimed values of $\Pi_{AC}$ were reduced to claims about $D+1$ oracles:
$\ora{\Vi[out]}$ for the claimed ciphertext outputs,
$\ora{\Vi[aux,2^u]}$ for all non-deterministic inputs from the prover that need to be proven smaller than $2^u$
and finally, for $j = 1, \ldots, D-1$, $\ora{\Vi[aux,j]}$ for non-deterministic inputs from the prover that need to be proven smaller than $q/q_{L-j-1}$. To those oracles, our richer arithmetization has added $\ora{\Vi[3,ptxt]}$ and $\ora{\Vi[3,copy]}$.

In the next section, we explain how the different oracles from our $\Pi_{AC}$ in Section~\ref{sec:PIOP_AC_progress} can be recombined so that we optimize our use of range proofs from Section~\ref{sec:range} and polynomial commitments.
In summary, $\ora{\Vi[aux,2^u]}$ can be implemented from $\ora{\Vi[3,rc_3]}$, $\ora{\Vi[aux,j]}$ from $\ora{\Vi[3,BD]}, \ora{\Vi[3,rc_3]}$ and $\ora{\Vi[out]}$ from $\ora{\Vi[3,rc_3]}, \ora{\Vi[3, copy]}$.

%% file: apx-sparsepack.tex
\section{Auxiliary Material on \spack}\label{sec:spack}

\subsection{Our \spack PIOP for Evaluating Sparse Polynomials}
In this section, we provide a full description of \spack. Precisely, we denote by \spack-$\piopiptree$ its version that internally relies on a PIOP $\piopiptree$.

\headingb{\spack Indexer.} On input the relation's index $({p_0}, \ldots, {p_{\np-1}})$, define the polynomial $\pol \in \Zqml[X_1, \ldots, X_{\ell}, T_1, T_2, Z_1, \ldots, Z_{\lnp + \mu}]$ such that, for all $\vec b \in \bin^{\ell}$, $\vec i \in \bin^{\lnp}$, $\vec j \in \bin^{\mu}$ and $\vec j' \in \bin^{\mu-1}$:
\begin{eqnarray*}
\pol(\vec b, 0, 0, \vec i, \vec 0 ) &=& \val(\vec b, \vec i) \\
\pol(\vec b, 1, 0, \vec i, \vec j ) &=& \spidx(\vec b, \vec i, \vec j) \\
\pol(\vec b, 1, 1, \vec i, \vec j',0) &=& \spidx(\vec b, \vec i, \vec j',0) \cdot \spidx(\vec b, \vec i, \vec j',1), \\
\pol(\vec b, 1, 1, \vec i, \vec j',1)  &=& 0
\end{eqnarray*}
and $\pol$ is $0$ on all the remaining points of $\bin^{\ell+2+\lnp+\mu}$.
Note that, by multilinearity, the following identities hold as polynomials, where $\spidx^{*}$ is defined as in Section~\ref{sec:sparse_PC}:
\[
\pol(\vec X, 1, 0, \vec Z) = \spidx(\vec X, \vec Z), \quad \pol(\vec X, 1, 1, \vec Z_{[1,\lnp+\mu-1]}, 0) = \spidx^{*}(\vec X, \vec Z_{[1,\lnp+\mu-1]}).
\]
The indexer returns $\ind=\ora{\packpol}$, where $\packpol = \pack(\pol) \in \Rqml[X_1, \ldots, X_{n}]$ and $n=\ell + 2+ \lnp + \mu - \nu$.

\headingb{\spack-$\piopiptree$ online phase.}
\begin{enumerate}
\item $\iopV$ sends $\vec t \sample \Eset^{\lnp}$.
\item  $\iopP$ and $\iopV$ run $\piopiptree$ for $((\tilde y(\vec t), \ora{\val}, \ora{f_2}, \ora{E^*_{\mu-1}}); \val, f_2, E^*_{\mu-1} ) \in \reliptree$, where $\pol = \unpack(\packpol)$, $\val=\pol(\vec X, 0,0,\vec Z_{[1,\lnp]}, \vec 0)$, $f_2=\eq(\vec X, \vec t)$, and $E^*_{\mu-1}$ is as in equation~\eqref{eq:Epoly}.
\item Let $\piopiptree.\iopV$'s queries be:
$(\vec r_1, \vec r_1')$ to $\ora{\val}$,
$\vec r_{2}$ to $\ora{f_2}$,
and $(\vec r, \vec s)$ to $\ora{E^{*}_{\mu-1}}$.

$\iopP$ sends $\psi_{\val} = \val(\vec r_1, \vec r_1')$, $\psi_0 = \spidx(\vec r, \vec s,0)$, $\psi_1=\spidx(\vec r, \vec s,1)$, and $\psi^{*}=\spidx^{*}(\vec r, \vec s)$.

$\iopV$ answers the query to $\ora{f_2}$ by computing $\eq(\vec r_2, \vec t)$ on its own, and the query to $\ora{E^{*}_{\mu-1}}$ by computing, according to equation~\eqref{eq:Epoly},
\[
\psi = \psi_0 \cdot \tilde x^{*}_0(\vec s) + \psi_1 \cdot \tilde x^{*}_1(\vec s) + \psi^{*} \cdot \tilde x^{*}_2(\vec s) + \tilde x^{*}_3(\vec s).
\]

\item $\iopV$ sends random $\tau, \zeta \sample \Eset$.

\item $\iopV$ computes
	$\psi_{\pol} = (\psi_0 \cdot (1-\zeta) + \psi_1 \cdot \zeta) \cdot (1-\tau) + \psi^* \cdot (1-\zeta) \cdot \tau$.
 Note that if $\psi_0, \psi_1, \psi^{*}$ are correct then $\psi_{\pol} = \pol(\vec r, 1, \tau, \vec s, \zeta)$, by the layout of $\pol$ fixed by the indexer and the multilinearity of $\pol$ in $T_2$ and $Z_{\lnp+\mu}$.

 $\iopP$ and $\iopV$ run the PIOP $\piopueval^{(\kappa)}$ on $((\psi_{\val},\vec x_{\val}, \psi_{\pol}, \vec x_{\pol},\ora{\packpol});\packpol) \in \relueval$ where $\vec x_{\val} = (\vec r_{1}, 0, 0, \vec r_{1}', \vec 0 )$ and $\vec x_{\pol} = (\vec r, 1, \tau, \vec s, \zeta)$.

 $\piopueval^{(\kappa)}$  ends with the verifier querying $\ora{\packpol}$ on $\vec\rho_1, \ldots, \vec\rho_\kappa \in \ZZ_q^{n}$.
\end{enumerate}

We refer to Section \ref{sec:sparse_PC} for an intuitive explanation of the protocol.
Here we provide an additional intuition on the last part of the protocol that deals with evaluating the unpacked polynomial $\pol$.
Our solution requires querying the unpacked $\ora{\pol}$ on points in $\Rq$, meaning that we cannot use the result of Lemma \ref{lemma:unpackedeval} to compute them from queries to $\ora{\packpol}$.
The last piece of the protocol, which solves this issue, is the use of our PIOP $\piopueval$ of Section \ref{sec:piopueval} for proving evaluations of unpacked polynomials on $\Rq$ points.

An alternative idea to avoid the use of $\piopueval$ and instead use Lemma \ref{lemma:unpackedeval} might have been to instantiate $\piopiptree$ so as to use random challenges in the exceptional set $\Esetz \subset \ZZ_q$. This change would lead to querying $\pol$ on points of $\Zq$. However, it would yield no gain in communication since the sumcheck of equation \eqref{eq:iptree} involves $\tilde x$, which encodes the evaluation inputs that are in $\Rq$. Thus using challenges in $\Esetz$ would still require the prover to send full $\Rq$ elements in every round of the sumcheck, in addition to the fact that such a sumcheck, with $\Rq$-dependent communication size, would need to be repeated several times to achieve a desired level of RBR-soundness.

Next, we state and prove the security of \spack.

\begin{theorem}\label{thm:spack}
Let $\piopueval$ be the PIOP for $\relueval$ of Section \ref{sec:piopueval}, and
let $\piopiptree$ be a PIOP for $\reliptree$ (with $\beta = \mu-1$) with $\delta$-knowledge-soundness. %
Then $\spack$-$\piopiptree$ is a PIOP for $\relspeval$ with $\delta_{speval}$-knowledge soundness where
$\delta_{speval} = \delta + (\lnp + 2)/|\Eset| + ((2n+5\nu+1) / |\Esetz|)^{\kappa}$, where $n = \ell+\lnp+\mu+2-\nu$ is the number of variables of $\packpol$.
\end{theorem}
\begin{proof}
Assume we start with a false statement $(\ind,\stm) \notin \lang(\relspeval)$.
Then the polynomial identity of equation \eqref{eq:speval-polystatement} does not hold and, since both its sides are multilinear in $\vec Z$, the probability over the choice of $\vec t$ that the equality of equation \eqref{eq:iptree} holds is $\leq \lnp/|\Eset|$.

Assume that equation \eqref{eq:iptree} does not hold, let $\psi_{\val}, \psi_0, \psi_1, \psi^{*}$ be the values sent by the prover in step 3, and let $\psi$ be the value computed by the verifier from them.
If $\psi_0 = \spidx(\vec r, \vec s,0)$, $\psi_1=\spidx(\vec r, \vec s,1)$, and $\psi^{*}=\spidx^{*}(\vec r, \vec s)$, then $\psi = E^{*}_{\mu-1}(\vec r, \vec s)$ by equation \eqref{eq:Epoly}.
Hence the probability that the verifier accepts in $\piopiptree$, conditioned on $\psi_0, \psi_1, \psi^{*}$ being correct and $\psi_{\val} = \val(\vec r_1, \vec r_1')$, is $\leq \delta$ by the soundness of $\piopiptree$.

If any of $\psi_0, \psi_1,\psi^{*}$ is wrong, then $\psi_{\pol} - \pol(\vec r, 1, \tau, \vec s, \zeta)$ is a nonzero polynomial of total degree $2$ in $(\tau, \zeta)$, hence $\psi_{\pol} = \pol(\vec r, 1, \tau, \vec s, \zeta)$ holds with probability at most $2/|\Eset|$ over the choice of $\tau, \zeta$.

Finally, if $\psi_{\pol} \neq \pol(\vec r, 1, \tau, \vec s, \zeta)$ or $\psi_{\val} \neq \pol(\vec r_{1}, 0, 0, \vec r_{1}', \vec 0 ) = \val(\vec r_1, \vec r_1')$ then the verifier accepts in $\piopueval^{(\kappa)}$ with probability at most $((2(\ell+\lnp+\mu+2)+3\nu+1) / |\Esetz|)^{\kappa} = ((2n+5\nu+1) / |\Esetz|)^{\kappa}$, by Theorem \ref{thm:piopuevalpacking} (applied to $\np=2$ evaluation points of a packed polynomial with $n$ variables) and Theorem \ref{thm:parallelrep}.

Furthermore, using a similar proof, if $\piopiptree$ has RBR-soundness $\delta'$, then we can argue the RBR-soundness of $\spack$-$\piopiptree$ to be the maximum of
\[
\delta'_1 = \frac{\lnp}{|\Eset|}, \; \delta', \; \delta'_{4}=\frac{2}{|\Eset|}, \; \left(\frac{2n+5\nu+1}{|\Esetz|}\right)^{\frac{\kappa}{n+\nu}}.
\]
\end{proof}

\subsection{Instantiations of \spack and Efficiency}\label{app:sparse_perf}
To obtain \spack we instantiate $\piopiptree$ with a PIOP that uses Thaler's approach for proving a tree of multiplications \cite{C:Thaler13}, adapted to work for $2^{\ell+\lnp}$ trees of multiplications of depth $\mu-1$, done in parallel, followed by an inner product. We refer to the next section for more details on this protocol. Here we recall that it can be proven by performing $\mu-1$ sequential sumchecks of increasing width $2^{\ell + \lnp+\eta}$, for $\eta=0$ to $\mu-2$, which, as required by $\piopiptree$, end up with the verifier querying the polynomials
$\val(\vec r_1, \vec r_1')$, $E^*_{\mu-1}(\vec r, \vec s)$, and computing $\eq(\vec r_1',\vec t)$.
The query to $E^*_{\mu-1}(\vec r, \vec s)$ is answered by the verifier itself from the values $\psi_0 = \spidx(\vec r, \vec s,0)$, $\psi_1=\spidx(\vec r, \vec s,1)$, and $\psi^{*}=\spidx^{*}(\vec r, \vec s)$ sent by the prover, so no additional sumcheck is needed.
Finally, all these queries are proven with an execution of $\piopueval^{(\kappa)}$.

In total, \spack's communication consists of
$4\ell(\mu -1) + (4\mu-1)\lnp +2\mu^2 -2\mu +2$ elements of $\Hq$ (see the paragraph on diagonal challenges below),
$\kappa(4(\ell + \lnp+\mu)+1)$ elements of $\ZZ_q$,
and one PC evaluation proof for a polynomial of $\ell+\lnp+\mu+2-\nu$ variables.

We compare \spack with Spark, as specified in Spartan~\cite{C:Setty20}, with the counter polynomials produced and committed in preprocessing, and with $c$ dimensions as in~\cite{EC:SetThaWah24}. We compile it with the PC transform for $\Rq$ of~\cite{C:CCCFGS25} over Basefold, i.e., one BaseFold PC on each CRT component of $\Rq$, without batching across them. As in Spartan, the grand products of all the multiset checks are proven as a single data-parallel layered circuit~\cite{C:Thaler13}, with one sumcheck per layer.
For $c = \lceil m / \ell\rceil$, this solution would require:
$3\cdot \np \cdot c$ committed polynomials of size $2^{\ell}$ in the preprocessing,
$\np \cdot c$ committed polynomials of size $2^{\ell}$ generated at proving time,
an overall communication of
$(c+2)\ell + 2(\lceil \log c \rceil + \lnp + 1)\ell(\ell-1) + 2\ell$ elements of $\Rq$,
and the PC evaluation proofs for $4\np c$ multilinear polynomials of $\ell$ variables, queried at two points, which we batch into two Basefold evaluation proofs~\cite{C:ZeiCheFis24}.

To obtain a concrete comparison, we consider the following parameters,
$N=2^{13}, \log q \approx135, \ell=15, m=34, \np=2$,
which correspond to the largest circuit of one of our neural networks applications (see Section \ref{sec:implementation}).

Even ignoring the size of PC evaluation proofs, \spack's proof is 25.4 KiB, against 235.3 MiB for Spark.
More interestingly, \spack avoids altogether running the PC on large polynomials: it does not generate any commitment during the proof, and it computes a {\em single} evaluation proof for the preprocessed (packed) polynomial $\packpol$ that, thanks to packing, has only $11$ variables; using our new BatchFold scheme this proof is $163$ MiB and, together with $\piopueval$, takes about two minutes to compute. In contrast, Spark prover has to commit to $6$ polynomials of size $2^{15}$ (which takes about 10 minutes) and prove evaluations of $4 \np c=24$ polynomials of size $2^{15}$. Using two batched Basefold proofs over $\Rq$ as in \cite{C:CCCFGS25}, these evaluations would take a total of about $7$ minutes, yielding a proof of $6.2$ GiB. %

\headingb{Diagonal challenges.}
The figures above and Table~\ref{tab:res_all} take the verifier's challenges in $\EsetN$ from its diagonal copy of $\Hq$, i.e., one element of $\FF_{p_i^d}$ repeated in the $N/d$ slots of each $p_i$. This was already the case in~\cite{AC:CCCFGS26}, but the protocol there did not exploit it. Since all the points at which the sparse polynomials are evaluated are then diagonal, every table of $\piopiptree$ lies in the diagonal subring of $\Rq$, which is isomorphic to $\Hq$: its sumchecks run over $\Hq$ and send elements of $\Hq$, $N/d$ times smaller than those of $\Rq$, and the evaluation proof of $\packpol$ needs no common-point sumcheck.
Without this, the sumchecks of \spack would take 28m13s, 3m28s and 1m44s for our three circuits (instead of about one second) and send 50.8, 35.5 and 33.4~MiB; \spack would then be $0.7\times$, $1.4\times$ and $1.3\times$ as fast as Spark (instead of $10.6\times$, $23.5\times$ and $11.2\times$), and still $28$--$31\times$ shorter.
The same choice gives the prover of $\piopueval$ the time $\BigO(\np 2^{\ell+\nu})$ of Theorem~\ref{thm:piopuevalpacking}: at points that are not diagonal, building its table $E$ directly could take $N$ times more operations in $\ZZ_q$.

\subsection{Details on our instantiation of $\piopiptree$}
In this section we give more details on how to instantiate our application of the $\piopiptree$ PIOP using Thaler's approach for trees of multiplications.

Consider the polynomial $E^*_{\mu-1}(\vec X, \vec Z, \vec Z')$ that is of degree 1 in $\vec X$ and degree 2 in $(\vec Z, \vec Z')$.
Let $\tilde E_{\mu-2}$ be the multilinear polynomial such that
\[
\forall (\vec b, \vec i, \vec j) \in \bin^{\ell+\lnp+\mu-2}: \tilde E_{\mu-2}(\vec b, \vec i, \vec j) = E^{*}_{\mu-1}(\vec b, \vec i, \vec j, 0) \cdot E^{*}_{\mu-1}(\vec b, \vec i, \vec j, 1)
\]
Next, for $\eta = \mu-3$ down to $1$, let $\tilde E_{\eta}$ be the multilinear polynomial such that
\[
\forall (\vec b, \vec i, \vec j) \in \bin^{\ell+\lnp+\eta}: \tilde E_{\eta}(\vec b, \vec i, \vec j) = \tilde E_{\eta+1}(\vec b, \vec i, \vec j, 0) \cdot \tilde E_{\eta+1}(\vec b, \vec i, \vec j, 1)
\]

Then $\piopiptree$ on $((\tilde y(\vec t), \ora{\val}, \ora{f_2}, \ora{E^*_{\mu-1}}); \val, f_2, E^*_{\mu-1} ) \in \reliptree$ runs $\mu-1$ sumchecks, sequentially, of increasing width $2^{\ell + \lnp+\eta}$, for $\eta=0$ to $\mu-2$. This proceeds as follows.

For $\eta = 0$, it starts with a sumcheck of width $2^{\ell+\lnp}$ and degree 3 in $\ell$ variables and $4$ in $\lnp$ variables, for the claim
\[
\tilde y(\vec t) =  \sum_{(\vec b, \vec i) \in \bin^{\ell + \lnp}} \val(\vec b, \vec i) f_2(\vec i) \cdot \tilde E_1(\vec b, \vec i, 0) \cdot \tilde E_1(\vec b, \vec i, 1)
\]
Its communication is $4\ell + 5\lnp + 2$ elements of $\Rq$, and it ends with a ``promise'' claim $v_1 = \tilde E_1(\vec r_1)$.

For $\eta=1$ to $\mu-3$, it runs a sumcheck of width $2^{\ell+\lnp+\eta}$ and degree 3 in all $\ell+\lnp+\eta$ variables, for the claim
\[
v_{\eta} =  \sum_{(\vec b, \vec i, \vec j) \in \bin^{\ell + \lnp + \eta}} \tilde E_{\eta+1}(\vec b, \vec i, \vec j, 0) \cdot \tilde E_{\eta+1}(\vec b, \vec i, \vec j, 1) \cdot \eq((\vec b, \vec i, \vec j), \vec r_{\eta})
\]
Its communication is $4(\ell + \lnp+\eta) + 2$ elements of $\Rq$, and it ends with a ``promise'' claim $v_{\eta+1} = \tilde E_{\eta+1}(\vec r_{\eta+1})$.

For $\eta=\mu-2$, it runs a sumcheck of width $2^{\ell+\lnp+\mu-2}$ of degree 3 in $\ell$ variables and 5 in $\lnp+\eta$ variables, for the claim
\[
v_{\mu-2} =  \sum_{(\vec b, \vec i, \vec j) \in \bin^{\ell + \lnp + \mu-2}}  E^*_{\mu-1}(\vec b, \vec i, \vec j, 0) \cdot E^*_{\mu-1}(\vec b, \vec i, \vec j, 1) \cdot \eq((\vec b, \vec i, \vec j), \vec r_{\eta})
\]
Its communication is $4\ell + 6(\lnp + \mu - 2) + 2$ elements of $\Rq$, and it ends with the claim $\psi = E^*_{\mu-1}(\vec r, \vec s)$.

Therefore the total communication of this protocol is $4\ell(\mu-1) + (4\mu-1)\lnp + 2\mu^2 - 2\mu - 2$ elements of $\Rq$.

%% file: appendix-basefold.tex
\section{Details about BaseFold}\label{app:basefold}

In this section we describe the BaseFold polynomial commitment, and our protocol $\Evalcommon_M$ for batched openings at a common point built on top on it, in detail.
Moreover, we show a slight improvement on the lower bounds for the minimum distance of the codes involved, that translates into a better soundness error.

\subsection{The BaseFold Error Correcting Codes}
The central component of the BaseFold polynomial commitment is a probabilistic construction of a family of linear codes. Starting from a fixed code $C_0$, the procedure constructs $\{C_i\}_{i\in\NN}$ such that the length and dimension of $C_i$ are double the ones from $C_{i-1}$. Thus, if $k_i$ and $n_i$ respectively denote the dimension and length of $C_i$,  we have that $k_i=2^i k_0$ and $n_i=2^i n_0$. For each $i$, a vector $\vec{t}_{i-1}$ is chosen uniformly at random in $(\FF^*)^{n_{i-1}}$(once and for all). The encoding function $\Enc_i$ with code $C_i$ is then defined from $\Enc_{i-1}$ and $\vec{t}_{i-1}$ as described in~\figref{fig:basefoldenc}.\\

\begin{Boxfig}[ht!]{Encoding algorithm for $C_i$ in Basefold}{fig:basefoldenc}{\underline{Encoding algorithm for $C_i$ in Basefold}}
	Recursive encoding function $\Enc_i$ for messages in $\FF^{k_i}$, with $k_i=2^i k_0$.
	It requires to have fixed a $[n_0,k_0]$-linear code $C_0$ with generator matrix $G_0\in\FF^{k_0\times n_0}$, and for each $j<i$, a vector $\vec{t}_j\in(\FF^*)^{n_j}$ (where $n_j=2^j n_0$).

	\begin{itemize}
		\item If $i=0$, then $\Enc_0(\vm)=\vm G_0\in\FF^{n_0}$
		\item Else if $i>0$ the encoding of $\vm=(\vm_L,\vm_R)\in\FF^{k_i}$ where $\vm_L,\vm_R\in\FF^{k_{i-1}}$ is given by

		$$
		\begin{array}{lll}
			\Enc_i(\vm)&=&\ \ \\
			&&(\Enc_{i-1}(\vm_L),\Enc_{i-1}(\vm_L))\  \\
			&+&(\Enc_{i-1}(\vm_R), \Enc_{i-1}(\vm_R))\ \circ\ (\vt_{i-1}, - \vt_{i-1})
		\end{array}
		$$
		where $\circ$ denotes the componentwise product.

		In other words $\Enc_i(\vm)=\vm G_i$, where the generator matrix $G_i$ of $C_i$ is
		$$G_i=\left(
		\begin{matrix}
			G_{i-1} 			& \ \ \ \ & G_{i-1}\\
			G_{i-1}\cdot T_{i-1}& \ \ \ \  & G_{i-1}\cdot -T_{i-1}
		\end{matrix}
		\right)
		$$

		where $T_{i-1}\in\FF^{n_{i-1}\times n_{i-1}}$ is the diagonal matrix with diagonal $\vt_{i-1}$.

	\end{itemize}

\end{Boxfig}

\subsection{The BaseFold Commitment and Evaluation}

In order to commit to a multilinear polynomial $f$ in $\FF[X_1,\dots,X_\ell]$, we set up a family of codes $\{C_i\}_{i=0}^d$ as above such that $k_0=2^{\ell_0}$ and $d=\ell-\ell_0$. Note that $k_d=2^\ell$, which is the number of coefficients of $f$.
We identify $f=\sum_{\vec{j}\in \{0,1\}^\ell} f_{\vec{j}} \vX^{\vec{j}}$ with the vector $\vec{f}\in\FF^{k_d}$ given by $\vec{f}[\toint(\vec{j})]=f_{\vec{j}}$.

The commitment to $f$ is then $\pi_d$, an oracle to $\Enc_d(\vec{f})$, which in practice will be instantiated with a vector commitment such as a Merkle-tree based commitment.

Rather than describing the protocol $\Eval$ from \cite{C:ZeiCheFis24} for a single opening of a committed polynomial, we already incorporate our additional steps to obtain $\Evalcommon_M$, the protocol that simultaneously proves the correct evaluation of $M$ committed polynomials at the same point; the original Basefold evaluation protocol can be retrieved as the case $M=1$.

Concretely the protocol $\PC.\Evalcommon_M$ is a proof for the following relation:
\begin{align*}
	\mathcal{R}_{\Evalcommon_M}=&\left\{((\vec{z},(y_m)_{m=0}^{M-1},(\pi_{d,f_m})_{m=0}^{M-1}); (f_m)_{m=0}^{M-1}):\right.\\ &\ \pi_{d,f_m} \textrm{ oracle to }\Enc_d(\vec{f}_m) \wedge
	\left.f_m(\vec{z})=y_m, \forall m\in [M] \right\}
\end{align*}

The protocol $\PC.\Evalcommon_M$ can be found in \figref{fig:batchfold_same_point}.

\begin{Boxfig}[ht!]{$\PC.\Evalcommon_M$: Batched opening proof of $M$ evaluations at common point for the Basefold Polynomial Commitment.}{fig:batchfold_same_point}{\underline{$\PC.\Evalcommon_M$: Batched opening proof at common point $\vec{z}$}}
	Inputs: oracles $\pi_{d,f_m}$, evaluations $y_m$, $m\in[M]$, and evaluation point $\vec{z}\in\FF^{\ell}$.
	Proof that $f_m(\vec{z})=y_m$, $m\in [M]$.
	\begin{enumerate}
		\item If $M>1$, the verifier selects $s\in\FF$, sends it to the prover. The prover sets $u(\vec X):=\sum_{m=0}^{M-1} s^{m} f_m (\vec X)$. This implicitly defines $\pi_d:=\pi_{d,u}=\sum_{m=0}^{M-1} s^{m} \pi_{d,f_m}$.
		Else, if $M=1$, define  $\pi_d:=\pi_{d,f_0}$, $u:=f_0$

		\item For $i=d-1,\dots,1$,
		\begin{enumerate}
			\item The verifier selects $\alpha_i\in\FF$ uniformly at random and sends it to the prover.
			\item The prover constructs   $\pi_i=(g_1(\alpha_i),\dots,g_{n_i}(\alpha_i))$ where
			$g_j(X)$ is the unique polynomial of degree $1$ with $g_j(\vec{t}_{i}[j])=\pi_{i+1}[j]$ and $g_j(-\vec{t}_{i}[j])=\pi_{i+1}[j+n_{i}]$.
			\item If $i>0$, the prover sends oracle $\pi_{i}$ to the verifier.
			\item In addition, the prover constructs the univariate polynomial $$h_i(X)=\sum_{\mathbf{b}\in\{0,1\}^{\ell_0+i-1}} u(\mathbf{b},X,\alpha_i,\dots,\alpha_{d-1})\cdot \teq ((\mathbf{b},X,\alpha_i,\dots,\alpha_{d-1}),\mathbf{z}) $$
			and sends $h_i(X)$ to the verifier.
		\end{enumerate}
		\item The verifier selects $\alpha_0\in\FF$ uniformly at random and sends it to the prover.
		\item The prover sends $\vec{m}$, the vector of coefficients of the $\ell_0$-variate polynomial $u(X_1,\dots,X_{\ell_0},\alpha_0,\dots,\alpha_{d-1})$.
		\item The verifier computes  $\pi_0=\Enc_0(\vec{m})$.
		\item The verifier now verifies the validity of the commitment by executing $\reps$ times the following check.
		For $\iota$ in $1,\dots,\reps$:
		\begin{enumerate}
			\item Samples $\mu:=\mu_{\iota}\in[1,\dots,n_{d-1}]$
			\item For each $i=d-1,\dots,0$:
			\begin{enumerate}
				\item Queries $\pi_{i+1}[\mu]$ and $\pi_{i+1}[\mu+n_{i}]$ if they have not been queried before.\\
				If $M>1$, the queries with $i=d-1$ are carried out by querying all  $(\pi_{d,f_m}[\mu],
				\pi_{d,f_m}[\mu+n_i])_{m=0}^{M-1}$ and defining $\pi_{d}[\mu]=\sum_{m=0}^{M-1} s^{m}\pi_{d,f_m}[\mu]$ and analogously with $\pi_{d}[\mu+n_i]$.
				\item Constructs $g_\mu(X)$ from $\pi_{i+1}[\mu]$ and $\pi_{i+1}[\mu+n_{i-1}]$ as above.
				\item Queries $\pi_{i}[\mu]$ and checks $g_\mu(\alpha_i)=^?\pi_{i}[\mu]$.
				\item Updates $\mu$ to $(\mu \bmod n_{i-1})$.
			\end{enumerate}
		\end{enumerate}
		\item The verifier checks $$h_1(\alpha_0)=^? \sum_{\mathbf{b}\in\{0,1\}^{\ell_0}} u(\mathbf{b},\alpha_0,\dots,\alpha_{d-1})\cdot \teq ((\mathbf{b},\alpha_0,\dots,\alpha_{d-1}),\mathbf{z}),$$
		using the received $\vec{m}$ to compute the evaluations of $f$ on the right hand side.
		\item The verifier also checks
		$$h_d(0)+h_d(1)=^?\sum_{m=0}^{M-1} s^{m} y_m$$
		and $$h_i(0)+h_i(1)=^? h_{i+1}(\alpha_i)$$ for $i\in [1,\dots,d-1]$
		\item The proof is accepted if all checks pass.
	\end{enumerate}
\end{Boxfig}

We now prove that the soundness of $\PC.\Evalcommon_M$.

\begin{lemma}[Lemma~\ref{lem:soundness_evalcom} restated]\label{lem:soundness_evalcom_restated}
	Fix a finite field $\FF$ an d let  $\{C_i\}_{i=0}^d$ linear codes over $\FF$ be  defined as above. If $\mathsf{error}_{\PC.\Evalcommon_M}$ and $\mathsf{error}_{\PC.\Eval}$ denote respectively the soundness error of $\PC.\Evalcommon_M$ and $\PC.\Eval$ then $\mathsf{error}_{\PC.\Evalcommon_M}\leq \frac{M-1}{|\FF|}+\mathsf{error}_{\PC.\Eval}$.
\end{lemma}

\begin{proof}

	We note that steps 2-9 are a proof $\PC.\Eval=\PC.\Evalcommon_1$ for the claim that $(\vec{z},y',\pi_{d,u})\in\mathcal{L}({\mathcal{R}_{\Evalcommon_1}})$
	where $y'=\sum_{m=0}^{M-1} s^{m} y_m$  and $\pi_{d,u}=\sum_{m=0}^{M-1} s^{m} \pi_{d,m}$ where $\pi_{d,m}$ are the oracles in the initial statement.

	Hence it is enough to show that if the received starting claim is false, i.e.   $(\vec{z},(y_m)_{m=0}^{M-1},(\pi_{d,m})_{m=0}^{M-1}) \notin \mathcal{L}({\mathcal{R}_{\Evalcommon_M}})$ then  $(\vec{z},y',\pi_{d,u})\in\mathcal{L}({\mathcal{R}_{\Evalcommon_1}})$ can happen with probability at most $(M-1)/|\FF|$ over the choice of $s$.

	Assume the claim is false, then there is some $m\in [M]$ such that either $\pi_{d,m}$ is not a valid oracle to $\Enc_d(\vec{f}_m)$ for some $\vec{f}_m\in \FF^{k_d}$ or $f_m(\vec{z})\neq y_m$ for the associated $f_m$.

	First suppose that for some $m$, $\pi_{d,f_m}$ is a oracle to a vector $\vec{v}_m\notin C_d$. Consider a parity check matrix $H\in\FF^{(n_d-k_d)\times n_d}$ i.e. a matrix $H$ such that for every $\vec{c}\in\FF^{n_d}$, $\vec{c}\in C_d$ iff $H\vec{c}=0$. Consequently since  $\vec{v}_m\notin C_d$ there is a row $H_i$ of $H$ with $H_i\vec{v}_m\neq 0$. Then the polynomial $h(T)=\sum_{m=0}^{M-1} (H_i\vec{v}_m) \cdot T^{m}$ is nonzero, so it has at most $M-1$ roots and hence $Pr[h(s)=0]\leq (M-1)/|\FF|$  when $s$ is sampled uniformly at random in $\FF$. But if $h(s)\neq 0$, then by linearity $H_i(\sum_{m=0}^{M-1} s^{m} \vec{v_m})\neq 0$ so $\sum_{m=0}^{M-1} s^{m} \vec{v_m}\notin C_d$ and therefore $\pi_{d,u}$ is not an oracle to a codeword in $C_d$, so $(\vec{z},y',\pi_{d,u})\notin\mathcal{L}({\mathcal{R}_{\Evalcommon_1}})$.

	Otherwise for all $m$, $\pi_{d,m}$ is a oracle to some $\Enc_d(\vec{f}_m)$, and clearly $\pi_{d,u}$ is an oracle to $\Enc_d(\vec{u})$ with $u=\sum_{m=0}^{M-1} s^{m}f_m$.
	Since the claim is false, $f_m(\vec{z})\neq y_m$ for some $m$. Then in an analogous way as before, the polynomial $h'(T):=\sum_{m=0}^{M-1} (f_m(\vec{z})-y_m)T^{m}$ is nonzero so $\Pr[h'(s)=0]\leq (M-1)/|\FF|$  when $s$ is sampled uniformly at random in $\FF$, while if $h'(s)\neq 0$ we have $u(\vec{z})=\sum_{m=0}^{M-1} s^{m} f_m(\vec{z})\neq y'$, so again  $(\vec{z},y',\pi_{d,u})\notin\mathcal{L}({\mathcal{R}_{\Evalcommon_1}})$.

\end{proof}

Now we retrieve the soundness result obtained in BaseFold. The soundness error depends on the relative minimum distance $\Delta_{C_i}$ of the codes $C_i$ defined as $\Delta(C_i)=\dmin(C_i)/n_i$

\begin{theorem}[Soundness of Basefold (case $M=1$), from \cite{C:ZeiCheFis24}]
	Let $\{C_i\}_{i=0}^d$, the linear codes constructed above, and $\gamma,\delta\in(0,1)$ satisfy the  conditions\footnote{Note that \cite{C:ZeiCheFis24} states $\Delta_{C_{i+1}} \leq \Delta_{C_i}$ for all $i\in[0,d-1]$ as an additional condition, but this is always satisfied in the stated code construction}
	$\delta < J_{\gamma}(J_\gamma(\Delta_{C_d}))$
	and $3\delta - d\gamma < \Delta_{C_d}$,
	where $J_\gamma(x):=1-\sqrt{1-x(1-\gamma)}$.
	Then the soundness error of $\PC.\Eval$ satisfies $$\mathsf{error}_{\PC.\Eval}\leq\frac{2d}{\gamma^3 |\FF|}+(1-\delta+\gamma d)^\reps $$
\end{theorem}

\begin{corollary}
In the same conditions as above, the soundness error of $\mathsf{error}_{\PC.\Evalcommon_M}$ satisfies
	$$\mathsf{error}_{\PC.\Evalcommon_M}\leq\frac{M-1}{|\FF|}+\frac{2d}{\gamma^3 |\FF|}+(1-\delta+\gamma d)^\reps $$

\end{corollary}

The size of the proof, assuming we instantiate the oracles $\pi_i$ with Merkle commitments to vectors of length $n_i=2^i\cdot n_0$ with hash functions of output size $\kappa$ bits, is: $$\left(2\reps\log n_0(M+1) + 2\reps M d+ \reps(d-1)d \right) \kappa + [3(d-1)+2\reps(M+d-1)+2^{\ell_0}]\log |\FF|$$ bits.
If we consider $\ell_0$ and $n_0$ small, so $d\sim\ell$, and considering that for security parameter
$\lambda$ we need $\log |\FF|,\kappa,\reps = \bigO{\lambda}$, this is
$$\bigO{(M\ell+\ell^2)\lambda^2}$$ communication.
Compare this with naively computing $M$ individual proofs for Basefold. The communication complexity in that case would be
 $$\bigO{M\ell^2\lambda^2}.$$

\subsection{Improvements on the minimum distance estimates of the BaseFold codes}
The security of the Basefold commitment depends on the relative minimum distance $\Delta_{C_i}=d_{\min}(C_i)/n_i$ of the codes $C_i$. We state next a lower bound on the minimum distance of these codes which is a slight improvement with respect to the one stated in \cite{C:ZeiCheFis24}.

\begin{theorem}\label{thm:codebound}
	Let $\lambda\in\mathbb{N}$.
	Define $\{C_i\}_{i\in \NN}$ as above, with the additional assumption that $C_0$ is a MDS code (i.e. $d_{\min} (C_0)=n_0-k_0+1$). For every $d\in\NN$,
	$$\Pr[d_{\min}(C_d)\leq n_d-t_d]\leq d\cdot 2^{-\lambda}$$
	where $t_0=k_0$, and for every $i>0$, $t_i=2{t_{i-1}}+\ell_i$ with
	$$\ell_i=\frac{\lambda +2t_{i-1}(1+\log(\frac{|\FF|}{|\FF|-1}))+0.585 n_i }{\log(|\FF|-1)-1}.$$
	The probability above is over the uniform choices of $\vec{t}_i\in(\FF^*)^{n_i}$, $i=0,\dots,d-1$, in the definition of the codes.

	For $|\FF|\geq 2^{11}$, we can set
	$$\ell_i=\frac{\lambda +2.002 t_{i-1}+0.585 n_i }{\log |\FF|-1.001}.$$
\end{theorem}

For comparison, in \cite[Theorem 2]{C:ZeiCheFis24}, the corresponding $\ell_i$ (for a sufficiently large field) is given as $$\ell_i=\frac{2(i-1)\log n_0+\lambda+2.002 t_{i-1}+0.6n_i}{\log |\FF|-1.001}$$ which is less tight.

In the remaining of this section, we prove this result. For our proof we will restate the results of \cite{C:ZeiCheFis24} using a more code-theoretic notation, that uses the notion of shortened code. This also simplifies somewhat the proofs.

For $n>0$ and $S\subseteq[n]$ we denote $\pi_{S}:\FF^n\rightarrow\FF^{|S|}$ the projection to the coordinates in $S$.
Given $S\in [n]$ we write $\bar{S}:=[n]\setminus S$ for its complementary.
We denote the Hamming weight of vector $\mathbf{v}$ as $\wHam(\mathbf{v})$. Recall that a $[n,k,d]_q$-linear code C is a linear code $C$ over $\FF_q$ of length $n$, dimension $k$ (i.e. cardinality $q^k$) and minimum distance $\dmin(C)=d$.

\begin{definition}[$C$ shortened at $S$]
	Given a code $C\subseteq \FF^n$, the \textit{shortening of $C$ at $S$} (or \textit{$C$ shortened at $S$}) is the code $C^S=\{\pi_{\bar {S}}(\vec c): c\in C, \pi_{S}(\vec c)=0 \}$. I.e. the code that we obtain by restricting to the codewords in $C$ that are $0$ in all the coordinates in $S$ and then removing these $0$-coordinates.
\end{definition}
\begin{lemma}
	If $C$ is a $[n,k,d]_q$-linear code, then if $C^S$ is a $[n-|S|,k_S,d_S]_q$-linear code such that $k_S\leq k$ and, if $C^S\neq\{0\}$, $d_S\geq d$.
\end{lemma}
\begin{proof}
	It is obvious that $C^S$ is of length $n-|S|$, and that $k_S\leq k$ because the amount of codewords can only decrease. If $C^S\neq\{0\}$, there is some nonzero codeword in $\vec w\in C^S$ which means there is a nonzero codeword $\vec c$ in $C$ with $\pi_S(\vec c)=0$, $\pi_{\bar{S}}(\vec c)=\vec w\neq \vec 0$. Then clearly $d\leq w_H(\vec c)=w_H(\pi_{\bar{S}}(\vec c))=w_H(\vec w)$ so indeed $d_S\geq d$.
\end{proof}

\begin{lemma}[Singleton bound]
	For any $[n,k,d]_q$-linear code	$C$, we have $k+d\leq n+1$
\end{lemma}
\begin{proof}
	Well known fact in code theory.
\end{proof}

\begin{definition}[MDS code]
	A MDS code is a $[n,k,d]_q$-linear code	$C$ with $k+d= n+1$.
\end{definition}

\begin{definition}
	Let $S\subseteq [n_{d-1}].$ We define $m_d(S)$ as the set of pair of codewords in $C_{d-1}$ such that they have common zeros exactly in the coordinates of $S$, i.e.
	\begin{align*}
		m_d(S):=\{(c,c')\in (C_{d-1})^{2} : &\ \forall i\in S,\ c[i]=c'[i]=0  \\
		&\ \forall i\notin S,\ c[i]\neq 0 \vee c'[i]\neq 0\}
	\end{align*}

	Note $(C_{d-1})^{2}$ is the disjoint union of $m_d(S)$ for $S\subseteq [n_{d-1}]$, as every pair of codewords belongs to exactly one $m_d(S)$.

\end{definition}

We now offer a proof of Lemma 2 in \cite{C:ZeiCheFis24} that we consider is slightly simpler, using the coding-theory definition and facts above.
\begin{lemma}\label{lem:sizeofmd}[Lemma 2 in \cite{C:ZeiCheFis24}].
	Suppose $d_{min}(C_{d-1})>n_{d-1}-t$ for some $t$. Let $S\subseteq [n_{d-1}]$. Then:
	\begin{itemize}
		\item If $t\leq |S|<n_{d-1}$, then $|m_d(S)|=0$.
		\item If $|S|<t$, then $|m_d(S)|\leq |\FF|^{2t-2|S|}$.
	\end{itemize}
\end{lemma}

\begin{proof}
	First, note $|m_d(S)|\leq (|(C_{d-1})^S|)^2$, where $(C_{d-1})^S$ is $C_{d-1}$ shortened at $S$.  It is not necessarily an equality, because a codeword in $C_{d-1}^S$ may be $0$ in some coordinates of $\bar{S}$.

	Then:
	\begin{itemize}
		\item If $|S|\geq t$, $m_d(S)=0$, unless $S=[n_{d-1}]$ for which $m_d(S)=1$\\
		Indeed, because $d_{min}(C_{d-1})>n_{d-1}-{t}$ there are no nonzero codewords in $C_{d-1}$ with $t$ zeros, and $(\vec{0},\vec{0})\in m_d([n_{d-1}])$.
		\item If $|S|<t$, $m_d(S)\leq |\FF|^{2t-2|S|}$.\\
		Indeed in this case, either $(C_{d-1})^S=\{0\}$ and then $m_d(S)=1$ or $(C_{d-1})^S\neq \{\vec 0\}$. Then $(C_{d-1})^S$ is a code of length $n_{d-1}-|S|$ and distance $d_{min}((C_{d-1})^S)\geq d_{min}(C_{d-1})\geq n_{d-1}-t+1$.\\
		Then, by the Singleton bound,
		$$\dim (C_{d-1})^S\leq (n_{d-1}-|S|)-d_{min}((C_{d-1})^S)+1\leq t-|S|.$$
		Consequently $$m_d(S)\leq |(C_{d-1})^S|^2\leq (|\FF|^{ t-|S|})^2.$$
	\end{itemize}

\end{proof}

Now we work towards the improvement on the lower bounds. The following lemma
shaves off a factor $n_{d-1}$ with respect to Lemma 3 in \cite{C:ZeiCheFis24}. The improvement stems from a better bound on the expression~\eqref{eq:bin}. Up to that point the proof is as in \cite{C:ZeiCheFis24}.

\begin{lemma}\label{lem:weightofwordinmd} [Improvement on Lemma 3 in \cite{C:ZeiCheFis24}]
	Let $S\subseteq [n_{d-1}]$, $(\vec{c},\vec{c'})\in m_d(S)$.
	Let $\hat{\vec{c}}=(\vec{c}+\vec{c'}\circ\vec{t}_{d-1},\vec{c}-\vec{c'}\circ\vec{t}_{d-1})$ where $\vec{t}_{d-1}$ is chosen uniformly at random in $(\FF\setminus\{0\})^{n_{d-1}}$.
	Then for $0<t\leq n_d$, $$\Pr[w_H(\hat{\vec{c}})\leq n_d-t]< 2^{n_{d-1}-|S|}\cdot \left(\frac{2}{|\FF|-1}\right)^{t-2|S|}$$
	where the probability is over the choice of $\vec{t}_{d-1}$
\end{lemma}

\begin{proof}
	For all $i\in [n_{d-1}\setminus S]$, either $\vec{c}[i]\neq 0$ or $\vec{c'}[i]\neq 0$.
	For the sake of notation let $a_i:=\vec{c}[i]$, $b_i=\vec{c'}[i]$, $u_i=\vec{t}_{d-1}[i]$

	Then the Hamming weight of the pair of coordinates
	$(\hat{\vec{c}}[i], \hat{\vec{c}}[n_{d-1}+i])$ is
	$$\wHam(a_i+b_i\cdot u_i, a_i-b_i\cdot u_i)=
	\begin{cases}
		1 & \textrm{ if } a_i\neq 0 \textrm{ and } u_i=\pm b_i/a_i\\
		2 & \textrm{ otherwise}
	\end{cases}$$
	Note that the weight cannot be 0 as this would imply $u_i=0$, but $\vec{t}_{d-1}$ is sampled from $(\FF\setminus\{0\})^{n_{d-1}}$.

	The case 1 above happens with probability at most $\frac{2}{|\FF|-1}$ over the random choice of $u_i$, and then case 2 happens with probability at least $1-\frac{2}{|\FF|-1}$.
	Since all $u_i$ for $i\notin S$ are chosen independently, we have that the Hamming weight
	$\wHam(\vec{\hat{c}})= \sum_{i\notin S} \wHam (\hat{\vec{c}}[i], \hat{\vec{c}}[n_{d-1}+i])$ is distributed according to a sum of  independently distributed Bernouilli variables $X_i$, with $\Pr[X_i=1]\leq \frac{2}{|\FF|-1}, \Pr[X_i=2]=1-\Pr[X_i=1]\geq \frac{2}{|\FF|-1}$. To simplify, and since we are upper bounding the probability of $\wHam(\vec{\hat{c}})$ being smaller than some quantity, we define $X'_i$ the Bernouilli variable with exact probabilities  $\Pr[X'_i=1]= \frac{2}{|\FF|-1}$, $\Pr[X'_i=2]= 1-\frac{2}{|\FF|-1}$
	and we have
	$$\Pr[w_H(\hat{\vec{c}})\leq n_d-t]\leq \Pr[\sum_{i\notin S} X'_i\leq n_d-t.]$$
	Moreover, we can write $X''_i=2-X'_i$ (hence   $\Pr[X''_i=1]= \frac{2}{|\FF|-1}$, $\Pr[X''_i=0]= 1-\frac{2}{|\FF|-1}$ ) and we have
	$\sum_{i\notin S} X'_i=2(n_{d-1}-|S|)-\sum_{i\notin S} X''_i$, so $\sum_{i\notin S} X'_i\leq n_d-t$ if and only if $\sum_{i\notin S} X''_i\geq 2(n_{d-1}-|S|)-(n_d-t)=t-2|S|$.
	Hence we want an upper bound for $\Pr[\sum_{i\notin S} X''_i\geq t-2|S|]$.
	Now
	\begin{align}
		\Pr[\sum_{i\notin S} X''_i\geq t-2|S|]&= \sum_{j=t-2|S|}^{n_{d-1}-|S|} \Pr[\sum_{i\notin S} X''_i=j]\\
		\label{eq:bin}
		&^=\sum_{j=t-2|S|}^{n_{d-1}-|S|} \binom{n_{d-1}-|S|}{j} \left(\frac{2}{|\FF|-1}\right)^j\left(1-\frac{2}{|\FF|-1}\right)^{n_{d-1}-|S|-j}
	\end{align}
	We now note that for any positive integers $0<a<b$, any real $0<p<1$, $$\sum_{j=a}^b \binom{b}{j}p^j(1-p)^{b-j}\leq \sum_{j=a}^b 2^{b} p^j(1-p)^{b-j}=2^b p^a \sum_{k=0}^{b-a} p^k (1-p)^{b-a-k}\leq 2^b p^a$$ since $$\sum_{k=0}^{b-a} p^k (1-p)^{b-a-k}\leq \sum_{k=0}^{b-a} \binom{b-a}{k} p^k (1-p)^{b-a-k},$$ which is the binomial expansion of $(p+(1-p))^{b-a}=1$

	Therefore
	$$\Pr[\sum_{i\notin S} X''_i\geq t-2|S|]\leq 2^{n_{d-1}-|S|} \left(\frac{2}{|\FF|-1}\right)^{t-2|S|}$$ which gives the result.
\end{proof}

In the following lemma again we can obtain further improvements by using the binomial theorem to bound some of the expressions of the proof (e.g. \eqref{eq:dminCd6} and below)

\begin{lemma}\label{lem:distanceonestep} [Improvement on Lemma 4 in \cite{C:ZeiCheFis24}]
	Suppose $\dmin(C_{d-1})\leq n_{d-1}-t_{d-1}$.
	Let $t_d=2t_{d-1}+\ell_d$ 	where $$\ell_d=\frac{\lambda +2t_{d-1}(1+\log(\frac{|\FF|}{|\FF|-1}))+0.585 n_d }{\log(|\FF|-1)-1}$$
	then
	$$\Pr[\dmin(C_d)\leq n_d-t_d]\leq 2^{-\lambda}$$

\end{lemma}
\begin{proof}
	Note that every $\vec{\hat{c}}\in C_d$ is of the form $\hat{\vec{c}}=(\vec{c}+\vec{c'}\circ\vec{t}_{d-1},\vec{c}-\vec{c'}\circ\vec{t}_{d-1})$ for a unique $(\vec{c},\vec{c'})\in C^{d-1}$. We use the shorthand $(\vec{c},\vec{c'})=F(\hat{\vec{c}})$   Moreover,  $F(\hat{\vec{c}})\in m_d(S)$ for a unique $S$.

	Then
	\begin{align}
		\Pr[\dmin(C_d)\leq n_d-t_d] & \leq \sum_{\vec{\hat{c}}\in C_d\setminus\{0\}} \Pr[w_H(\vec{\hat{c}})\leq n_d-t_d]\\
		& = \sum_{S\subsetneq [n_{d-1}]} \sum_{\substack{\vec{\hat{c}}\in C_d:\\ f(\vec{\hat{c}})\in m_d(S)}} \Pr[w_H(\vec{\hat{c}})\leq n_d-t_d]\label{eq:dminCd2}\\
		&\leq \sum_{S\subsetneq [n_{d-1}]} |m_d(S)|\cdot 2^{n_{d-1}-|S|}\left(\frac{2}{|\FF|-1}\right)^{t_d-2|S|}\label{eq:dminCd3}\\
		&\leq\sum_{j=0}^{t_{d-1}-1} |\FF|^{2t_{d-1}-2j} \binom{n_{d-1}}{j} 2^{n_{d-1}-j}\left(\frac{2}{|\FF|-1}\right)^{t_d-2j}\label{eq:dminCd4}	\\
		&=\sum_{j=0}^{t_{d-1}-1}\left(\frac{|\FF|}{|\FF|-1}\right)^{2t_{d-1}-2j}\binom{n_{d-1}}{j} 2^{n_{d-1}+2t_{d-1}-3j}\left(\frac{2}{|\FF|-1}\right)^{\ell_d}\label{eq:dminCd5}.
	\end{align}
	where we have used Lemma~\ref{lem:weightofwordinmd} for the inequality \eqref{eq:dminCd2}$\leq$\eqref{eq:dminCd3}, Lemma~\ref{lem:sizeofmd} for the inequality \eqref{eq:dminCd3}$\leq$\eqref{eq:dminCd4} and $t_d=2t_{d-1}+\ell_{d}$ and rearranging terms for the last equality

	Now \eqref{eq:dminCd5} can be rearranged as
	\begin{equation}\label{eq:dminCd6}
		2^{n_{d-1}+2t_{d-1}}\cdot \left(\frac{|\FF|}{|\FF|-1}\right)^{2t_{d-1}}\cdot \left(\frac{2}{|\FF|-1}\right)^{\ell_d}\cdot \sum_{j=0}^{t_{d-1}-1}\binom{n_{d-1}}{j} \left(1-\frac{1}{|\FF|}\right)^{2j}2^{-3j}
	\end{equation}

	We focus on the sum, upper bounding $(1-\frac{1}{|\FF|})$ by $1$, which  we can write as $1^{n_{d-1}-j}$ by reasons that will become clear next.
	We get
	$$\sum_{j=0}^{t_{d-1}-1}\binom{n_{d-1}}{j} \left(1-\frac{1}{|\FF|}\right)^{2j} 2^{-3j}\leq \sum_{j=0}^{n_{d-1}} \binom{n_{d-1}}{j}\cdot \left(\frac{1}{8}\right)^j 1^{n_{d-1}-j} $$
	and we get a binomial sum in the right hand side which equals $$\left(\frac{1}{8}+1\right)^{n_{d-1}}=\left(\frac{9}{8}\right)^{n_{d-1}} $$

	Coming back to the full expression in \eqref{eq:dminCd6} (and merging the terms $2^{n_{d-1}}$ and the term $\left(\frac{9}{8}\right)^{n_{d-1}}$ we have obtained as bound for the sum),
	we get
	$$\Pr[\dmin(C_d)\leq n_d-t_d]\leq 2^{2t_{d-1}}\cdot \left(\frac{|\FF|}{|\FF|-1}\right)^{2t_{d-1}}\cdot \left(\frac{2}{|\FF|-1}\right)^{\ell_d}\cdot \l
	\left(\frac{9}{4}\right)^{n_{d-1}}$$

	For this to be smaller than $2^{-\lambda}$
	we need to set $\ell_d$ as

	$$\ell_d\geq \frac{\lambda +2t_{d-1}(1+\log(\frac{|\FF|}{|\FF|-1})+ n_{d-1}\log(9/4) }{\log(\frac{|\FF|-1}{2})}$$
	Using that $\log(9/4)= 1.1699...\leq 1.17$ and that $n_{d-1}=n_d/2$, we have that it is enough to set

	$$\ell_d= \frac{\lambda +2t_{d-1}(1+\log(\frac{|\FF|}{|\FF|-1}))+0.585 n_d }{\log(|\FF|-1)-1}$$
\end{proof}

The final bound can be now be obtained as in \cite{C:ZeiCheFis24}
\begin{proof}[Proof of Theorem~\ref{thm:codebound}]

	Induction.
	For $d=0$ it comes from the assumption that $C_0$ is an MDS code. In this case $d_{min}(C_0)=n_0-k_0+1>n_0-t_0$, so clearly the stated probability is $0$.\\

	If $d>0$, we assume the property is true for every $i\leq d-1$, i.e. $\Pr[d_{min}(C_i)\leq n_i-t_i]\leq i\cdot 2^{-\lambda}$.

	Let $E_i$ be the event ``$d_{min}(C_i)\leq n_i-t_i$''. From the above we are assuming $Pr[E_i]\leq i\cdot 2^{-\lambda}$ for $i\leq d-1$.

	Now note
	\begin{align*}
		\Pr[E_d]&=\Pr[E_d|E_{d-1}]\cdot \Pr [ E_{d-1}]+\Pr[E_d|\neg E_{d-1}]\cdot \Pr [\neg E_{d-1}]\\ &\leq  \Pr [ E_{d-1}]+ \Pr[E_d|\neg E_{d-1}]\\
		&\leq (d-1)\cdot 2^{-\lambda}+	\Pr[E_d|\neg E_{d-1}]
	\end{align*}
	But $\Pr[E_d|\neg E_{d-1}]$ is precisely the probability analyzed in Lemma~\ref{lem:distanceonestep} and it is at most $ 2^{-\lambda}$.
	Hence $\Pr[E_d]\leq d\cdot 2^{-\lambda}$ and we are done.

	Finally for $|\FF|\geq 2^{11}$, $\log(|\FF|)-\log(|\FF|-1)\leq 0.001$, which implies both an upper bound $\log(\frac{|\FF|}{|\FF|-1})\leq 0.001$ for the expression in the enumerator and a lower bound $\log(|\FF|-1)-1\geq \log |\FF|-1.001$ for the one in the denominator.

\end{proof}

%% file: extra-results.tex
\section{Implementation and Evaluation additional details}
\label{sec:exp-results}
\subsection{Neural Networks}
We evaluate the efficiency of our method on a generic neural networks architecture. The aim of our
implementation is to demonstrate the feasability of our approach, and not to optimise the machine
learning aspect of our solution. For this reason, we do not focus on a particular dataset. We note
however, that the neural network architecture that we use could be used, for example, for the income
prediction dataset~\cite{income} or the Wisconsin breast
cancer~\cite{wisconsin}
one.

We focus on a small neural network for a binary classification task, and make use of the following
architecture.

\begin{description}
	\label{des:nn}
	\item \textbf{Fully connected layer: } this layer connects the input nodes to the output nodes. Here,
	we
	have 30 inputs for $n_0$ outputs;
	\item \textbf{Square activation: } this is an activation layer; the function $f(x) = x^2$ is applied to all $n_0$
	input nodes;
	\item \textbf{Fully connected layer: } another fully connected layer, connecting $n_0$ inputs to
	$n_1$ outputs;
	\item \textbf{Square activation: } this activation layer applies $f(x) = x^2$ to all $n_1$
	inputs;
	\item \textbf{Fully connected layer: } the final fully connected layer, taking $n_1$ inputs into 2
	outputs, corresponding to the two classes.
\end{description}

\paragraph{CKKS constraints}
We use the following CKKS parameters for all our experiments. We let the ring degree $N=2^{13}$,
$L=2$ and $\log_2(p_i) = 45$, for all the $p_i$ in the modulus chain, so that our top modulus
is the product of three primes $q = q_0 = p_0p_1p_2$. Given these parameters, we
choose the splitting degree $d=4$ to achieve soundness. Both the secret key and the error
distributions
are Gaussians of width $\sigma = 3.2$.
The parameters achieve at least 128 bits of security, as estimated by the lattice
estimator~\cite{estimator}.

\subsection{Additional experimental results}\label{app:additional_results}

Table~\ref{tab:res_range_proof} shows the complete measurements for our range proofs. Column ``Other'' gathers the rest of the prover: the digit decomposition, the partial evaluations $\tilde\psi'_\eta$, the counters and the fingerprints of the lookup. The first-layer quotients, which are split into $c = 8$ chunks, are timed with the measurement for $c = 4$ at the same $\ell^*$, i.e., the same number of chunks and the same table.

\begin{table}[h]
	\centering
	\caption{Timing performance of our range proofs: the optimized variant $\pioprangeopt$ over $\FF_p$ (208-bit $p$, $(\sigma, m, \kappa) = (2^{11}, 15, 20)$), measured.}
	\label{tab:res_range_proof}
	\begin{tabular}{@{\hskip 8pt}c@{\hskip 8pt}|@{\hskip 8pt}c@{\hskip 12pt}c@{\hskip 12pt}c@{\hskip 12pt}c@{\hskip 12pt}c@{\hskip 8pt}}
		\hline
		Model $(n_0, n_1)$
		& $\mathtt{SumCheck}$
		& $\mathtt{Com}$
		& $\mathtt{Eval}$
		& Other
		& Total \\
		\hline
		$(128, 64)$ & 13s & 3m47s & 1m30s & 24s & 5m54s \\
		$(64, 16)$ & 6s & 1m47s & 43s & 11s & 2m47s \\
		$(32, 16)$ & 3s & 52s & 22s & 6s & 1m24s \\
		\hline
	\end{tabular}
\end{table}
\begin{landscape}
	\begin{table}[h]
		\centering
		\renewcommand{\arraystretch}{2.5}
		\begin{tabular}{|c|c|c|c|c|c|c|c|c|c|c|}%
			\hline
			& & & & & & &   & \multicolumn{2}{|c|}{Non-zero} & \\
			\cline{9-10}
			$(n_0, n_1)$&$\pXmult$ & $\tcopycon$ & $\Vi[3,det]$ & $\Vi[3,copy]$ &
			$\ceil{\log(D)}$ & $\Vi[3,BD]$ & $\Vi[3, quot]$ & $\pXmult$ &
			$\tcopycon$	 & $\#\tcopycon'$ variables \\ \hline
			$(128, 64)$ & 768 & 24320& 12160 & 388 & 2 & 11 &
			10 & 10 & 15 & 34 \\ \hline
			$(64, 16)$ &320 & 5952 & 2976 & 164 & 2 & 10 &
			9 & 9 & 13 & 30 \\ \hline
			$(32, 16)$ & 192 &3008 & 1504 & 100 & 2 & 9 &
			8 & 8 & 12 & 27 \\ \hline
		\end{tabular}
		\caption{Circuit sizes for the neural network architectures considered. The entries in the columns
		$(n_0, n_1)$ represent the sizes of the hidden layers. The entries in the columns $\pXmult$
		and $\tcopycon$ are reported in concrete values.
		The columns $\Vi[3,BD]$ and $\Vi[3, quot]$ report the number of PC variables needed to
		represent the wires as multilinear polynomials.
		Entries in the columns ``Non-zero'' are reported in $\log_2$ values. }
		\label{tab:full-results}
	\end{table}
\end{landscape}